\documentclass[
aps,
a4paper,
longbibliography,
nofootinbib
]{revtex4-2}
\usepackage{amssymb}
\usepackage{amsmath,amssymb,graphicx,color,epsfig}
\usepackage{pstricks,float}
\usepackage[toc,page]{appendix}
\usepackage{tensor}
\usepackage{graphicx}
\usepackage{tabularx}
\usepackage{caption}
\usepackage{subcaption}
\usepackage{xcolor}

\usepackage[
colorlinks=true,
citecolor=blue,
linkcolor=blue,
urlcolor=blue
]{hyperref}
\begin{document}

\title{Probing Two Dark Dimensions through Primordial Black Holes, Gravitational Waves, and Colliders}
\author{Waqas Ahmed$^{1}$\footnote{E-mail: \texttt{\href{mailto:waqasmit@hbpu.edu.cn}{waqasmit@hbpu.edu.cn}}},
George K. Leontaris$^{2}$\footnote{E-mail: \texttt{\href{mailto:mjunaid@ualberta.ca}{leonta@uoi.gr}}} 
}

\affiliation{$^1$ Center for Fundamental Physics, School of Artificial Intelligence, Hubei Polytechnic University, Huangshi 435003, China.\\
$^2$ Physics Department, University of Ioannina, 45110, Ioannina, Greece.}

\begin{abstract}

We study primordial-black-hole (PBH) dark matter in the two-dark-dimensions (2DD) framework, a six-dimensional brane-world scenario with two compact extra dimensions and a fundamental gravity scale of order $10\,\mathrm{TeV}$. We calculate the evolution of higher-dimensional PBHs including the recently proposed quantum-gravitational memory-burden effect. For a memory exponent $p=2$, the evaporation rate is strongly suppressed, allowing PBHs with initial masses as small as $\sim10^{-3}\,\mathrm{g}$ to survive until the present epoch. Consequently, PBHs can account for the observed dark matter over a mass range extending from $10^{-3}\,\mathrm{g}$ to $10^{21}\,\mathrm{g}$. We further compute the stochastic gravitational-wave background generated at second order by the primordial curvature perturbations responsible for PBH formation. We show that the conventional four-dimensional formalism for scalar-induced gravitational waves remains applicable throughout the mass range accessible to current and future gravitational wave experiments. The resulting signals can be probed by LISA, DECIGO, and pulsar timing arrays. Using Fisher forecasts, we find that these observations can constrain the PBH mass, dark-matter fraction, and width of the primordial curvature spectrum with high precision. The low fundamental gravity scale of the 2DD framework also permits the production of microscopic black holes at future high-energy colliders. Their decay signatures, together with gravitational-wave measurements, provide complementary tests of higher-dimensional gravity, the memory-burden mechanism, and primordial-black-hole dark matter.

\end{abstract}

\maketitle

\section{Introduction}
\label{sec:intro}

The large hierarchy between the electroweak scale,
$M_{\rm EW}\sim 100~{\rm GeV}$, and the four-dimensional
Planck scale, $M_{\rm Pl}\sim 10^{18}~{\rm GeV}$, remains one of the
central puzzles in fundamental physics.  A well-known way to address
this problem is provided by large-extra-dimensional scenarios
\cite{Arkani-Hamed:1998jmv,Antoniadis:1998ig,Arkani-Hamed:1998sfv}, in which the Standard
Model (SM) fields are localized on a three-brane while gravity
propagates in a higher-dimensional bulk.  If the compact space has
volume $\mathcal V_{\mathfrak n}$, the fundamental quantum-gravity scale
$M_*$ is related to the four-dimensional Planck scale by
\begin{equation}
M_{\rm Pl}^{2}=M_*^{2+\mathfrak n}\mathcal V_{\mathfrak n}.
\end{equation}
For sufficiently large compact dimensions, $M_*$ can be lowered to the
multi-TeV range, thereby softening the electroweak hierarchy problem.

At the same time, the Swampland programme
\cite{Vafa:2005ui,Ooguri:2006in} (see also reviews \cite{Palti:2019pca,vanBeest:2021lhn,Agmon:2022thq,Anchordoqui:2026hys}) provides a framework for
identifying low-energy effective field theories that may consistently
couple to quantum gravity.  One of its striking implications is the
Dark Dimension proposal \cite{Montero:2022prj}, in which the observed
smallness of the cosmological constant is related to the existence of
a large compact dimension.  The corresponding Kaluza--Klein (KK) scale
is naturally very light, which can have potentially important consequences
for cosmology, neutrino physics, and primordial black holes (PBHs)
\cite{Anchordoqui:2024akj,Leontaris:2025piz,Ahmed:2026pjd}.

Recently, this idea was extended to the case of two compact dark
dimensions, $\mathfrak n=2$ \cite{Anchordoqui:2025nmb,Leontaris:2026kvu}.
\footnote{For a recent review, see Ref.~\cite{Anchordoqui:2026hys}.}
In this six-dimensional (6d) two-dark-dimensions framework, the
fundamental gravity scale is reduced to
\begin{equation}
M_*\sim 10~{\rm TeV},
\end{equation}
making the scenario potentially testable at future high-energy
colliders such as the Future Circular Collider (FCC)
\cite{FCC:2025lpp, FCC:2018vvp, Benedikt:2022kan}.  For a square two-torus with compactification
length $L=2\pi R$, the reduced Planck mass satisfies
\begin{equation}
M_{\rm Pl}^2=M_*^4L^2,
\end{equation}
and the characteristic KK mass gap is
\begin{equation}
\Delta m_{\rm KK}\simeq \frac{1}{L}
=\frac{M_*^2}{M_{\rm Pl}}
\simeq 3.1\times10^{-2}~{\rm eV}
\left(\frac{M_*}{8.7~{\rm TeV}}\right)^2 .
\end{equation}
Equivalently, for $M_*\simeq10~{\rm TeV}$ the gap is of order
$4\times10^{-2}~{\rm eV}$, close to the atmospheric neutrino mass scale
\cite{King:2014nza}.  Thus the two-dark-dimensions scenario links the
hierarchy problem, the dark-energy scale, and neutrino physics while
remaining phenomenologically testable.

Primordial black holes, formed through the gravitational collapse of
large primordial curvature perturbations
\cite{Carr:1974nx,Hawking:1971ei}, have long been studied as possible
dark-matter candidates.  In standard four-dimensional cosmology, PBHs
lighter than approximately $5\times10^{14}~{\rm g}$ evaporate before the
present epoch, while heavier PBHs are strongly constrained by
microlensing, gamma-ray observations, and cosmic microwave background
energy-injection bounds \cite{Carr:2020gox}. However, in higher-dimensional
theories, the evaporation law is modified.  In the five dark
dimensions, rotating PBHs may survive to masses as low as 
$M\sim10^{10}~{\rm g}$ \cite{Leontaris:2025piz}.  In six dimensions the
effect is even more striking. As shown in Ref.~\cite{Leontaris:2026kvu}
the 6d PBHs evaporate more slowly, and when the quantum-gravitational
memory-burden effect \cite{Dvali:2011aa,Dvali:2020wft} is included with
exponent $p=2$, the lifetime is enhanced enough to allow PBHs as light
as
\begin{equation}
M_{\rm PBH}\sim10^{-3}~{\rm g},
\end{equation}
to survive until today.  This opens a broad dark-matter window,
\begin{equation}
10^{-3}~{\rm g}\lesssim M_{\rm PBH}\lesssim10^{21}~{\rm g},
\end{equation}
subject to the usual cosmological and astrophysical constraints.

The same primordial curvature perturbations responsible for PBH
formation inevitably generate a stochastic background of gravitational
waves at second order in cosmological perturbation theory
\cite{Ananda:2006af,Baumann:2007zm}.  These scalar-induced
gravitational waves (SIGWs) have been extensively studied in the
four-dimensional setting \cite{Kohri:2024qpd,Inomata:2018epa} and
provide an independent probe of the PBH formation mechanism. 
%
%
In the present six dimensional scenario, however, one must distinguish carefully between two regimes. The relevant comparison is between two fundamental scales:
\begin{enumerate}
    \item The physical momentum of the scalar mode at horizon re-entry, which is equivalent to the Hubble scale at PBH formation,
    \begin{equation}
        H_{\rm form} \simeq \frac{4 \pi\gamma M_{\rm Pl}^{2}}{M_{\rm PBH}}\,,
        \qquad \gamma \simeq 0.2\,,
    \end{equation}
    where \(M_{\rm PBH}\) is the PBH mass and $\gamma$ is the efficiency factor for gravitational collapse.
    
    \item The KK mass gap, set by the compactification radius $R$ of the extra dimensions:
    \begin{equation}
        m_{\rm KK} \equiv \frac{1}{2\pi R}\,.
    \end{equation}
\end{enumerate}

This threshold determines whether KK modes can be excited. The boundary between the two regimes occurs when these scales coincide, $H_{\rm form} = m_{\rm KK}$. Solving for the corresponding PBH mass defines the critical mass:
\begin{equation}
    M_{\rm KK} \equiv \frac{\gamma M_{\rm Pl}^{2}}{2m_{\rm KK}} \simeq 8.5 \times 10^{23}~\mathrm{g}~.
\end{equation}
This yields the following physical picture:

\begin{itemize}
\item \textbf{Regime I:} $M_{\rm PBH}\gtrsim M_{\rm KK}$ (heavy PBHs).  In this
case $H_{\rm form}\lesssim m_{\rm KK}$, massive KK gravitons are not
kinematically excited, the extra dimensions are not probed, and the usual four-dimensional SIGW formalism is
a rigorous effective description.

\item \textbf{Regime II:} $M_{\rm PBH}\lesssim M_{\rm KK}$ (light PBHs).  In this
case $H_{\rm form}\gtrsim m_{\rm KK}$, the KK tower can be excited, and
a complete treatment requires the full six-dimensional tensor dynamics.
The standard four-dimensional SIGW result should then be interpreted as
a zero-mode benchmark or exploratory estimate rather than a complete
prediction.
\end{itemize}
Since the full PBH dark-matter window considered in this work,
$10^{-3}~{\rm g}\lesssim M_{\rm PBH}\lesssim10^{21}~{\rm g}$, lies
below $M_{\rm KK}$, the SIGW signals associated with this window belong
to Regime~II.  Accordingly, throughout this paper we use the standard
four-dimensional SIGW spectrum as a benchmark for the massless
zero-mode contribution, while emphasizing that the full six-dimensional
KK contribution is model-dependent and requires a dedicated calculation.

The low fundamental scale $M_*\sim10~{\rm TeV}$ also allows microscopic
black holes to be produced at future colliders  if the partonic
centre-of-mass energy exceeds the fundamental gravity scale
\cite{Giudice:1998ck,Anchordoqui:2003ug}.  Such black holes would decay
through Hawking radiation into high-multiplicity final states \cite{Dimopoulos:2001hw,Giddings:2001bu,Anchordoqui:2003ug,Kanti:2004nr}, with
possible missing energy carried by bulk gravitational modes.  Their
mass temperature relation, event multiplicity, and missing-energy
pattern would provide complementary probes of the number of extra
dimensions and of the fundamental scale of gravity.

In this work we study PBH dark matter, scalar-induced gravitational
waves, and collider signatures in the 6d two-dark-dimensions scenario.
We first review the evaporation of six-dimensional PBHs, including the
memory-burden effect~\cite{Dvali:2011aa,Dvali:2020wft,
Alexandre:2024nuo,Dvali:2024hsb}, and identify the broad dark-matter
window \(10^{-3}~{\rm g}\lesssim M_{\rm PBH}\lesssim10^{21}~{\rm g}\)
for \(p=2\).  We then compute the associated zero-mode benchmark SIGW
spectra for a log-normal primordial curvature power spectrum, following
the standard scalar-induced gravitational-wave formalism
~\cite{Ananda:2006af,Baumann:2007zm,Saito:2008jc,Inomata:2018epa,
Domenech:2021ztg}, and discuss the validity of the four-dimensional
approximation in terms of the Regime~I and Regime~II classification
above.  We also perform Fisher forecasts for future gravitational-wave
observatories~\cite{Cutler:1994ys,Tegmark:1996bz,LISA:2017pwj,
Seto:2001qf,Ando:2010zz,Corbin:2005ny}, treating the benchmark spectrum
as a phenomenological probe of the PBH mass, abundance, and spectral
width.  Finally, we discuss the collider phenomenology of
six-dimensional microscopic black holes, including their production
cross section, Hawking temperature, multiplicity, and possible
missing-energy signatures~\cite{Dimopoulos:2001hw,Giddings:2001bu,
Giudice:2001ce,Yoshino:2002tx,Yoshino:2005hi,Kanti:2004nr,
Kanti:2014dxa}.

The paper is organised as follows. Section~\ref{sec:6D} introduces the two-dark-dimensions framework and the evaporation of six-dimensional PBHs. Section~\ref{sec:memory_burden} discusses the memory-burden effect, near-extremal extensions, and the resulting dark-matter window. Section~\ref{sec:SIGW} presents the scalar-induced gravitational-wave calculation and the Regime~I/Regime~II validity analysis. The Fisher forecast methodology and projected constraints are presented next. Section~\ref{sec:collider} discusses collider signatures of microscopic six-dimensional black holes, mono-jet events, and virtual graviton exchange. We conclude in Sec.~\ref{sec:conclusions}.

\section{The Two$-$Dark$-$Dimensions Scenario}
\label{sec:6D}
We consider a six-dimensional spacetime of the form
$\mathcal M_4\times T^2$, where the two additional spatial dimensions
are flat and compact.  Gravity propagates in the full six-dimensional
bulk, while the Standard Model fields are localized on a four-dimensional
brane \cite{Arkani-Hamed:1998jmv,Antoniadis:1998ig, Arkani-Hamed:1998sfv}.  The six-dimensional gravitational dynamics are governed by
\begin{equation}
S_6=\frac{M_*^4}{2}\int d^4x\,d^2y\sqrt{-G}\,\mathcal R(G),
\label{eq:S6}
\end{equation}
where $M_*$ is the reduced six-dimensional Planck scale and
$\mathcal R(G)$ is the Ricci scalar constructed from the
six-dimensional metric $G_{MN}$. After compactification, the four-dimensional reduced Planck mass is
related to the fundamental scale through the volume of the compact
space,
\begin{equation}
M_{\rm Pl}^2=M_*^4 V_2 ~,
\label{eq:MPl_general}
\end{equation}
where $V_2$ refers to the `volume' of the two extra dimensions and for a square two-torus with geometric radius $R$ and periods
$2\pi R$, one has
\begin{equation}
V_2=(2\pi R)^2 ,
\qquad
M_{\rm Pl}^2=M_*^4(2\pi R)^2 .
\label{eq:MPl}
\end{equation}
With the benchmark value $R\simeq1~\mu{\rm m}$, this gives
\begin{equation}
M_*=
\left(
\frac{M_{\rm Pl}^2}{(2\pi R)^2}
\right)^{1/4}
\simeq 8.7\times10^3~{\rm GeV}
\sim10~{\rm TeV}.
\label{eq:Mstar}
\end{equation}
Thus, the fundamental gravity scale lies close to the energy range
targeted by future high-energy colliders \cite{FCC:2018vvp,Benedikt:2022kan}.

%

Compactification gives rise to a tower of Kaluza--Klein excitations \cite{Kaluza:1921tu,Klein:1926tv,Appelquist:1987nr}.
For a two-torus with periods $2\pi R$, the wavefunctions read
$\exp(i n_i y_i / R)$ and the KK mass spectrum is
\begin{equation}
m_{\vec n}
= \frac{\sqrt{n_1^{2} + n_2^{2}}}{R}\,,
\qquad
\vec n = (n_1,n_2) \in \mathbb{Z}^{2}.
\label{eq:KK_spectrum}
\end{equation}
The lightest non-zero KK mass -- {\it i.e.} the physical mass gap of the tower -- is therefore
\begin{equation}
\Delta m_{\rm KK}
\equiv m_{(1,0)} - m_{(0,0)}
= \frac{1}{R}
= 2\pi\,\frac{M_{*}^{2}}{M_{\rm Pl}}
\simeq 1.97 \times 10^{-1}~\mathrm{eV}
\left(\frac{1~\mu\mathrm{m}}{R}\right).
\label{eq:deltam_standard}
\end{equation}
It is sometimes convenient to introduce the reduced length
\begin{equation}
L \equiv 2\pi R\,,
\end{equation}
so that the Planck/string relation becomes
\begin{equation}
\frac{1}{L}
= \frac{M_{*}^{2}}{M_{\rm Pl}}
\simeq 3.1 \times 10^{-2}~\mathrm{eV}
\left(\frac{M_{*}}{8.7~\mathrm{TeV}}\right)^{2}.
\label{eq:deltam_reduced}
\end{equation}
Note that $1/L$ is \emph{not} the physical KK mass gap; it is related to it by
\begin{equation}
\Delta m_{\rm KK} = \frac{1}{R} = \frac{2\pi}{L}~.
\end{equation}
In what follows we adopt the strict geometric torus convention,
{\it i.e.}~we always take
$\Delta m_{\rm KK} = 1/R$ as the physical gap.

The presence of extra dimensions modifies the properties of black holes
formed in the early Universe.  For black holes whose horizon radius is
smaller than the compactification scale, the appropriate non-rotating
solution is the six-dimensional Schwarzschild--Tangherlini geometry \cite{Tangherlini:1963bw,Emparan:2008eg}.
With normalization in Eq.~(\ref{eq:S6}), the horizon radius satisfies
\begin{equation}
r_h^3
=
\frac{3M_{\rm BH}}{16\pi^2 M_*^4},
\end{equation}
or equivalently
\begin{equation}
r_h
=
k_6\,\frac{1}{M_*}
\left(
\frac{M_{\rm BH}}{M_*}
\right)^{1/3},
\qquad
k_6=
\left(\frac{3}{16\pi^2}\right)^{1/3}.
\label{eq:rh_6d}
\end{equation}
Up to the numerical coefficient $k_6$, this gives the scaling
$r_h\propto M_{\rm BH}^{1/3}$.  The Hawking temperature is \cite{Hawking:1975vcx},
\begin{equation}
T_{\rm BH}
=
\frac{d-3}{4\pi r_h}
=
\frac{3}{4\pi r_h}
\propto M_{\rm BH}^{-1/3},
\label{eq:TBH}
\end{equation}
and the Bekenstein--Hawking entropy scales as \cite{Bekenstein:1973ur},
\begin{equation}
	S_{\rm BH}
	= \frac{A_4}{4G_6}
	\propto M_{*}^{4} r_h^{4}
	\propto \left(\frac{M_{\rm BH}}{M_{*}}\right)^{4/3},
	\label{eq:entropy}
\end{equation}
where $A_4$ is the horizon $4$-area and $G_6 = (8\pi M_{*}^{4})^{-1}$.
The six-dimensional Hawking luminosity is set by the 
Stefan-Boltzmann scaling law for $d=6$: 
\begin{equation}
-\frac{dM}{dt}
=
\frac{C_M}{r_h^2},
\label{eq:Mdot_6d}
\end{equation}
where $C_M$ is a dimensionless coefficient which encodes the greybody factors
and the number of available particle species \cite{Kanti:2004nr,Cardoso:2005vb,Kanti:2014dxa}. Integrating the above, gives the lifetime
\begin{equation}
	\tau_0(M)
	\propto \int^{M} dM'\, r_h^{2}(M')
	\propto M^{5/3} M_{*}^{-8/3}.
	\label{eq:tau_scaling_6d}
\end{equation}

A convenient phenomenological normalization is
\begin{equation}
\tau_0(M)
\simeq
13.7
\left(
\frac{M}{10^8~{\rm g}}
\right)^{5/3}
\left(
\frac{10~{\rm TeV}}{M_*}
\right)^{8/3}
\left(
\frac{C_{M,0}}{C_M}
\right)
{\rm Gyr},
\label{eq:tau6D}
\end{equation}
where $C_{M,0}$ denotes the reference greybody coefficient used to
obtain the quoted numerical normalization.  Thus, in the conventional
six-dimensional Hawking picture, PBHs with masses of order
$M\gtrsim10^8~{\rm g}$ survive until the present epoch, while lighter
PBHs require an additional lifetime-enhancing mechanism. If the PBH is initially rotating, its early evolution is described by
the six-dimensional Myers--Perry solution~\cite{Myers:1986un} with a single spin parameter.
The horizon radius is related to the non-rotating (Schwarzschild) radius 
$r_s$ ( see Eq.~\eqref{eq:rh_6d}) by
\begin{equation}
r_h
=
\frac{r_s}{(1+a_*^2)^{1/3}},
\qquad
r_s
=
k_6\,\frac{1}{M_*}
\left(
\frac{M_{\rm BH}}{M_*}
\right)^{1/3},
\label{eq:rh_rot}
\end{equation}
where $a_*=a/r_h$ is the dimensionless spin parameter.  The corresponding
Hawking temperature for ${\mathfrak n}=2$ extra dimensions is
\begin{equation}
T_{\rm MP}
=
\frac{3+a_*^2}{4\pi r_h(1+a_*^2)}.
\label{eq:TMP}
\end{equation}
For $a_{*} \to 0$ one recovers the Schwarzschild result 
$T_{\rm MP} = 3/(4\pi r_s)$, as required.

Rotation modifies the evaporation through coupled mass and angular
momentum loss rates,
\begin{equation}
\frac{dM}{dt}
=
-\frac{C_M(a_*)}{r_h^2},
\qquad
\frac{dJ}{dt}
=
-\frac{C_J(a_*)}{r_h},
\label{eq:coupled}
\end{equation}
where $C_M(a_*)$ and $C_J(a_*)$ encode spin-dependent greybody factors
and the particle spectrum \cite{Ida:2002ez,Ida:2005ax,Casals:2006xp,Kanti:2014dxa}.  During the spin-down stage the black hole
loses angular momentum efficiently and subsequently approaches a
Schwarzschild--Tangherlini configuration.  The later evolution can then
be treated using the non-rotating scaling laws above.  If the
memory-burden effect becomes important at late times, the evaporation
history separates into three stages: an initial spin-down phase, an
approximately Schwarzschild-like Hawking phase, and a final
memory-burdened phase in which the emission rate is suppressed.

\section{Memory burden, near-extremality, and the dark-matter window} 
\label{sec:memory_burden}

The evaporation history of a six-dimensional primordial black hole
(PBH) depends on whether the black hole is non-rotating, rotating, or
near-extremal.  In this section we clarify these regimes and derive the
corresponding lifetime scalings.  This distinction is important because
ordinary non-extremal PBHs and near-extremal charged PBHs have different
temperature suppressions and therefore different mass dependences.

\subsection{Ordinary six-dimensional evaporation}

We first recall the ordinary non-extremal evaporation regime.  For a
six-dimensional Schwarzschild--Tangherlini PBH with horizon radius below
the compactification scale, the horizon radius is given by
Eq.~\eqref{eq:rh_6d}.  The corresponding entropy scales as
Eq.~\eqref{eq:entropy}, namely
$S_{\rm BH}\propto (M_{\rm BH}/M_*)^{4/3}$.

The standard semiclassical mass-loss rate is given by
Eq.~\eqref{eq:Mdot_6d}.  Since the horizon radius scales as
$r_h\propto M_{\rm BH}^{1/3}$, the evaporation time follows the
six-dimensional scaling derived in Eq.~\eqref{eq:tau_scaling_6d},
\[
\tau_0(M_{\rm BH})\propto M_{\rm BH}^{5/3}M_*^{-8/3}.
\]
Using the numerical normalization in Eq.~\eqref{eq:tau6D}, ordinary
six-dimensional PBHs survive until the present epoch only for masses of
order
\[
M_{\rm BH}\gtrsim 10^8~{\rm g},
\]
up to the uncertainty associated with the greybody coefficient and the
available particle spectrum. 
Lighter PBHs would have 
evaporated completely and therefore cannot constitute a viable 
dark matter candidate unless their lifetime is extended beyond 
the standard Hawking estimate.  This could be achieved, for 
instance, by the memory-burden effect~\cite{Dvali:2011aa,Dvali:2020wft}  or by 
near-extremal temperature suppression, both of which 
would allow the mass range to be extended downward.


If the PBH is initially rotating, the early evolution is instead
described by the six-dimensional Myers--Perry solution.  The relation
between the rotating horizon radius and the non-rotating radius was given
in Eq.~\eqref{eq:rh_rot}, while the spin-dependent temperature and the
coupled mass/angular-momentum loss equations were given in
Eqs.~\eqref{eq:TMP} and \eqref{eq:coupled}.  The spin-down stage is
typically efficient, so the PBH subsequently approaches a nearly
Schwarzschild--Tangherlini configuration.  Consequently, the late-time
ordinary evaporation of an initially rotating PBH is controlled by the
same six-dimensional scaling,
\[
\tau_0(M_{\rm BH})\propto M_{\rm BH}^{5/3},
\]
with only spin-dependent numerical corrections inherited from the early
spin-down phase.

\subsection{Memory-burdened evaporation}

The memory-burden mechanism modifies the late-time evaporation law of a
black hole.  The basic idea is that Hawking emission does not only reduce
the black-hole mass, but also changes the state of a large number of
internal memory modes.  Once the number of emitted Hawking quanta becomes
comparable to the Bekenstein--Hawking entropy, these modes can backreact
on the evaporation process and suppress further emission.

 This effect is  parametrized phenomenologically by multiplying the standard
Hawking rate by an entropy-suppression factor,
\begin{equation}
-\dot M_{\rm BH}^{\rm MB}
\equiv
-\frac{dM_{\rm BH}}{dt}\bigg|_{\rm MB}
=
\frac{1}{S_{\rm BH}^{p}}
\left(
-\frac{dM_{\rm BH}}{dt}
\right)_{\rm std},
\label{eq:Mdot_MB_general}
\end{equation}
where \(p>0\) is a model-dependent memory-burden exponent.  For an
ordinary non-extremal PBH after the spin-down stage, the geometry is
well approximated by the six-dimensional Schwarzschild--Tangherlini
solution.  Using Eqs.~\eqref{eq:Mdot_6d} and \eqref{eq:entropy}, one
therefore obtains
\begin{equation}
-\dot M_{\rm BH}^{\rm MB}
=
\frac{C_M^{(0)}}{r_s^2 S_{\rm BH}^{p}},
\label{eq:Mdot_MB_schw}
\end{equation}
where \(r_s\) is the non-rotating six-dimensional horizon radius and
\(C_M^{(0)}\) denotes the corresponding greybody coefficient.  In this
phenomenological treatment, the greybody coefficient is assumed to be
unchanged by the memory burden; the suppression is encoded entirely in
the factor \(S_{\rm BH}^{-p}\).

A useful way to motivate the onset of the memory-burdened phase is to
introduce a critical occupation number \(N_c\simeq S_{\rm BH}\).  In a
schematic microscopic description, the energy gap of the memory modes may
be written as
\begin{equation}
E_K(n_0)
=
\left(
1-\frac{n_0}{N_c}
\right)^p
\epsilon_K,
\qquad
N_c\simeq S_{\rm BH},
\qquad
\epsilon_K\simeq \sqrt{S_{\rm BH}}\,r_s^{-1}.
\label{eq:memory_gap}
\end{equation}
The memory modes correspond to soft excitations that store information about the black 
hole's quantum state. The occupation number $n_0$ counts how many Hawking quanta have been emitted.
At early times \(n_0\ll N_c\), so the memory modes are energetically
costly to excite and the black hole evaporates approximately at the
standard Hawking rate.  As evaporation proceeds, \(n_0\) increases and
the gap decreases.  When \(n_0\) becomes comparable to \(N_c\), the memory
modes become nearly degenerate and the evaporation rate is suppressed.

The onset of the memory-burdened phase may be characterized by the
fractional mass loss
\begin{equation}
q
\equiv
\frac{\Delta M_{\rm crit}}{M_0}
\sim
\left[
p\,S_{\rm BH}(M_0)
\right]^{-1/p},
\label{eq:qcrit_MB}
\end{equation}
where \(M_0\) is the initial PBH mass.  Thus,
\begin{equation}
p=1:\quad q\sim S_{\rm BH}^{-1},
\qquad
p=2:\quad q\sim \left(2S_{\rm BH}\right)^{-1/2}.
\label{eq:qcrit_special}
\end{equation}
Since macroscopic PBHs have \(S_{\rm BH}\gg1\), the memory-burdened phase
can begin after only a small fractional mass loss.  The subsequent
lifetime, however, depends sensitively on the exponent \(p\).

Using Eq.~\eqref{eq:Mdot_MB_schw}, the memory-burdened lifetime scales as
\begin{equation}
\tau_{\rm MB}^{(p)}(M_{\rm BH})
\sim
\int^{M_{\rm BH}} dM'\,
r_s^2(M')\,S_{\rm BH}^{p}(M').
\label{eq:tau_MB_integral}
\end{equation}
The required mass dependences are already given in
Eqs.~\eqref{eq:rh_6d} and \eqref{eq:entropy}:
\[
r_s^2\propto M_{\rm BH}^{2/3}M_*^{-8/3},
\qquad
S_{\rm BH}^{p}\propto
\left(
\frac{M_{\rm BH}}{M_*}
\right)^{4p/3}.
\]
Therefore,
\begin{align}
\tau_{\rm MB}^{(p)}(M_{\rm BH})
&\propto
\int^{M_{\rm BH}} dM'\,
(M')^{2/3}M_*^{-8/3}
\left[
(M')^{4/3}M_*^{-4/3}
\right]^p
\nonumber\\
&\propto
\frac{1}{M_*}
\left(
\frac{M_{\rm BH}}{M_*}
\right)^{(4p+5)/3}.
\label{eq:tau_MB_scaling}
\end{align}
Equivalently, at fixed \(M_*\),
\begin{equation}
\tau_{\rm MB}^{(p)}(M_{\rm BH})
\propto
M_{\rm BH}^{(4p+5)/3}.
\label{eq:tau_MB_fixedMstar}
\end{equation}
The important special cases are
\begin{equation}
p=0:\quad
\tau\propto M_{\rm BH}^{5/3},
\qquad
p=1:\quad
\tau\propto M_{\rm BH}^{3},
\qquad
p=2:\quad
\tau\propto M_{\rm BH}^{13/3}.
\label{eq:ordinary_MB_exponents}
\end{equation}
The case \(p=0\) reproduces the standard six-dimensional Hawking result,
while \(p=1\) and \(p=2\) correspond to increasingly strong entropy
suppression and therefore to substantially longer lifetimes for light
PBHs.

\subsection{Near-extremal charged PBHs}

The scaling derived above applies to ordinary non-extremal PBHs.  A
different behaviour may arise for near-extremal charged black holes,
because their Hawking temperature is suppressed relative to the
Schwarzschild--Tangherlini temperature of a black hole with the same
mass.  We denote by \(\beta\) the deviation from extremality, with
\(\beta\ll1\) corresponding to a nearly extremal configuration. Compared with the static BH temperature,  the near-extremal temperature can be expressed 
as~\cite{Leontaris:2026kvu}
\begin{equation}
T_{\rm ne}
\simeq
T_0
\sqrt{\frac{\beta}{S_{\rm BH}}},
\label{eq:Tne_scaling_MB}
\end{equation}
where \(T_0\) is the ordinary six-dimensional Schwarzschild--Tangherlini
temperature given in Eq.~\eqref{eq:TBH}, and \(S_{\rm BH}\) is the
corresponding Bekenstein--Hawking entropy given in Eq.~\eqref{eq:entropy}.
Equivalently,
\begin{equation}
\frac{T_0}{T_{\rm ne}}
\simeq
\frac{S_{\rm BH}^{1/2}}{\sqrt{\beta}}.
\label{eq:near_ext_temp_ratio_MB}
\end{equation}

In what follows, we use this temperature suppression as a
phenomenological parametrization of the near-extremal lifetime
enhancement.  Including the memory-burden suppression factor
\(S_{\rm BH}^{p}\), we write
\begin{equation}
\tau_{\rm ne}^{(p)}
\sim
\frac{S_{\rm BH}^{p+1/2}}{\sqrt{\beta}}\,
\tau_0,
\label{eq:tau_ne_general_MB}
\end{equation}
where \(\tau_0\) is the ordinary six-dimensional
Schwarzschild--Tangherlini lifetime defined in Eq.~\eqref{eq:tau6D}.
This expression should be understood as a phenomenological scaling Ansatz
for near-extremal charged PBHs.  It should not be used for ordinary
non-extremal Schwarzschild--Tangherlini PBHs, nor for generic rotating
PBHs after spin-down.

Using the entropy scaling in Eq.~\eqref{eq:entropy} and the ordinary
lifetime scaling in Eq.~\eqref{eq:tau_scaling_6d}, Eq.~\eqref{eq:tau_ne_general_MB}
gives
\begin{align}
\tau_{\rm ne}^{(p)}(M_{\rm BH})
&\propto
\frac{1}{\sqrt{\beta}}
\left(
\frac{M_{\rm BH}}{M_*}
\right)^{\frac{4}{3}\left(p+\frac12\right)}
\frac{1}{M_*}
\left(
\frac{M_{\rm BH}}{M_*}
\right)^{5/3}
\nonumber\\
&\propto
\frac{1}{\sqrt{\beta}}\,
\frac{1}{M_*}
\left(
\frac{M_{\rm BH}}{M_*}
\right)^{\frac{4p+7}{3}} .
\label{eq:tau_ne_scaling_MB}
\end{align}
At fixed \(M_*\), this becomes
\begin{equation}
\tau_{\rm ne}^{(p)}
\propto
\beta^{-1/2}
M_{\rm BH}^{(4p+7)/3}.
\label{eq:tau_ne_mass_scaling_MB}
\end{equation}
Therefore, within this near-extremal parametrization,
\begin{equation}
p=0:\quad
\tau_{\rm ne}\propto
\beta^{-1/2}M_{\rm BH}^{7/3},
\qquad
p=1:\quad
\tau_{\rm ne}\propto
\beta^{-1/2}M_{\rm BH}^{11/3},
\qquad
p=2:\quad
\tau_{\rm ne}\propto
\beta^{-1/2}M_{\rm BH}^{5}.
\label{eq:near_ext_MB_exponents}
\end{equation}

For illustration, we adopt the following benchmark normalization for
near-extremal charged six-dimensional PBHs:
\begin{align}
\tau_{\rm ne}^{(p=0)}(M_{\rm BH})
&\simeq
\frac{53.5}{\sqrt{\beta}}
\left(
\frac{M_{\rm BH}}{{\rm g}}
\right)^{7/3}
{\rm Gyr},
\nonumber\\[1mm]
\tau_{\rm ne}^{(p=1)}(M_{\rm BH})
&\simeq
\frac{2.5\times10^{27}}{\sqrt{\beta}}
\left(
\frac{M_{\rm BH}}{{\rm g}}
\right)^{11/3}
{\rm Gyr},
\nonumber\\[1mm]
\tau_{\rm ne}^{(p=2)}(M_{\rm BH})
&\simeq
\frac{1.1\times10^{54}}{\sqrt{\beta}}
\left(
\frac{M_{\rm BH}}{{\rm g}}
\right)^5
{\rm Gyr}.
\label{eq:tau_ne_numbers_MB}
\end{align}
These numerical expressions are model-dependent and apply only to the
near-extremal charged PBH scenario.  They should not be confused with
the ordinary six-dimensional Hawking lifetime in Eq.~\eqref{eq:tau6D}.

Table~\ref{tab:beta_windows} summarizes the resulting viable PBH
dark-matter windows for representative values of \(\beta\).  As the
black hole approaches extremality, \(\beta\ll1\), the lifetime is
enhanced by the factor \(\beta^{-1/2}\), and the allowed mass range can
extend to smaller masses.  Such an extension, however, requires a
formation mechanism capable of producing a population of nearly
extremal charged PBHs.  In the absence of such a mechanism, the
near-extremal case should be regarded as an illustrative possibility
rather than the conservative baseline.

\begin{table}[h]
\centering
\caption{
Illustrative lower mass bounds for six-dimensional PBH dark matter.
The column labelled \(p=0\) corresponds to the standard Hawking
evaporation case, while the column labelled \(p=2\) includes the
memory-burden enhancement.  The entries for \(p=2\) should be understood
as conservative phenomenological lower limits, including the expected
effect of pre-burden energy injection and BBN constraints.  The parameter
\(\beta\) denotes the deviation from extremality; \(\beta=1\) corresponds
to the ordinary non-extremal limit, while \(\beta\ll1\) corresponds to a
near-extremal charged configuration.
}
\label{tab:beta_windows}
\begin{tabular}{|c|c||c|}
\hline
\(\beta\) &
Standard evaporation &
Memory-burdened evaporation\\
$\downarrow$ &$p=0$&$p=2$\\
\hline
\(1\) &
\(M_{\rm BH}\gtrsim 10^{8}\,{\rm g}\) &
\(M_{\rm BH}\gtrsim 10^{-3}\,{\rm g}\) \\
\(10^{-4}\) &
\(M_{\rm BH}\gtrsim 10^{6}\,{\rm g}\) &
\(M_{\rm BH}\gtrsim 10^{-4}\,{\rm g}\) \\
\(10^{-8}\) &
\(M_{\rm BH}\gtrsim 10^{4}\,{\rm g}\) &
\(M_{\rm BH}\gtrsim 10^{-5}\,{\rm g}\) \\
\hline
\end{tabular}
\end{table}

\begin{figure}[t]
\centering
\includegraphics[width=0.7\columnwidth]{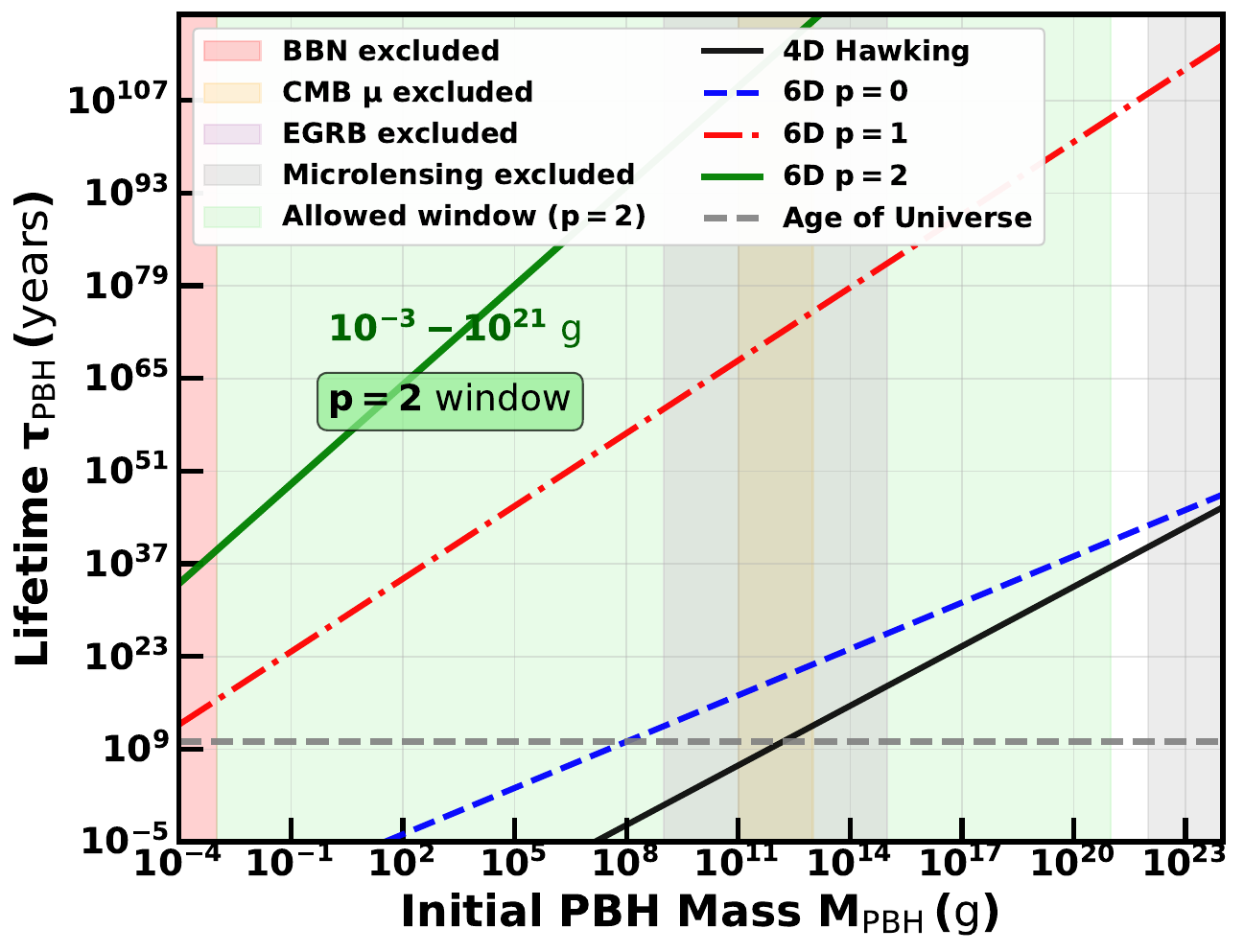}
\caption{
Lifetime of primordial black holes as a function of the initial PBH mass in four and six dimensions. The black solid curve shows the standard 4d Hawking evaporation result, while the blue dashed, red dash-dotted, and green solid curves correspond to the 6d cases with memory-burden parameters $p=0$, $p=1$, and $p=2$, respectively. The horizontal grey dashed line marks the age of the Universe ($13.8\ \mathrm{Gyr}$). Shaded regions indicate observational constraints from BBN (red), CMB $\mu$-distortions (orange), the extragalactic gamma-ray background (purple), and microlensing (grey). The light-green region highlights the viable dark-matter window for the delayed memory-burden scenario ($p=2$), namely $10^{-3}\,\mathrm{g}\lesssim M_{\rm PBH}\lesssim10^{21}\,\mathrm{g}$.
}
\label{fig:lifetime}
\end{figure}

\subsection{Energy-injection constraints and the viable dark-matter window}
\label{sec:DM_window}

The viability of a PBH population as dark matter rests on two basic
requirements.  First, the individual black holes must survive until the
present epoch.  Second, their evaporative emission must not violate
observational bounds on energy injection during BBN, the CMB
\(\mu\)-distortion era, or the later diffuse gamma-ray background.  In
the memory-burdened picture, the evaporation history separates into a
short pre-burden phase, during which the PBH evaporates approximately at
the standard Hawking rate, and a much longer post-burden phase in which
the emission rate is suppressed.  The relevant scaling laws differ for
ordinary non-extremal six-dimensional PBHs and for near-extremal charged
PBHs, which have an additional temperature suppression.

\medskip
\noindent\textbf{Ordinary non-extremal six-dimensional PBHs.}
For an ordinary non-extremal black hole that has spun down to a
Schwarzschild--Tangherlini configuration, the memory-burdened lifetime
was derived in Eq.~\eqref{eq:tau_MB_scaling}.  Equivalently,
\begin{equation}
\tau_{\rm MB}^{(p)}(M_{\rm BH})
\propto
M_{\rm BH}^{(4p+5)/3}
M_*^{-(4p+8)/3}.
\label{eq:tau_ord}
\end{equation}
At fixed \(M_*\), this gives
\begin{equation}
p=0:\quad
\tau\propto M_{\rm BH}^{5/3},
\qquad
p=1:\quad
\tau\propto M_{\rm BH}^{3},
\qquad
p=2:\quad
\tau\propto M_{\rm BH}^{13/3}.
\end{equation}
The post-burden luminosity is suppressed by one entropy factor for each
power of the memory exponent,
\begin{equation}
L_{\rm MB}^{(p)}
\sim
L_{\rm std}\,S_{\rm BH}^{-p},
\label{eq:L_ord}
\end{equation}
where \(L_{\rm std}\) is the standard six-dimensional Hawking luminosity
at the same mass and \(S_{\rm BH}\) is the Bekenstein--Hawking entropy.

\medskip
\noindent\textbf{Near-extremal charged six-dimensional PBHs.}
For a charged black hole close to extremality, the Hawking temperature is
further reduced relative to the Schwarzschild--Tangherlini temperature.
Using the near-extremal parametrization introduced in
Eq.~\eqref{eq:Tne_scaling_MB}, and including the memory-burden factor,
the lifetime scales as derived in Eq.~\eqref{eq:tau_ne_scaling_MB}:
\begin{equation}
\tau_{\rm ne}^{(p)}(M_{\rm BH})
\propto
\beta^{-1/2}
M_{\rm BH}^{(4p+7)/3}
M_*^{-(4p+10)/3}.
\label{eq:tau_ne}
\end{equation}
At fixed \(M_*\), this becomes
\begin{equation}
p=0:\quad
\tau_{\rm ne}\propto
\beta^{-1/2}M_{\rm BH}^{7/3},
\qquad
p=1:\quad
\tau_{\rm ne}\propto
\beta^{-1/2}M_{\rm BH}^{11/3},
\qquad
p=2:\quad
\tau_{\rm ne}\propto
\beta^{-1/2}M_{\rm BH}^{5}.
\end{equation}
Consistently with the same phenomenological temperature and entropy
suppression, the post-burden luminosity may be estimated as
\begin{equation}
L_{\rm ne}^{(p)}
\sim
L_{\rm std}\,
\frac{\sqrt{\beta}}{S_{\rm BH}^{p+1/2}}.
\label{eq:Lne}
\end{equation}
This near-extremal scaling should not be interpreted as the ordinary
non-extremal limit at \(\beta=1\).  It applies only to the charged
near-extremal regime.

\medskip
\noindent\textbf{Quantitative comparison: \(p=1\) versus \(p=2\).}
To estimate the size of the post-burden emission, consider a
representative PBH mass
\[
M_{\rm BH}=10^{11}\,{\rm g}.
\]
Using the reference six-dimensional lifetime normalization in
Eq.~\eqref{eq:tau6D}, with \(M_*\simeq10~{\rm TeV}\), one obtains
\[
\tau_0(10^{11}{\rm g})
\simeq
1.37\times10^6~{\rm Gyr}.
\]
The corresponding standard luminosity is roughly
\begin{equation}
L_{\rm std}
\sim
\frac{M_{\rm BH}c^2}{\tau_0}
\approx
2\times10^{9}\,{\rm erg/s}.
\end{equation}
The six-dimensional entropy is
\begin{equation}
S_{\rm BH}
\sim
\left(
\frac{M_{\rm BH}}{M_*}
\right)^{4/3}
\approx
10^{41},
\qquad
M_*\simeq10~{\rm TeV}
\simeq1.78\times10^{-20}\,{\rm g}.
\label{eq:numbers_11}
\end{equation}
Substituting these values into Eqs.~\eqref{eq:L_ord} and \eqref{eq:Lne}
gives the order-of-magnitude estimates presented in Table~\ref{tab:lumi}.

\begin{table}[h]
\centering
\caption{Luminocities for ordinary non-extremal, and near-extremal Primordial Black Holes}
\label{tab:lumi}
\begin{tabular}{c|c|c}
\hline
 & Ordinary non-extremal & Near-extremal reference \\
\hline
\(p=1\) &
\(L_{\rm MB}^{(1)}\sim10^{-32}\,{\rm erg/s}\) &
\(L_{\rm ne}^{(1)}\sim10^{-52}\sqrt{\beta}\,{\rm erg/s}\) \\
\(p=2\) &
\(L_{\rm MB}^{(2)}\sim10^{-73}\,{\rm erg/s}\) &
\(L_{\rm ne}^{(2)}\sim10^{-93}\sqrt{\beta}\,{\rm erg/s}\) \\
\hline
\end{tabular}
\end{table}
Even the largest of these post-burden luminosities is negligible on
cosmological scales.  Therefore, the dominant phenomenological
constraint does not come from the late post-burden luminosity itself,
but from the short pre-burden phase, during which the PBH still
evaporates approximately at the standard Hawking rate.  The relevant
energy-injection bound is then controlled by the amount of energy emitted
before the memory burden becomes effective, rather than by the highly
suppressed post-burden luminosity.

\medskip
\noindent\textbf{Pre-burden energy injection and the status of the \(p=1\) scenario.}
Before the memory burden becomes effective, the black hole loses a fractional mass through approximately standard Hawking radiation. Associated with this, we introduce the following definition
\begin{equation}
q
\equiv
\frac{\Delta M_{\rm crit}}{M_0}\label{q_ratio}
\end{equation}
 Using the onset
criterion in Eq.~\eqref{eq:qcrit_MB}, one has
\[
q\sim
\left[
p\,S_{\rm BH}(M_0)
\right]^{-1/p},
\]
where \(M_0\) is the initial PBH mass.  Therefore,
\begin{equation}
p=1:\quad q\sim S_{\rm BH}^{-1},
\qquad
p=2:\quad q\sim (2S_{\rm BH})^{-1/2}.
\label{eq:q_special_DM}
\end{equation}
For macroscopic black holes, \(S_{\rm BH}\gg1\), so the fractional mass
loss before the onset of the burden is small.

For example, for \(M_0=10^{11}\,{\rm g}\) and
\(M_*\simeq10\,{\rm TeV}\), Eq.~\eqref{eq:numbers_11} gives
\[
S_{\rm BH}(M_0)\sim10^{41}.
\]
Thus, for \(p=1\),
\begin{equation}
q\sim10^{-41},
\qquad
\Delta E_{\rm pre}
\sim qM_0c^2
\sim10^{-9}\,{\rm erg},
\end{equation}
which is negligible per black hole.  For \(p=2\), the onset fraction is
larger,
\[
q\sim(2\times10^{41})^{-1/2}\sim10^{-21},
\]
but it is still extremely small compared with unity.

The smallness of the pre-burden fractional mass loss does not by itself
guarantee that the \(p=1\) scenario is automatically safe.  After the
onset of the burden, the ordinary non-extremal luminosity is suppressed
only by one entropy factor, \(S_{\rm BH}^{-1}\), whereas for \(p=2\) it
is suppressed by \(S_{\rm BH}^{-2}\).  Consequently, the \(p=1\) case is
more sensitive to the precise onset prescription, the emitted spectrum,
the PBH abundance, and the redshift at which the emission occurs.  A
dedicated redshift-dependent energy-injection calculation is therefore
required before a robust \(p=1\) dark-matter window can be claimed.

For near-extremal charged PBHs with \(p=1\), the additional temperature
suppression reduces the post-burden luminosity further, as shown in
Eq.~\eqref{eq:Lne}.  However, the onset of the burden and the
cosmological impact of the pre-burden emission may depend on the
near-extremality parameter \(\beta\), the formation mechanism, and the
PBH abundance.  We therefore treat the \(p=1\) near-extremal case as a
possible but model-dependent extension, rather than as the conservative
baseline.  By contrast, the \(p=2\) scenario is more robust because its
post-burden luminosity is suppressed by an additional power of the
entropy.

\medskip

\noindent\textbf{Phenomenological lower bound from the pre-burden phase for \(p=2\).} 
Even for \(p=2\), the pre-burden phase can impose a lower limit on the
PBH mass.  Taking the benchmark value
\[
M_*
\simeq
10\,{\rm TeV}
\simeq
1.78\times10^{-20}\,{\rm g},
\]
and using the entropy scaling in Eq.~\eqref{eq:entropy}, one finds for
\(M_{\rm BH}=10^{-3}\,{\rm g}\)
\begin{equation}
S_{\rm BH}
\sim
\left(
\frac{10^{-3}}{1.78\times10^{-20}}
\right)^{4/3}
\approx
2\times10^{22}.
\label{eq:S_light}
\end{equation}
For \(p=2\), the critical fractional mass loss is therefore
\begin{equation}
q
\sim
(2S_{\rm BH})^{-1/2}
\approx
5\times10^{-12}.
\label{eq:q_light}
\end{equation}
The corresponding pre-burden energy release per PBH is
\begin{equation}
\Delta E_{\rm pre}
\simeq
qM_{\rm BH}c^2
\approx
5\times10^{-12}
\times10^{-3}\,{\rm g}
\times9\times10^{20}\,{\rm erg/g}
\approx
4.5\times10^{6}\,{\rm erg}.
\label{eq:Epre_light}
\end{equation}

For \(p=2\), the onset fraction scales as
\begin{equation}
q\propto S_{\rm BH}^{-1/2}\propto M_{\rm BH}^{-2/3}.
\end{equation}
Thus, smaller PBHs lose a larger fraction of their mass before the memory
burden becomes effective.  The energy released per individual PBH scales
as
\[
\Delta E_{\rm pre}\sim qM_{\rm BH}c^2\propto M_{\rm BH}^{1/3},
\]
whereas the injected energy density at fixed dark-matter fraction scales
as
\[
\rho_{\rm inj}^{\rm pre}
\sim
q\rho_{\rm PBH}c^2
\propto
M_{\rm BH}^{-2/3}.
\]
Therefore, the cosmological impact of the pre-burden phase becomes more
important for lighter PBHs even though the energy released by each
individual PBH decreases.

The precise lower bound cannot be obtained from the energy per black
hole alone.  It requires the PBH abundance, the emitted particle
spectrum, the redshift of emission, and the relevant BBN, CMB
\(\mu\)-distortion, and diffuse-background energy-injection bounds.
Nevertheless, for a PBH population making up all of the dark matter, we
adopt
\begin{equation}
M_{\rm min}
\sim
10^{-3}\,{\rm g}
\label{eq:Mmin}
\end{equation}
as a conservative phenomenological lower edge in the \(p=2\)
memory-burdened scenario.

\medskip
\noindent\textbf{Resulting dark-matter windows.}
Combining the survival requirement with the expected constraints from
pre-burden energy injection, BBN, CMB \(\mu\)-distortions, the diffuse
gamma-ray background, and compact-object searches, we obtain the
conservative mass interval summarized in Table~\ref{tab:windows}.  For
the ordinary non-extremal \(p=2\) scenario, the viable baseline window is
taken to be
\begin{equation}
10^{-3}\,{\rm g}
\lesssim
M_{\rm PBH}
\lesssim
10^{21}\,{\rm g}.
\label{eq:window_final}
\end{equation}
The lower edge is phenomenological and is controlled by pre-burden
energy injection, while the upper edge is set by the usual compact-object
dark-matter constraints, especially microlensing.  Near-extremal charged
PBHs can have even longer lifetimes because of the additional
\(\beta^{-1/2}\) enhancement, but such scenarios require a formation
mechanism capable of producing a near-extremal PBH population.  We
therefore treat them as illustrative extensions rather than as the
baseline dark-matter window.

\begin{table}[h]
\centering
\caption{Conservative PBH dark-matter windows in the six-dimensional scenario for \(f_{\rm PBH}=1\).}
\label{tab:windows}
\renewcommand{\arraystretch}{1.25}
\begin{tabularx}{\textwidth}{lcc}
\hline
Scenario & Standard evaporation, \(p=0\) & Memory burden, \(p=2\) \\
\hline
Ordinary non-extremal &
\(M_{\rm PBH}\gtrsim10^{8}\,{\rm g}\) &
\(10^{-3}\,{\rm g}\lesssim M_{\rm PBH}\lesssim10^{21}\,{\rm g}\) \\
Near-extremal charged, \(\beta\ll1\) &
Enhanced by \(\beta^{-1/2}\) &
Formation-dependent extension \\
\hline
\end{tabularx}
\end{table}

\noindent
Thus the ordinary non-extremal \(p=2\) memory-burdened scenario provides
a conservative broad PBH dark-matter window.  Near-extremality can extend
the lifetime further, but the corresponding mass range is not quoted as
a model-independent bound because it depends on the charge distribution,
the value of \(\beta\), and the formation mechanism.

\subsection{The Festina--Lente bound and near-extremal charged PBHs}
\label{sec:FL_bound}

The preceding discussion of near-extremal charged PBHs raises an
important consistency question: can the electric field of a nearly
extremal black hole be made arbitrarily small, corresponding to
\(\beta\to0\), without conflicting with quantum-gravity constraints?
In de Sitter space, the Festina--Lente (FL) bound provides a criterion
for whether charged particles are sufficiently light to allow efficient
Schwinger discharge of near-extremal charged black holes
\cite{Montero:2019ekk,Montero:2021otb}.  Parametrically, the bound
requires the existence of charged states light enough that Schwinger
pair production is not exponentially suppressed in the relevant
background field.  In four-dimensional notation this is often written,
up to order-one factors, as an upper bound of the form
\begin{equation}
m^2 \lesssim qg\,H\,M_{\rm Pl}.
\end{equation}
In higher dimensions the precise expression depends on the normalization
of the gauge coupling, the Planck scale, and the near-Nariai geometry.
Below we use the FL bound only as a consistency check on the
near-extremal charged PBH interpretation; it is not needed for the
ordinary non-extremal \(p=2\) memory-burdened dark-matter window.

\subsubsection{The Nariai limit in \(d\) dimensions}

Consider the \(d\)-dimensional Reissner--Nordström--de Sitter (RNdS)
geometry \cite{Gibbons:1977mu,Bousso:1996au},
\begin{equation}
ds^2
=
-U(r)dt^2
+
\frac{dr^2}{U(r)}
+
r^2 d\Omega_{d-2}^2 ,
\end{equation}
with
\begin{equation}
U(r)
=
1
-
\frac{\mu}{r^{d-3}}
+
\frac{Q_d^2}{r^{2d-6}}
-
\frac{r^2}{\ell^2},
\label{eq:UdRNdS_full}
\end{equation}
where
\begin{equation}
\mu
\equiv
\frac{2M}{M_P^{d-2}},
\qquad
Q_d^2
\equiv
\frac{g^2q^2}{4\pi M_P^{d-2}}.
\end{equation}
Here \(M\) is the black-hole mass, \(q\) is the charge quantum number,
\(g\) is the gauge coupling, \(M_P\) is the \(d\)-dimensional Planck
scale, and \(\ell\) is the de Sitter radius.

The Nariai limit corresponds to the degeneracy of the black-hole and
cosmological horizons ~\cite{Nariai:1999iok,Bousso:1996au}.  At the degenerate horizon \(r=r_N\), one has
\begin{equation}
U(r_N)=0,
\qquad
U'(r_N)=0.
\label{eq:Nariai_conditions_full}
\end{equation}
It is useful to define
\begin{equation}
n\equiv d-3.
\end{equation}
Then
\begin{equation}
U(r)
=
1
-
\frac{\mu}{r^n}
+
\frac{Q_d^2}{r^{2n}}
-
\frac{r^2}{\ell^2}.
\end{equation}
The derivative is
\begin{equation}
U'(r)
=
\frac{n\mu}{r^{n+1}}
-
\frac{2nQ_d^2}{r^{2n+1}}
-
\frac{2r}{\ell^2}.
\label{eq:Uprime}
\end{equation}
Imposing \(U'(r_N)=0\) gives
\begin{equation}
\frac{\mu}{r_N^n}
=
2\frac{Q_d^2}{r_N^{2n}}
+
\frac{2}{n}\frac{r_N^2}{\ell^2}.
\label{eq:mu_relation_Nariai}
\end{equation}
Substituting this into \(U(r_N)=0\), we obtain
\begin{equation}
0
=
1
-
\frac{Q_d^2}{r_N^{2n}}
-
\left(1+\frac{2}{n}\right)
\frac{r_N^2}{\ell^2}.
\end{equation}
Therefore,
\begin{equation}
\frac{Q_d^2}{r_N^{2n}}
=
1
-
\frac{n+2}{n}
\frac{r_N^2}{\ell^2}.
\end{equation}
Restoring \(n=d-3\), this becomes
\begin{equation}
\frac{Q_d^2}{r_N^{2d-6}}
=
1
-
\frac{d-1}{d-3}
\frac{r_N^2}{\ell^2}
\label{eq:ChargeRelation_full}
\end{equation}
or, in terms of the original charge parameter,
\begin{equation}
\frac{g^2q^2}
{4\pi M_P^{d-2}r_N^{2d-6}}
=
1
-
\frac{d-1}{d-3}
\frac{r_N^2}{\ell^2}.
\label{eq:ChargeRelation_full_original}
\end{equation}

The mass parameter is obtained from Eq.~(\ref{eq:mu_relation_Nariai}):
\begin{equation}
\frac{\mu}{r_N^n}
=
2
-
2\frac{n+1}{n}
\frac{r_N^2}{\ell^2}.
\end{equation}
Thus,
\begin{equation}
M
=
M_P^{d-2}r_N^{d-3}
\left[
1
-
\frac{d-2}{d-3}
\frac{r_N^2}{\ell^2}
\right].
\label{eq:Mass_Nariai_full}
\end{equation}

Equations~(\ref{eq:ChargeRelation_full_original}) and
(\ref{eq:Mass_Nariai_full}) show that, for charged RNdS black holes, the
Nariai radius is not fixed uniquely by \(d\) and \(\ell\).  Instead,
\(r_N\) parametrizes a family of charged Nariai solutions.  The charge
relation implies
\begin{equation}
\frac{r_N^2}{\ell^2}
\leq
\frac{d-3}{d-1}.
\label{eq:Nariai_bound}
\end{equation}
The equality corresponds to the neutral Nariai limit \(q=0\), for which
\begin{equation}
r_N^2
=
\frac{d-3}{d-1}\ell^2.
\label{eq:Neutral_Nariai_radius}
\end{equation}

For nonzero charge, \(q\neq0\), the left-hand side of
Eq.~(\ref{eq:ChargeRelation_full_original}) is strictly positive.
Consequently,
\(1-\frac{d-1}{d-3}\frac{r_N^2}{\ell^2}>0\), which implies
\(\frac{r_N^2}{\ell^2}<\frac{d-3}{d-1}\), or equivalently
\(r_N<\ell\sqrt{(d-3)/(d-1)}=r_N^{(q=0)}\).  Hence, the Nariai
radius of a charged solution is strictly smaller than its neutral
value and is determined by
Eq.~(\ref{eq:ChargeRelation_full_original}).


\subsubsection{Electric field on the Nariai horizon}

The discharge of a charged near-Nariai black hole proceeds through
Schwinger pair production in the background electric field
\cite{Schwinger:1951nm,Garriga:1994bm,Montero:2019ekk}.  With the same
normalization used in Eq.~(\ref{eq:UdRNdS_full}), we define
\begin{equation}
Q_d^2
\equiv
\frac{g^2q^2}{4\pi M_P^{d-2}},
\end{equation}
so that the electric field may be written as
\begin{equation}
E(r)
=
\frac{Q_d}{2\sqrt{\pi}\,r^{d-2}}.
\label{eq:ElectricField_general_full}
\end{equation}
At the Nariai horizon,
\begin{equation}
E_N
=
\frac{Q_d}{2\sqrt{\pi}\,r_N^{d-2}}.
\label{eq:EN_def}
\end{equation}

Using the charge--radius relation derived in
Eq.~(\ref{eq:ChargeRelation_full}), the Nariai charge parameter is
\begin{equation}
Q_d
=
r_N^{d-3}
\left[
1-\frac{d-1}{d-3}\frac{r_N^2}{\ell^2}
\right]^{1/2}.
\label{eq:Qd_Nariai}
\end{equation}
Substituting this into Eq.~(\ref{eq:EN_def}) gives
\begin{equation}
E_N
=
\frac{1}{2\sqrt{\pi}\,r_N}
\left[
1
-
\frac{d-1}{d-3}
\frac{r_N^2}{\ell^2}
\right]^{1/2}.
\label{eq:EN_correct}
\end{equation}
Equivalently, defining
\begin{equation}
x_N\equiv \frac{r_N}{\ell},
\qquad
H\equiv \ell^{-1},
\end{equation}
$E_N$ can be re-written as follows:
\begin{equation}
E_N
=
\frac{H}{2\sqrt{\pi}\,x_N}
\left[
1
-
\frac{d-1}{d-3}x_N^2
\right]^{1/2}~,
\label{eq:EN_xN}
\end{equation}
which for $d=6$, takes the form
\begin{equation}
E_N^{(6)}
=
\frac{H}{2\sqrt{\pi}\,x_N}
\left(
1-\frac{5}{3}x_N^2
\right)^{1/2},
\qquad
0<x_N^2\leq\frac35 .
\label{eq:EN_6D_correct}
\end{equation}
Thus the Nariai electric field is not fixed solely by the spacetime
dimension.  It depends on the charged Nariai branch through
\(x_N=r_N/\ell\).  In the neutral Nariai limit,
\(x_N^2=(d-3)/(d-1)\), the electric field correctly vanishes.

\subsubsection{Schwinger pair production and the Festina--Lente bound}

In a locally constant electric field \(E\), the Schwinger pair-production
rate for a particle of mass \(m\) and physical charge \(e_{\rm phys}\)
has the schematic form~\cite{Schwinger:1951nm}
\begin{equation}
\Gamma_{\rm Schw}
\sim
\frac{(e_{\rm phys}E)^{d/2}}{(2\pi)^{d-1}}
\exp\!\left[
-\frac{\pi m^2}{e_{\rm phys}E}
\right],
\label{eq:Schwinger_rate}
\end{equation}
up to spin-dependent, curvature-dependent, and geometry-dependent
prefactors, the Schwinger exponent separates two parametric regimes:
\begin{equation}
\frac{\pi m^2}{e_{\rm phys}E_N}
\begin{cases}
\lesssim 1,
& \text{efficient Schwinger discharge},\\[1mm]
\gtrsim 1,
& \text{exponentially suppressed discharge (FL regime)}.
\end{cases}
\label{eq:Schwinger_FL_combined}
\end{equation}
Thus, avoiding rapid Schwinger discharge of charged Nariai black holes
requires parametrically
\(m^2\gtrsim e_{\rm phys}E_N/\pi\), up to order-one corrections.  This
constitutes the physical content of the Festina--Lente bound
\cite{Montero:2019ekk,Montero:2021otb}. In \(d\) spacetime dimensions, the FL bound can be written, up to
order-one numerical factors, as
\begin{equation}
m^2
\gtrsim
g_D q
\left[
\frac{(d-1)(d-2)}{2}
\right]^{1/2}
M_D^{(d-2)/2}H ,
\label{eq:FL_D_lower_m2}
\end{equation}
or equivalently
\begin{equation}
m
\gtrsim
(g_D q)^{1/2}
\left[
\frac{(d-1)(d-2)}{2}
\right]^{1/4}
M_D^{(d-2)/4}H^{1/2}.
\label{eq:FL_D_lower}
\end{equation}
Thus the FL condition is a lower bound on the mass of charged states.
For the two-dark-dimensions scenario, \(d=6\), \(M_D=M_6=M_*\), and the
six-dimensional gauge coupling has mass dimension
\begin{equation}
[g_6]=-1 .
\end{equation}
Equation~(\ref{eq:FL_D_lower}) then gives
\begin{equation}
m
\gtrsim
(g_6 q)^{1/2}
10^{1/4}
M_*H^{1/2}.
\label{eq:FL6D_start}
\end{equation}
If a charged field propagates in the six-dimensional bulk, its
six-dimensional gauge coupling is related to the four-dimensional
coupling by
\begin{equation}
e
=
\frac{g_6}{\sqrt{V_2}}
=
\frac{g_6}{2\pi R},
\qquad
g_6=e(2\pi R).
\label{eq:g6}
\end{equation}
Using the compactification relation in Eq.~(\ref{eq:MPl}), one obtains
\begin{equation}
(g_6 q)^{1/2}
=
(e q)^{1/2}
\frac{M_{\rm Pl}^{1/2}}{M_*}.
\end{equation}
Substituting this into Eq.~(\ref{eq:FL6D_start}) gives
\begin{equation}
m
\gtrsim
(e q)^{1/2}
10^{1/4}
\left(
M_{\rm Pl}H
\right)^{1/2}.
\label{eq:FL6D_final}
\end{equation}
This is the useful four-dimensional form of the six-dimensional FL
condition after dimensional reduction.

For \(q=1\), \(e\simeq0.3\), and
\(M_{\rm Pl}=2.4\times10^{18}\,{\rm GeV}\), Eq.~(\ref{eq:FL6D_final})
becomes
\begin{equation}
m
\gtrsim
1.5\times10^{16}\,{\rm GeV}
\left(
\frac{H}{10^{14}\,{\rm GeV}}
\right)^{1/2},
\label{eq:FL6D_numeric}
\end{equation}
up to order-one geometric factors.  This scale is vastly larger than
the KK mass gap.  Therefore, if electrically charged Standard Model
fields were allowed to propagate in the bulk during a high-scale
de Sitter phase, their light KK excitations would violate the FL
criterion.  This provides additional motivation for localizing charged
Standard Model fields on the brane, or alternatively for requiring a
much lower inflationary Hubble scale.

For the present de Sitter epoch,
\begin{equation}
H_0\simeq1.5\times10^{-42}\,{\rm GeV},
\end{equation}
Eq.~(\ref{eq:FL6D_final}) gives
\begin{equation}
m
\gtrsim
2\times10^{-3}\,{\rm eV},
\label{eq:FL_today}
\end{equation}
again up to order-one factors.  Thus the present-day FL bound is mild
for ordinary charged Standard Model particles, but it can constrain
ultralight charged states.

\begin{figure}[t]
\centering
\includegraphics[width=\textwidth]{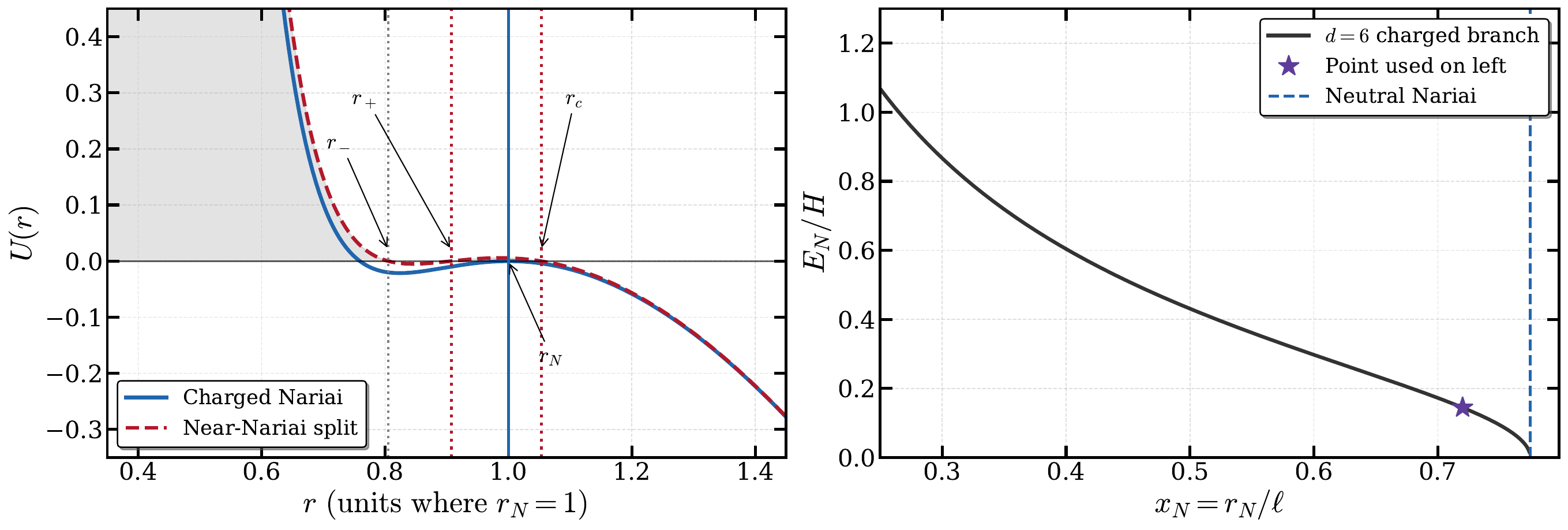}
\caption{
Illustration of the charged Nariai limit in six-dimensional
Reissner--Nordstr\"om--de Sitter geometry.
Left: metric function \(U(r)\) for the charged Nariai solution and a
near-Nariai split solution.  In the Nariai limit the black-hole and
cosmological horizons merge at the degenerate radius \(r_N\), while the
near-Nariai solution displays separated horizons \(r_+\) and \(r_c\),
together with the inner horizon \(r_-\).
Right: electric field on the charged Nariai horizon, normalized by the
de Sitter scale \(H=\ell^{-1}\), as a function of
\(x_N=r_N/\ell\) for \(d=6\).  The electric field is not a universal
constant times \(H\); it depends on the charged Nariai branch and
vanishes at the neutral Nariai endpoint \(x_N^2=3/5\).
}
\label{fig:nariai_6d}
\end{figure}

Figure~\ref{fig:nariai_6d} illustrates the charged Nariai structure in
the six-dimensional case.  The left panel shows how the black-hole and
cosmological horizons merge in the Nariai limit, while a small
deformation away from this limit splits the degenerate horizon into
separate horizons.  The right panel shows the corresponding electric
field on the charged Nariai branch.  Unlike the neutral Nariai radius,
the charged Nariai electric field is not fixed only by the spacetime
dimension; it depends on the branch parameter \(x_N=r_N/\ell\).  This
is why the FL bound should be applied as a consistency condition on
charged states rather than as a universal fixed numerical bound on the
near-extremality parameter.

We emphasize, however, that the FL bound does not give a
model-independent constraint on the near-extremality parameter
\(\beta\) of a primordial black hole.  Such a bound would require the
explicit charged black-hole solution, the relation between \(\beta\) and
the horizon electric field, and the spectrum of charged particles
available for Schwinger discharge.  Therefore, in this work we use the
FL bound only as a consistency check on charged bulk states.  It does
not modify the conservative ordinary non-extremal \(p=2\) PBH
dark-matter window discussed above.

\section{Scalar-Induced Gravitational Waves in the 6d Scenario}
\label{sec:SIGW}

Primordial black holes are produced when large curvature perturbations
generated during inflation re-enter the Hubble horizon in the
radiation-dominated epoch and undergo gravitational collapse
\cite{Carr:1974nx,Hawking:1971ei}.  A perturbation mode with comoving
wavenumber \(k\) re-enters the horizon when \(k=aH\).  The mass of the
resulting black hole is approximately a fraction \(\gamma\) of the
horizon mass at that time,
\begin{equation}
M_{\rm PBH}
\simeq
\gamma M_H
=
\gamma\,\frac{4\pi}{3}\rho H^{-3}\bigg|_{\rm re-entry},
\label{eq:MPBH_horizon}
\end{equation}
where \(\gamma\simeq0.2\) characterizes the collapse efficiency.  During
radiation domination, \(\rho=3M_{\rm Pl}^2H^2\), so
\begin{equation}
M_{\rm PBH}
\simeq
4\pi\gamma\frac{M_{\rm Pl}^2}{H_{\rm form}} .
\label{eq:MPBH_Hform}
\end{equation}

Using entropy conservation, the PBH mass is related to the comoving
scale by
\begin{equation}
M_{\rm PBH}(k)
\simeq
10^{18}\,{\rm g}
\left(\frac{\gamma}{0.2}\right)
\left(\frac{g_*}{106.75}\right)^{-1/6}
\left(\frac{k}{7\times10^{13}\,{\rm Mpc}^{-1}}\right)^{-2}.
\label{eq:M_k_relation}
\end{equation}
Equivalently,
\begin{equation}
k
\simeq
7\times10^{13}\,{\rm Mpc}^{-1}
\left(\frac{\gamma}{0.2}\right)^{1/2}
\left(\frac{g_*}{106.75}\right)^{-1/12}
\left(\frac{M_{\rm PBH}}{10^{18}\,{\rm g}}\right)^{-1/2}.
\label{eq:kstar}
\end{equation}
The corresponding present-day frequency is
\begin{equation}
f
\simeq
0.11\,{\rm Hz}
\left(\frac{\gamma}{0.2}\right)^{1/2}
\left(\frac{g_*}{106.75}\right)^{-1/12}
\left(\frac{M_{\rm PBH}}{10^{18}\,{\rm g}}\right)^{-1/2}.
\label{eq:f_M_relation}
\end{equation}

The primordial curvature power spectrum is modelled as a log-normal 
enhancement at the scale \(k_*\) corresponding to the characteristic PBH
mass \(M_{PBH}\) \cite{Inomata:2018epa,Inomata:2019ivs,Pi:2020otn}:
\begin{equation}
\mathcal P_{\mathcal R}(k)
=
\frac{A_{\mathcal R}}{\sqrt{2\pi}\,\sigma_k}
\exp\left[
-\frac{\ln^2(k/k_*)}{2\sigma_k^2}
\right],
\label{eq:pzeta}
\end{equation}
where \(A_{\mathcal R}\) sets the integrated power and \(\sigma_k\)
controls the logarithmic width of the peak. For the PBH abundance we use a phenomenological log-normal mass function per logarithmic mass interval,
\begin{equation}
\frac{df_{\rm PBH}}{d\ln M}
=
\frac{f_{\rm PBH}}{\sqrt{2\pi}\,\sigma_M}
\exp\left[
-\frac{\ln^2(M/M_p)}{2\sigma_M^2}
\right],
\label{eq:lognorm_dlnM}
\end{equation}
which satisfies
\begin{equation}
\int d\ln M\,
\frac{df_{\rm PBH}}{d\ln M}
=
f_{\rm PBH}.
\end{equation}
Equivalently, if one defines
\begin{equation}
\psi(M)
\equiv
\frac{df_{\rm PBH}}{dM},
\end{equation}
then
\begin{equation}
\psi(M)
=
\frac{f_{\rm PBH}}{\sqrt{2\pi}\,\sigma_M M}
\exp\left[
-\frac{\ln^2(M/M_p)}{2\sigma_M^2}
\right],
\label{eq:lognorm_dM}
\end{equation}
so that \(\int dM\,\psi(M)=f_{\rm PBH}\).

In this work we use the log-normal mass function as a phenomenological
description of the PBH population.  We consider representative
benchmarks with \(f_{\rm PBH}=1\).  A broad benchmark with
\(\sigma_M=1\) and \(M_p=10^{10}\,{\rm g}\) corresponds to very high
frequencies, \(f\sim10^3\,{\rm Hz}\).  The memory-burdened lower edge,
\(M_p=10^{-3}\,{\rm g}\), corresponds to GHz frequencies and is therefore
far above the standard LISA/DECIGO/BBO/PTA bands.  Masses in the range
\(M_p\sim10^{18}\)--\(10^{21}\,{\rm g}\) correspond instead to the
deci-Hz to mHz bands relevant for future space-based gravitational-wave
observatories.

\begin{figure}[t]
\centering
\includegraphics[width=0.7\columnwidth]{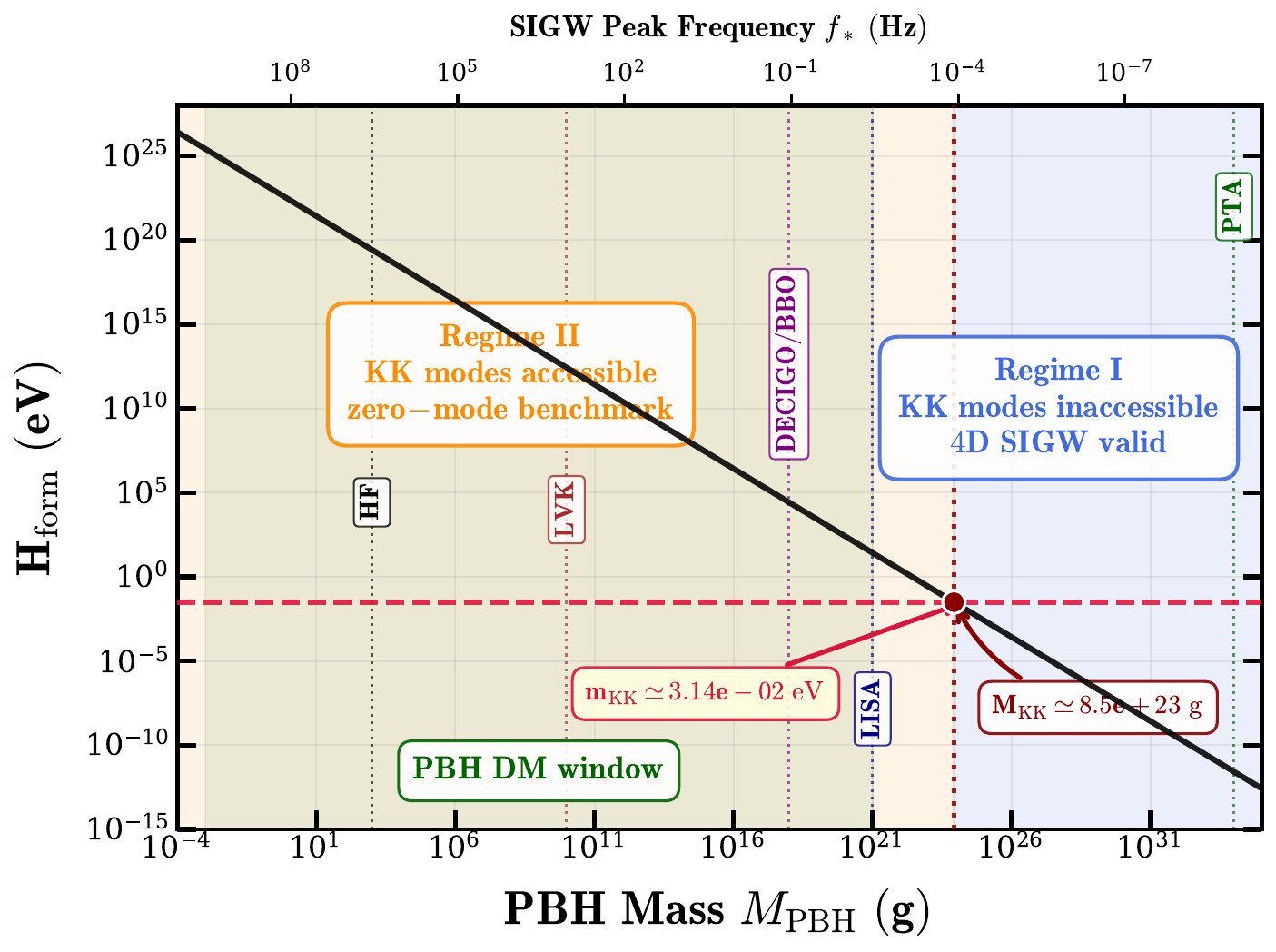}
\caption{
Regimes of validity for scalar-induced gravitational waves in the
six-dimensional two-dark-dimensions scenario.  The black curve shows the
Hubble scale at PBH formation, \(H_{\rm form}\), as a function of the
PBH mass.  The horizontal dashed line denotes the KK mass gap
\(m_{\rm KK}\).  For \(M_{\rm PBH}>M_{\rm KK}\), massive KK gravitons are
kinematically inaccessible and the standard four-dimensional SIGW
formalism is a valid effective description.  For
\(M_{\rm PBH}<M_{\rm KK}\), the KK tower is accessible and the 4D SIGW
spectrum should be interpreted as a zero-mode benchmark rather than the
complete six-dimensional prediction.  The green shaded region indicates
the memory-burdened PBH dark-matter window.
}
\label{fig:regimes}
\end{figure}

\subsection{Regimes of validity}
\label{sec:SIGW_validity}

The standard scalar-induced gravitational-wave calculation assumes a
four-dimensional massless tensor sector.  In the present six-dimensional
scenario this approximation must be checked against the KK mass gap.  At
horizon re-entry, the physical momentum of the scalar mode is of order
the Hubble scale, \(k/a\simeq H_{\rm form}\).  Therefore massive KK
tensor modes can be kinematically excited only if
\begin{equation}
H_{\rm form}\gtrsim m_{\rm KK}.
\label{eq:KK_threshold}
\end{equation}
This condition separates the parameter space into a four-dimensional
regime and a KK-sensitive regime. Using the PBH mass--Hubble relation in Eq.~(\ref{eq:MPBH_Hform}), the
formation scale can be written as
\begin{equation}
H_{\rm form}
\simeq
4\pi\gamma
\frac{M_{\rm Pl}^2}{M_{\rm PBH}},
\label{eq:Hform_validity}
\end{equation}
where \(M_{\rm Pl}\) denotes the reduced Planck mass.  With the
reduced-length convention \(L=2\pi R\), the KK threshold used in this
work is
\begin{equation}
m_{\rm KK}
=
\frac{1}{L}
=
\frac{1}{2\pi R}
\simeq
3.1\times10^{-2}\,{\rm eV}.
\label{eq:mKK_validity}
\end{equation}
Combining Eqs.~(\ref{eq:KK_threshold})--(\ref{eq:mKK_validity}), the
boundary mass is
\begin{equation}
M_{\rm KK}
\equiv
4\pi\gamma
\frac{M_{\rm Pl}^2}{m_{\rm KK}}
\simeq
8.5\times10^{23}\,{\rm g}
\left(\frac{\gamma}{0.2}\right)
\left(\frac{3.1\times10^{-2}\,{\rm eV}}{m_{\rm KK}}\right).
\label{eq:MKK}
\end{equation}
Equivalently, using the mass--frequency relation in
Eq.~(\ref{eq:f_M_relation}), this corresponds to
\begin{equation}
f_{\rm KK}
\simeq
1.2\times10^{-4}\,{\rm Hz}.
\label{eq:fKK}
\end{equation}

We therefore distinguish two regimes:
\begin{itemize}
\item \textbf{Regime I:} \(M_{\rm PBH}\gtrsim M_{\rm KK}\), or
equivalently \(f_*\lesssim f_{\rm KK}\).  In this case
\(H_{\rm form}\lesssim m_{\rm KK}\), so the massive KK tower is not
kinematically accessible.  The usual four-dimensional SIGW formalism is
then a self-consistent effective description
\cite{Ananda:2006af,Baumann:2007zm,Domenech:2021ztg}.

\item \textbf{Regime II:} \(M_{\rm PBH}\lesssim M_{\rm KK}\), or
equivalently \(f_*\gtrsim f_{\rm KK}\).  In this case
\(H_{\rm form}\gtrsim m_{\rm KK}\), so massive KK tensor modes can in
principle be excited.  The four-dimensional SIGW spectrum should then
be interpreted as the massless zero-mode benchmark rather than the full
six-dimensional prediction.
\end{itemize}

The memory-burdened PBH dark-matter window considered in this work,
\begin{equation}
10^{-3}\,{\rm g}
\lesssim
M_{\rm PBH}
\lesssim
10^{21}\,{\rm g},
\end{equation}
lies entirely below \(M_{\rm KK}\simeq8.5\times10^{23}\,{\rm g}\).
Therefore these PBH masses belong to Regime~II, and the spectra shown
for this window should be read as zero-mode benchmarks.  A complete
six-dimensional prediction would require the massive tensor Green
functions, brane-to-bulk overlap coefficients, massive-mode transfer
functions, and detector response to massive tensor polarizations
\cite{Fierz:1939ix,Hinterbichler:2011tt,deRham:2014zqa,
Nishizawa:2009bf}.

By contrast, PTA-band SIGWs correspond to much smaller frequencies and
therefore to much larger PBH masses.  For example,
\(f_*\sim10^{-9}\,{\rm Hz}\) corresponds to
\(M_{\rm PBH}\sim10^{34}\,{\rm g}\), which is far above
\(M_{\rm KK}\).  PTA-band SIGWs therefore lie in Regime~I, where the
standard four-dimensional SIGW calculation is self-consistent
\cite{NANOGrav:2023gor,EPTA:2023fyk}.

\subsection{Full six-dimensional tensor dynamics}
\label{sec:full6D}

In the six-dimensional theory the tensor perturbation decomposes into a
massless four-dimensional zero mode \(h_{ij}^{(0)}\) and a tower of
massive KK tensor modes \(h_{ij}^{(\vec n)}\) \cite{Kaluza:1921tu,Klein:1926tv,Overduin:1997sri,Maartens:2010ar}, with
\begin{equation}
m_{\vec n}
=
m_{\rm KK}\sqrt{n_1^2+n_2^2},
\qquad
\vec n=(n_1,n_2).
\end{equation}
At the level of the four-dimensional effective theory, each KK mode
obeys a massive tensor equation of the schematic form \cite{Fierz:1939ix,Hinterbichler:2011tt,deRham:2014zqa},
\begin{equation}
{h_{\vec n}^{\lambda}}''
+
2\mathcal H {h_{\vec n}^{\lambda}}'
+
\left(k^2+a^2m_{\vec n}^2\right)
h_{\vec n}^{\lambda}
=
S_{\vec n}^{\lambda}(k,\eta),
\label{eq:KK_tensor}
\end{equation}
where \(S_{\vec n}^{\lambda}\) denotes the projection of the quadratic
scalar source onto the corresponding KK wavefunction.  This projection
contains a model-dependent overlap coefficient, denoted here by
\(\mathcal C_{\vec n}\).  The solution can be written formally in terms
of the massive Green function \(G_{\vec n}(k;\eta,\tilde\eta)\), which
then determines the tensor power spectrum
\(\mathcal P_h^{(\vec n)}(k,\eta)\).

The zero mode, \(\vec n=0\), is massless and reproduces the standard
four-dimensional scalar-induced gravitational-wave result \cite{Ananda:2006af,Baumann:2007zm,Kohri:2018awv,Domenech:2021ztg}. Using the
log-normal curvature spectrum from Eq \eqref{eq:pzeta} the present-day zero-mode spectrum is
\begin{align}
\Omega_{\rm GW,0}^{(0)}(k)
&=
\frac{\Omega_{r,0}}{24}
\left(\frac{g_{*,0}}{g_{*,c}}\right)^{1/3}
\int_{0}^{\infty} dv
\int_{|1-v|}^{1+v} du\,
\mathcal K(u,v)\,
\overline{I^2(u,v)}
\nonumber\\
&\hspace{3.0cm}\times
\mathcal P_{\mathcal R}(ku)
\mathcal P_{\mathcal R}(kv),
\label{eq:Omega_4D}
\end{align}
where \(\mathcal K(u,v)\) is the standard scalar-induced tensor kernel
and \(\overline{I^2(u,v)}\) is the radiation-era time-averaged kernel \cite{Ananda:2006af,Baumann:2007zm,Domenech:2021ztg}. For a massive KK mode, the corresponding expression is modified by the
massive Green function and by the overlap with the source.  Schematically
one may write
\begin{align}
\Omega_{\rm GW,0}^{(\vec n)}(k)
&\sim
\frac{\Omega_{r,0}}{24}
\left(\frac{g_{*,0}}{g_{*,c}}\right)^{1/3}
\int_{0}^{\infty} dv
\int_{|1-v|}^{1+v} du\,
\mathcal K(u,v)\,
\mathcal C_{\vec n}^2\,
\overline{I_{\vec n}^{\,2}(u,v,k)}
\nonumber\\
&\hspace{3.0cm}\times
\mathcal P_{\mathcal R}(ku)
\mathcal P_{\mathcal R}(kv),
\label{eq:Omega_massive_integral}
\end{align}
where \(\overline{I_{\vec n}^{\,2}}\) is obtained from the convolution of
the scalar source with the massive Green function.  In the limit
\(m_{\vec n}\to0\) and \(\mathcal C_{\vec n}\to1\), the standard
massless kernel is recovered.

The full six-dimensional tensor spectrum can be written formally as
\begin{equation}
\Omega_{\rm GW}^{(6d)}(k)
=
\Omega_{\rm GW}^{(0)}(k)
+
\sum_{\vec n\neq0}
\Omega_{\rm GW}^{(\vec n)}(k),
\label{eq:total_6D}
\end{equation}
where \(\Omega_{\rm GW}^{(0)}\) denotes the massless zero-mode
contribution and \(\Omega_{\rm GW}^{(\vec n)}\) denotes the contribution
from the massive KK tensor mode. A first-principles evaluation of the massive contribution requires the overlap coefficients \(\mathcal C_{\vec n}\), the massive tensor Green
functions \(G_{\vec n}(k;\eta,\tilde\eta)\), the subsequent transfer of
massive tensor excitations to the present epoch, and the response of a
given detector to massive tensor polarizations \cite{Nishizawa:2009bf,deRham:2014zqa}.  These ingredients
depend on the localization of the scalar source and on the
brane-to-bulk coupling, and are therefore not fixed uniquely by the
low-energy two-dark-dimensions setup.

To illustrate the possible size of KK effects in the regime where the
tower is kinematically accessible, we introduce a phenomenological
six-dimensional extension of the zero-mode result.  We write
\begin{equation}
\Omega_{\rm GW}^{6d,{\rm ph}}(k)
=
\Omega_{\rm GW}^{(0)}(k)
\left[
1+
\xi_{\rm KK}
\sum_{\vec n\neq0}
\mathcal C_{\vec n}^{2}\,
\Theta\!\left(H_{\rm form}-m_{\vec n}\right)
\left(
1-\frac{m_{\vec n}^{2}}{H_{\rm form}^{2}}
\right)^p
\mathcal T_{\vec n}^{2}(k)
\right].
\label{eq:Omega_6D_pheno}
\end{equation}
Here \(\xi_{\rm KK}\) parametrizes the efficiency with which the scalar
source excites massive tensor modes, \(\mathcal C_{\vec n}\) is the
overlap of the source with the KK wavefunction, the step function
implements the kinematic threshold \(H_{\rm form}>m_{\vec n}\), and the
power \(p\) controls the threshold behaviour.  The factor
\(\mathcal T_{\vec n}(k)\) represents the transfer of the massive mode
from production to observation.  In the illustrative curves shown below
we use this expression only as a phenomenological model, taking a smooth
overlap-suppressed tower, for example
\begin{equation}
\mathcal C_{\vec n}^{2}
=
\exp\!\left[-\frac{n_1^2+n_2^2}{N_{\rm loc}^{2}}\right],
\label{eq:overlap_toy}
\end{equation}
with \(N_{\rm loc}\) controlling how many KK levels couple efficiently
to the source.  Eq~\eqref{eq:Omega_6D_pheno} should therefore not
be interpreted as an exact six-dimensional prediction; rather, it is a
controlled phenomenological parameterization of possible KK-tower
effects.

The relevance of the massive tower depends on the formation scale.  The
boundary between the four-dimensional and six-dimensional regimes is
defined by \(H_{\rm form}=m_{\rm KK}\).  For the KK gap used in this
work this corresponds to
\begin{equation}
M_{\rm KK}
\simeq
8.5\times10^{23}\,{\rm g}.
\end{equation}
Equivalently, using Eq \eqref{eq:f_M_relation} this boundary corresponds to
\begin{equation}
f_{\rm KK}
\simeq
1.2\times10^{-4}\,{\rm Hz}.
\end{equation}

For \(M_{\rm PBH}>M_{\rm KK}\), or equivalently
\(f_*<f_{\rm KK}\), the formation scale satisfies
\(H_{\rm form}<m_{\rm KK}\). 
The massive KK tower is then
kinematically inaccessible, and the standard four-dimensional SIGW
formalism is a valid effective description.  This is Regime~I.  In
particular, PTA-band frequencies correspond to very large PBH masses,
for example

\begin{equation}
f_*\sim10^{-9}\,{\rm Hz}
\quad\Longleftrightarrow\quad
M_{\rm PBH}\sim10^{34}\,{\rm g},
\end{equation}
which lies far above \(M_{\rm KK}\).  Therefore PTA-band SIGWs belong to
Regime~I and can be treated self-consistently within the standard
four-dimensional SIGW calculation \cite{NANOGrav:2023gor,EPTA:2023fyk}.

For \(M_{\rm PBH}<M_{\rm KK}\), or equivalently \(f_*>f_{\rm KK}\), the
formation scale satisfies \(H_{\rm form}>m_{\rm KK}\).  The KK tower is
then kinematically accessible, and the four-dimensional result should no
longer be interpreted as the complete six-dimensional prediction.  This
is Regime~II.  The memory-burdened PBH dark-matter window considered in
this work,
\begin{equation}
10^{-3}\,{\rm g}
\lesssim
M_{\rm PBH}
\lesssim
10^{21}\,{\rm g},
\end{equation}
lies entirely below \(M_{\rm KK}\), and therefore belongs to Regime~II.

As a useful benchmark inside Regime~II, consider
\begin{equation}
M_{\rm PBH}=10^{10}\,{\rm g}.
\end{equation}
Using Eq.~\eqref{eq:Hform_validity}, one finds
\begin{equation}
H_{\rm form}
\simeq
2.7~{\rm TeV}
\left(
\frac{10^{10}\,{\rm g}}{M_{\rm PBH}}
\right).
\end{equation}
This is many orders of magnitude larger than the KK gap used in this
work, \(m_{\rm KK}\simeq3.1\times10^{-2}\,{\rm eV}\).  Thus the
\(10^{10}\,{\rm g}\) benchmark lies deep in the KK-accessible regime.
If many KK tensor modes have unsuppressed overlap coefficients
\(\mathcal C_{\vec n}\) and efficient transfer factors
\(\mathcal T_{\vec n}\), their summed contribution can be comparable to,
or even larger than, the zero-mode contribution.  In two compact
dimensions the possible enhancement is controlled parametrically by the
number of accessible KK states.  The actual size of the correction,
however, is model-dependent because it depends on the source
localization, the brane-to-bulk coupling, and the massive-mode transfer
functions.  Therefore, in Regime~II we keep the four-dimensional SIGW
spectrum as the massless zero-mode benchmark and use the KK-tower
extension only phenomenologically.

For experimental comparisons in Regime~II, we use the massless zero-mode
spectrum as the robust benchmark,
\begin{equation}
\Omega_{\rm GW}^{\rm bench}(k)
\equiv
\Omega_{\rm GW,0}^{(0)}(k).
\label{eq:bench}
\end{equation}
This benchmark represents the part of the signal continuously connected
to the standard four-dimensional scalar-induced gravitational-wave
background.  The phenomenological six-dimensional expression in
Eq.~\eqref{eq:Omega_6D_pheno} is used only to illustrate how the signal
could be modified if a non-negligible fraction of the tensor power is
transferred into the accessible KK tower.

\begin{figure}[t]
\centering
\includegraphics[width=0.8\columnwidth]{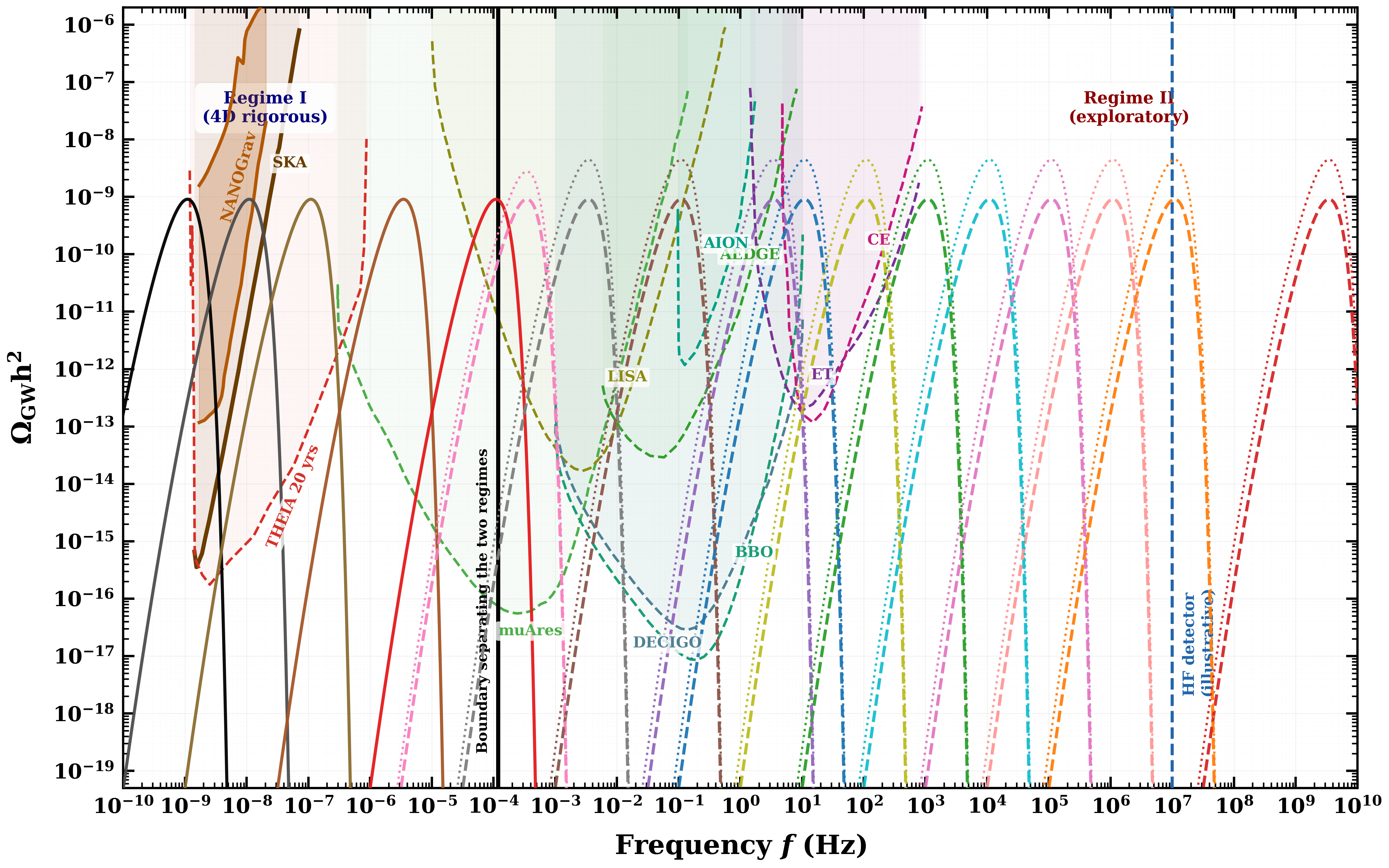}
\caption{Scalar-induced gravitational-wave spectra associated with primordial
black holes in the six-dimensional two-dark-dimensions scenario. The colored
solid and dashed curves show the standard four-dimensional zero-mode benchmark,
$\Omega_{\rm GW}^{(0)}h^2$, for representative PBH masses. Solid model curves
correspond to Regime~I, $M_{\rm PBH}>M_{\rm KK}$, where the four-dimensional
calculation is self-consistent, whereas dashed model curves correspond to
Regime~II, $M_{\rm PBH}<M_{\rm KK}$, where the Kaluza--Klein tower is
kinematically accessible. The colored dotted curves, when present, illustrate
a phenomenological six-dimensional KK-tower extension and should not be
interpreted as a first-principles six-dimensional prediction. The vertical
black line marks the boundary between the two regimes,
$f_{\rm KK}^{\rm regime}\simeq1.2\times10^{-4}\,{\rm Hz}$, corresponding to
$M_{\rm KK}\simeq8.5\times10^{23}\,{\rm g}$. The experimental sensitivity
curves are obtained from the corresponding numerical datasets. SKA is shown by a solid boundary, while
THEIA (20 yr), $\mu$Ares, LISA, AION, AEDGE, DECIGO, BBO, ET, and CE are shown
by dashed boundaries. The filled NANOGrav region denotes the interval between
its two numerical confidence boundaries. The blue dashed vertical line
indicates an illustrative high-frequency detector scale.}
\label{fig:SIGW_spectra}
\end{figure}

Figure~\ref{fig:SIGW_spectra} displays the SIGW spectra for
representative PBH masses.  The solid curves denote masses in
Regime~I, where the four-dimensional SIGW calculation is self-consistent.
The dashed curves denote masses in Regime~II, where the plotted
zero-mode spectra should be read as benchmarks rather than complete
six-dimensional predictions.  When shown, the phenomenological 6D curve
is an illustrative KK-tower extension based on
Eq.~\eqref{eq:Omega_6D_pheno}, not a first-principles prediction of the
full compactified theory.

\section{Fisher Forecasts}
\label{sec:fisher}

To estimate how well future space-based gravitational-wave experiments
could reconstruct the parameters of the benchmark scalar-induced
gravitational-wave spectrum, we perform a Fisher matrix analysis
\cite{Cutler:1994ys,Tegmark:1996bz}.  Since the full six-dimensional
massive KK contribution is model-dependent, the forecast is applied only
to the massless zero-mode benchmark \(\Omega_{\rm GW}^{\rm bench}\)
defined in Eq.~(\ref{eq:bench}).  The results should therefore be
interpreted as projected sensitivities to the four-dimensional zero-mode
component rather than to the complete six-dimensional tensor spectrum.

We parametrize the benchmark model by
\begin{equation}
\boldsymbol{\theta}
=
\left(
\log_{10}\frac{M_{\rm PBH}}{{\rm g}},
\log_{10} f_{\rm PBH},
\sigma
\right),
\end{equation}
where \(M_{\rm PBH}\) fixes the peak frequency through
Eq.~(\ref{eq:f_M_relation}), \(f_{\rm PBH}\) controls the overall PBH
abundance, and \(\sigma\) determines the width of the log-normal mass
function.  We use a narrow fiducial width \(\sigma=0.1\), for which the
correspondence between the PBH mass and the peak frequency is sharp.
Using Eq.~(\ref{eq:f_M_relation}), the LISA-band benchmark is represented
by
\[
M_{\rm PBH}=10^{21}\,{\rm g},
\qquad
f_*\simeq3.5\times10^{-3}\,{\rm Hz},
\]
while the DECIGO/BBO-band benchmark is represented by
\[
M_{\rm PBH}=10^{18}\,{\rm g},
\qquad
f_*\simeq0.11\,{\rm Hz}.
\]
Both benchmarks lie below the KK-threshold mass
\(M_{\rm KK}\simeq8.5\times10^{23}\,{\rm g}\), and hence belong to
Regime~II.  Therefore the Fisher analysis probes the zero-mode benchmark
spectrum in the regime where a complete six-dimensional prediction would
require additional information about the massive KK tower.

For an experiment \(X\), we denote the effective noise energy density by
\begin{equation}
\Omega_{{\rm noise},X}(f)
=
\frac{2\pi^2}{3H_0^2}
f^3 S_{n,X}(f),
\label{eq:Omega_noise}
\end{equation}
where \(S_{n,X}(f)\) is the strain-noise power spectral density and
\(H_0=67.4\,{\rm km\,s^{-1}\,Mpc^{-1}}\).  We use representative
analytic noise models for LISA, DECIGO, and BBO
\cite{Robson:2018ifk,Yagi:2011wg,Seto:2001qf,Corbin:2005ny,
Thrane:2013oya}.  In the simplified notation used here,
\begin{align}
S_n^{\rm LISA}(f)
&=
\frac{10}{3}
\left[
\frac{4\times10^{-42}}{f^{4}}
+
\frac{1.6\times10^{-41}}{f^{2}}
+
1.2\times10^{-43}
\right]
\,{\rm Hz}^{-1},
\\
S_n^{\rm DECIGO}(f)
&=
1.0\times10^{-46}
+
2.5\times10^{-49}f^{-2}
+
6.0\times10^{-51}f^{-4}
\,{\rm Hz}^{-1},
\\
S_n^{\rm BBO}(f)
&=
4.0\times10^{-48}
+
4.0\times10^{-49}f^{-2}
+
1.0\times10^{-49}f^{-4}
\,{\rm Hz}^{-1}.
\end{align}
These expressions are used as approximate sensitivity models.  A
mission-specific forecast would require the full detector response
functions, observation time, foreground modelling, and cross-correlation
strategy.

For logarithmically spaced frequency bins, we compute the Fisher matrix
as
\begin{equation}
F_{ij}
=
T_{\rm obs}
\sum_n
\frac{f_n\,\Delta\ln f}
{\Omega_{\rm noise}^2(f_n)}
\left.
\frac{\partial \Omega_{\rm GW}^{\rm bench}(f_n)}
{\partial\theta_i}
\right|_{\boldsymbol{\theta}_0}
\left.
\frac{\partial \Omega_{\rm GW}^{\rm bench}(f_n)}
{\partial\theta_j}
\right|_{\boldsymbol{\theta}_0},
\label{eq:fisher_matrix}
\end{equation}
where \(T_{\rm obs}\) is the observation time and the derivatives are
evaluated numerically around the fiducial parameter point
\(\boldsymbol{\theta}_0\).  In the numerical implementation we use a
five-point finite-difference stencil.  The covariance matrix is then
\begin{equation}
C_{ij}
=
(F^{-1})_{ij},
\end{equation}
and the marginalized \(1\sigma\) uncertainty on parameter \(\theta_i\)
is
\begin{equation}
\sigma(\theta_i)
=
\sqrt{C_{ii}}.
\end{equation}

The fiducial values and resulting illustrative uncertainties are
summarized in Table~\ref{tab:fisher}.  Since the PBH abundance satisfies
\(f_{\rm PBH}\leq1\), the Fisher expansion in
\(\log_{10}f_{\rm PBH}\) should be interpreted as a local sensitivity
estimate around the fiducial point rather than as a global posterior
constraint.

\begin{table}[h]
\centering
\caption{
Illustrative Fisher forecast \(1\sigma\) uncertainties for the
zero-mode benchmark SIGW spectrum.  The LISA fiducial mass is
\(M_{\rm PBH}=10^{21}\,{\rm g}\), corresponding to
\(f_*\simeq3.5\times10^{-3}\,{\rm Hz}\), while the DECIGO/BBO fiducial
mass is \(M_{\rm PBH}=10^{18}\,{\rm g}\), corresponding to
\(f_*\simeq0.11\,{\rm Hz}\).  We take \(f_{\rm PBH}=1\) and
\(\sigma=0.1\).  These forecasts apply to the zero-mode benchmark in
Regime~II and should not be interpreted as constraints on the full
six-dimensional KK tower.
}
\label{tab:fisher}
\begin{tabular}{lccc}
\hline
Parameter & LISA & DECIGO & BBO \\
\hline
\(\log_{10}(M_{\rm PBH}/{\rm g})\)
&
\(21.00\pm0.15\)
&
\(18.00\pm0.12\)
&
\(18.00\pm0.08\)
\\
\(\log_{10}f_{\rm PBH}\)
&
\(0.00\pm0.10\)
&
\(0.00\pm0.08\)
&
\(0.00\pm0.05\)
\\
\(\sigma\)
&
\(0.10\pm0.02\)
&
\(0.10\pm0.02\)
&
\(0.10\pm0.01\)
\\
\hline
\end{tabular}
\end{table}

\begin{figure}[t]
\centering
\includegraphics[width=\columnwidth]{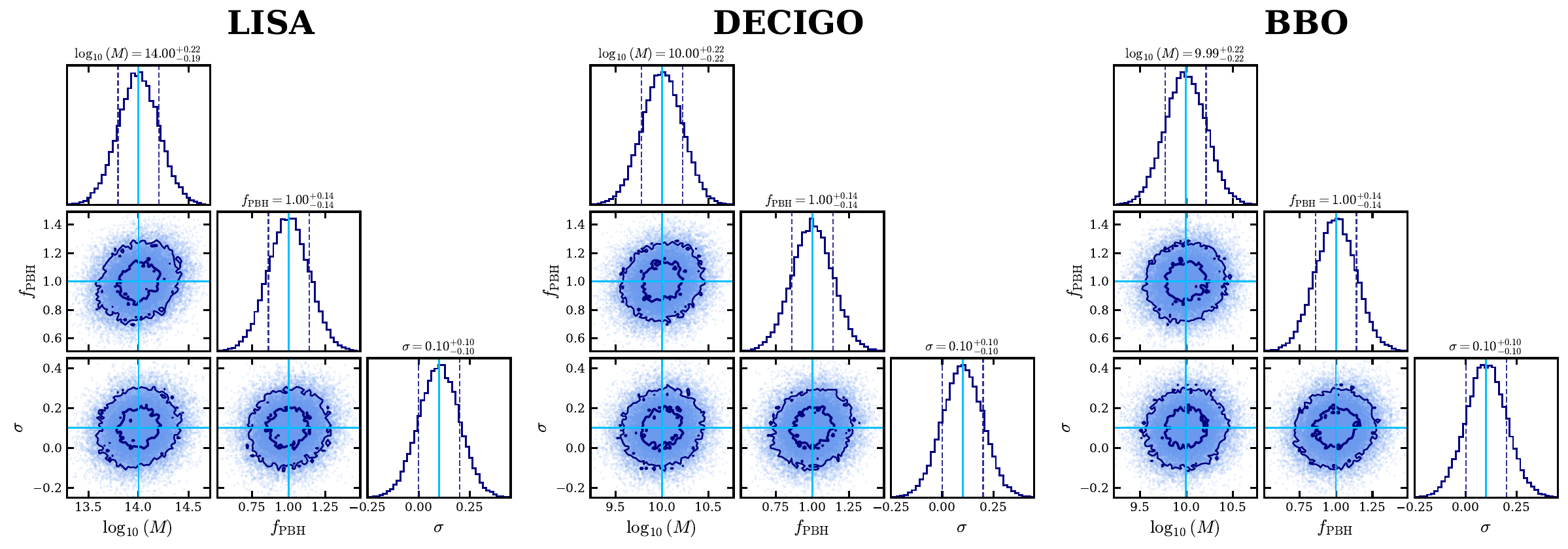}
\caption{
Corner plots for the Fisher forecasts of LISA, DECIGO, and BBO using
the zero-mode benchmark SIGW spectrum.  The diagonal panels show the
marginalized one-dimensional constraints, while the off-diagonal panels
show the corresponding \(1\sigma\) and \(2\sigma\) confidence regions.
}
\label{fig:corner}
\end{figure}

The corresponding corner plots are displayed in Fig.~\ref{fig:corner}.
BBO gives the tightest constraints among the three benchmark experiments,
mainly because its sensitivity overlaps well with the deci-Hz peak
associated with \(M_{\rm PBH}\sim10^{18}\,{\rm g}\).  The forecasted
contours quantify the ability to reconstruct the peak position,
amplitude, and width of the zero-mode SIGW spectrum.  They do not include
the memory-burden exponent \(p\), which affects the allowed PBH mass
window rather than directly modifying the zero-mode SIGW kernel used in
the forecast.  A future extension could include \(p\), together with
phenomenological KK parameters such as \(\xi_{\rm KK}\) and \(N_{\rm loc}\),
once a concrete six-dimensional massive-mode model is specified.

\section{Collider implications and multimessenger consistency tests}
\label{sec:collider}

The fundamental gravitational scale in the two-dark-dimensions scenario,
\(M_*\sim10~\mathrm{TeV}\), is sufficiently low that partonic collisions at a
future \(100~\mathrm{TeV}\) proton--proton collider could, in principle, probe
higher-dimensional gravity.  Relevant signatures include real and virtual
Kaluza--Klein (KK) graviton effects and, if the partonic energy is sufficiently
above \(M_*\), the production of microscopic black holes
\cite{Arkani-Hamed:1998jmv,Dimopoulos:2001hw,Giddings:2001bu}.  The latter are
localized near the Standard-Model brane at formation but have horizon radii
much smaller than the compactification radius.  They therefore behave as
genuinely six-dimensional objects.

The collider discussion below is intended as a study of parametric scaling and
benchmark sensitivity.  A quantitative discovery forecast would additionally
require a specified parton distribution function (PDF) set, the treatment of
energy lost during black-hole formation, spin-dependent greybody factors,
parton showering, hadronization, and detector response.  We therefore
distinguish the robust six-dimensional mass scalings from normalization- and
simulation-dependent rate estimates.

\subsection{Semiclassical black-hole production}
\label{sec:production}

In the semiclassical picture, a microscopic black hole can form in a
trans-Planckian parton--parton collision when the impact parameter is smaller
than the corresponding higher-dimensional horizon radius
\cite{Dimopoulos:2001hw,Giddings:2001bu,Giudice:2001ce}.  Neglecting the energy
radiated during formation, the partonic cross section is commonly approximated
by the geometric expression
\begin{equation}
 \hat\sigma_{ij\to{\rm BH}}(\hat s)
 \simeq
 F_{\rm BH}\,\pi r_h^2(\sqrt{\hat s})
 \Theta(\sqrt{\hat s}-M_{\rm min}),
 \label{eq:sigmahat}
\end{equation}
where \(F_{\rm BH}=\mathcal O(1)\) parametrizes formation effects.  The
semiclassical threshold is written as
\begin{equation}
 M_{\rm min}=x_{\rm min}M_* .
 \label{eq:Mmincoll}
\end{equation}
We take \(x_{\rm min}=5\) as a conservative benchmark.  This choice reduces
the importance of the quantum-gravity regime, but it does not eliminate the
theoretical uncertainty associated with the transition from a few-body
collision to a semiclassical black hole
\cite{Yoshino:2002tx,Yoshino:2005hi,Kanti:2004nr,Kanti:2014dxa}.

For two extra dimensions, the Schwarzschild--Tangherlini radius may be written
as
\begin{equation}
 r_h(M_{\rm BH})
 =
 \frac{k_{\rm BH}}{M_*}
 \left(\frac{M_{\rm BH}}{M_*}\right)^{1/3},
 \label{eq:rh_prod}
\end{equation}
where \(k_{\rm BH}\) is an order-one constant fixed by the convention used to
define the higher-dimensional Planck scale.  With the benchmark identification
\(M_{\rm BH}=\sqrt{\hat s}\), Eqs.~\eqref{eq:sigmahat} and
\eqref{eq:rh_prod} give
\begin{equation}
 \hat\sigma_{ij\to{\rm BH}}(\hat s)
 \simeq
 \frac{F_{\rm BH}\pi k_{\rm BH}^2}{M_*^2}
 \left(\frac{\sqrt{\hat s}}{M_*}\right)^{2/3}
 \Theta(\sqrt{\hat s}-M_{\rm min}).
 \label{eq:sigma_explicit}
\end{equation}
The scaling \(\hat\sigma_{\rm BH}\propto\hat s^{1/3}\) follows directly from
the six-dimensional horizon relation and is independent of the overall
Planck-scale convention.  In a more complete treatment, only a fraction
\(y<1\) of the partonic energy may be trapped behind the horizon, in which case
\(M_{\rm BH}=y\sqrt{\hat s}\) and the effective partonic threshold is raised.
Consequently, Eq.~\eqref{eq:sigma_explicit} should be regarded as a benchmark
geometric estimate rather than a precision production cross section.

\subsubsection{Proton--proton production rate}

The proton--proton cross section is obtained by convoluting the partonic result
with the parton luminosities,
\begin{equation}
 \sigma_{pp\to{\rm BH}}(s)
 =
 \sum_{i,j}\int_{\tau_{\rm min}}^1 d\tau\,
 \frac{d{\cal L}_{ij}}{d\tau}\,
 \hat\sigma_{ij\to{\rm BH}}(\tau s),
 \label{eq:ppBH}
\end{equation}
where
\begin{equation}
 \tau_{\rm min}=\frac{M_{\rm min}^2}{s},
 \qquad
 \frac{d{\cal L}_{ij}}{d\tau}
 =
 \int_\tau^1\frac{dx}{x}
 f_i(x,Q)f_j\!\left(\frac{\tau}{x},Q\right),
 \label{eq:partonlum}
\end{equation}
and \(Q\) is the factorization scale.  For
\(M_*=10~\mathrm{TeV}\), \(x_{\rm min}=5\), and
\(\sqrt{s}=100~\mathrm{TeV}\), one obtains
\begin{equation}
 M_{\rm min}=50~\mathrm{TeV},
 \qquad
 \tau_{\rm min}=0.25.
 \label{eq:FCCthreshold}
\end{equation}
Thus the threshold region typically probes momentum fractions of order
\(x_1\sim x_2\sim0.5\), where the parton luminosities are already strongly
suppressed.  The fact that \(M_{\rm min}<\sqrt{s}\) therefore establishes only
kinematic accessibility; it does not guarantee an observable event rate.
Moreover, a black-hole mass of \(100~\mathrm{TeV}\) lies at the hadronic
endpoint, where the PDF-convoluted rate vanishes.  Observable events, if
present, would be concentrated closer to the lower end of the allowed mass
interval.

For an integrated luminosity \({\cal L}_{\rm int}\), the expected number of
produced black holes is
\begin{equation}
 N_{\rm BH}
 =
 \sigma_{pp\to{\rm BH}}{\cal L}_{\rm int}
 =
 2.0\times10^4
 \left(\frac{\sigma_{pp\to{\rm BH}}}{1~\mathrm{fb}}\right)
 \left(\frac{{\cal L}_{\rm int}}{20~\mathrm{ab}^{-1}}\right).
 \label{eq:NBH}
\end{equation}
Equation~\eqref{eq:NBH} is a conversion between cross section and event yield,
not a prediction that the cross section is of femtobarn size.  Establishing
such a rate requires evaluating Eq.~\eqref{eq:ppBH} with modern PDFs and
propagating the uncertainties in \(x_{\rm min}\), \(F_{\rm BH}\), the trapped
energy fraction, and the Planck-scale convention.

\begin{figure}[t]
 \centering
 \begin{subfigure}{0.48\columnwidth}
  \centering
  \includegraphics[width=\linewidth]{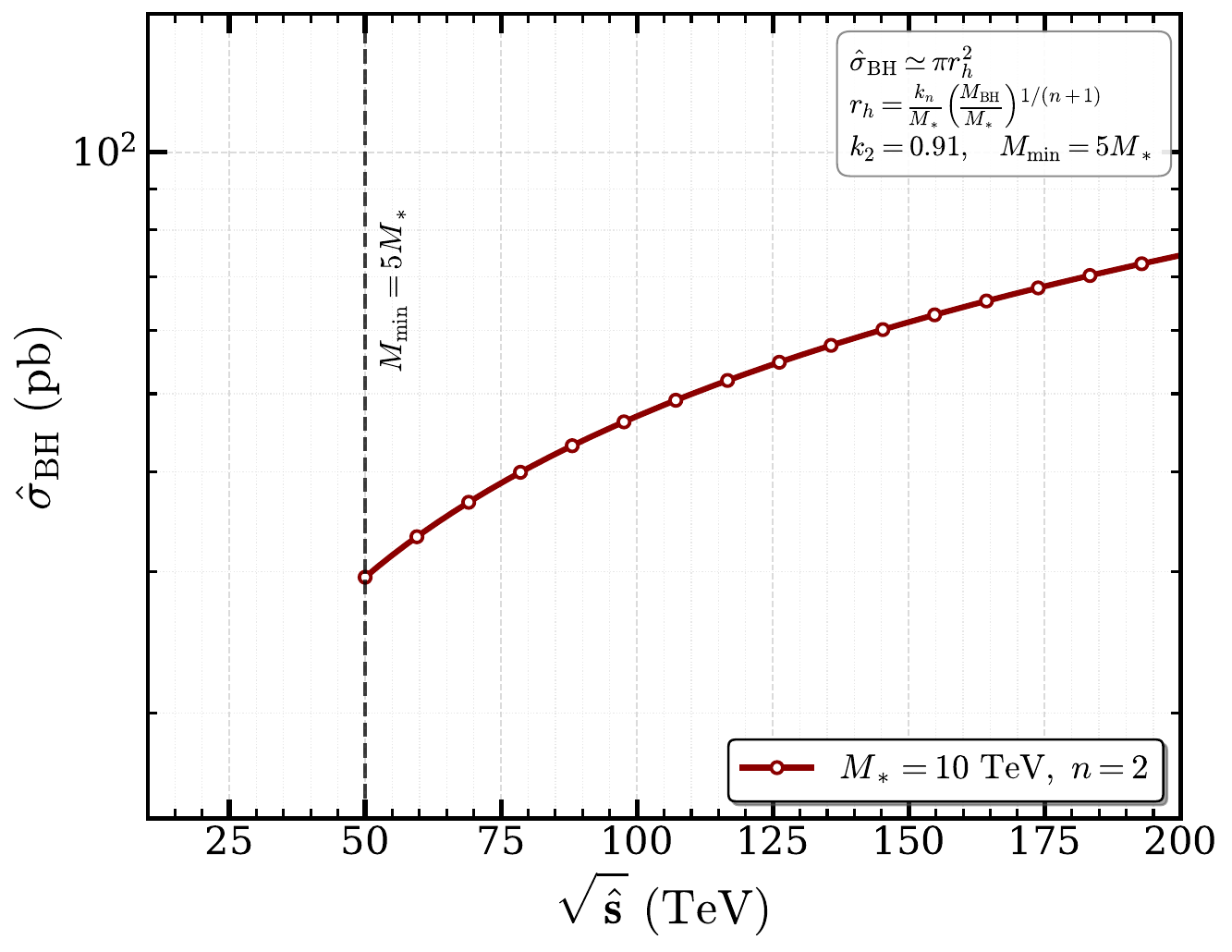}
 \end{subfigure}
 \hfill
 \begin{subfigure}{0.48\columnwidth}
  \centering
  \includegraphics[width=\linewidth]{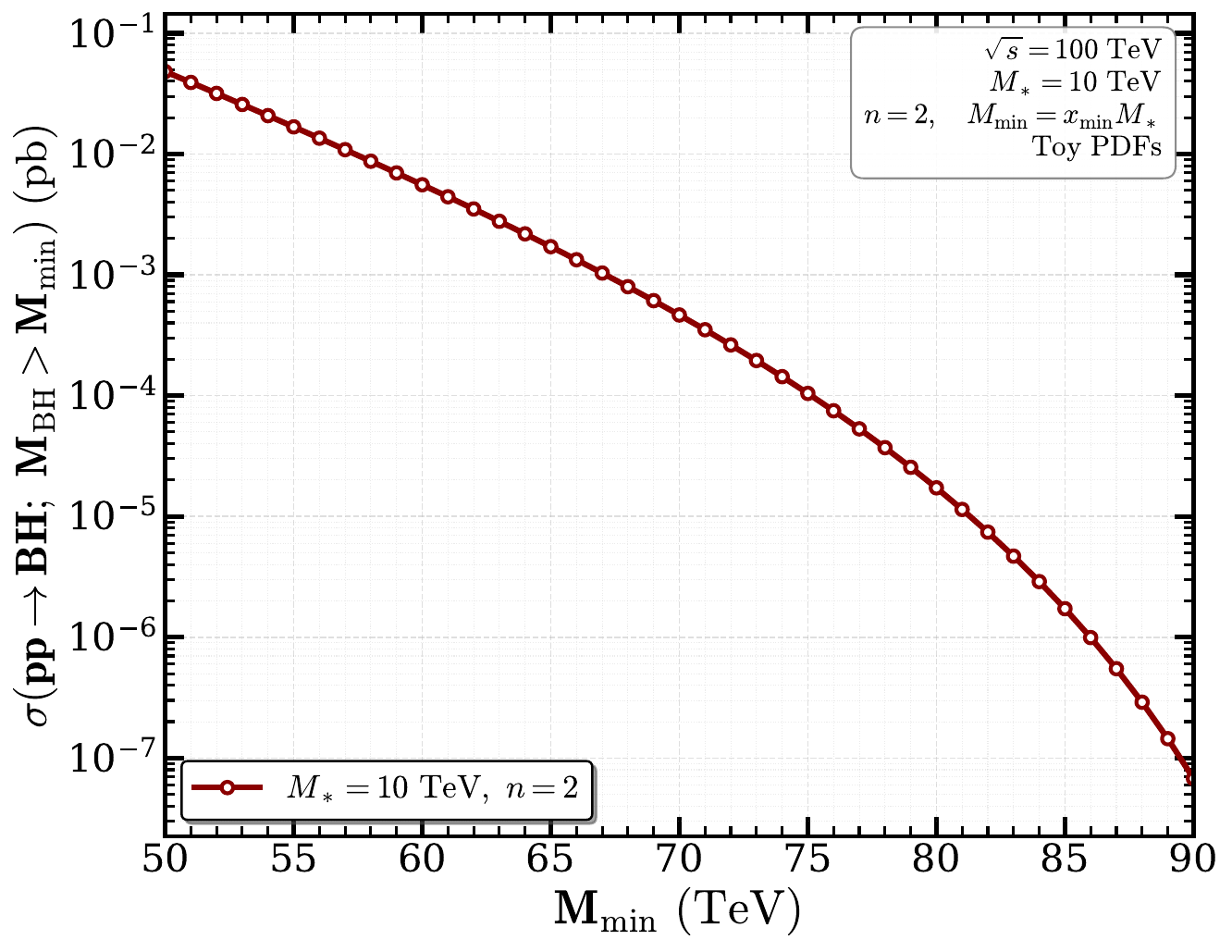}
 \end{subfigure}
 \caption{Benchmark production cross sections for microscopic
 six-dimensional black holes.  The left panel shows the geometric partonic
 estimate, while the right panel shows the corresponding PDF-convoluted
 proton--proton rate at \(\sqrt{s}=100~\mathrm{TeV}\).  We take
 \(M_*=10~\mathrm{TeV}\), \(x_{\rm min}=5\), and \(n=2\).  The robust
 geometric scaling is \(\hat\sigma_{\rm BH}\propto\hat s^{1/3}\); the
 normalization and hadronic rate depend on the Planck-scale convention,
 formation efficiency, PDF choice, factorization scale, and threshold
 prescription.}
 \label{fig:BH_production_summary}
\end{figure}

These two levels of the calculation are compared in
Fig.~\ref{fig:BH_production_summary}.  The left panel isolates the geometric
partonic result in Eq.~\eqref{eq:sigma_explicit}.  Above the imposed threshold,
the curve rises slowly with \(\sqrt{\hat s}\) because a heavier
six-dimensional black hole has a larger horizon,
\(r_h\propto M_{\rm BH}^{1/3}\), and hence a larger capture area,
\(\pi r_h^2\propto M_{\rm BH}^{2/3}\).  The sharp onset at
\(M_{\rm min}=50~\mathrm{TeV}\) is produced by the step-function threshold and
should not be interpreted as a physical discontinuity.  Quantum formation
effects would generally smooth this transition.

The right panel includes the proton structure through
Eq.~\eqref{eq:ppBH}.  It consequently exhibits a qualitatively different
behavior: the growth of the geometric area is overwhelmed by the rapid fall
of the large-\(x\) parton luminosities.  This contrast is important.  The
partonic curve describes the strength of a collision at fixed partonic
energy, whereas the hadronic curve also includes the probability of finding
the required partons inside the two protons.  The right panel is therefore the
relevant one for estimating an FCC-hh yield, subject to the theoretical and
PDF assumptions stated above.

Figure~\ref{fig:BH_production_summary} also shows why the nominal collider
energy should not be identified with a typical black-hole mass.  Moving from
the threshold toward \(100~\mathrm{TeV}\) increases the elementary geometric
cross section only as a mild power, while forcing \(x_1x_2\) toward unity.
The resulting loss of parton luminosity strongly favors black holes produced
near the lowest trusted mass.  Consequently, the phenomenological reach is
controlled jointly by \(M_*\) and \(x_{\rm min}\), and a modest change in the
threshold can produce a large change in the predicted event rate.

\subsection{Hawking evaporation and inclusive event signatures}
\label{sec:evaporation}

Following formation and an initial balding and spin-down stage, a sufficiently
semiclassical black hole is expected to enter a quasi-stationary Hawking
evaporation phase \cite{Hawking:1975vcx,Dimopoulos:2001hw,Giddings:2001bu,
Kanti:2004nr}.  Combining the six-dimensional temperature relation with
Eq.~\eqref{eq:rh_prod} gives
\begin{equation}
 T_{\rm BH}
 =
 \frac{3}{4\pi k_{\rm BH}}M_*
 \left(\frac{M_{\rm BH}}{M_*}\right)^{-1/3}.
 \label{eq:Tcoll}
\end{equation}
The scaling \(T_{\rm BH}\propto M_{\rm BH}^{-1/3}\) is fixed by the number of
extra dimensions, whereas its normalization depends on \(k_{\rm BH}\).  For
\(M_*=10~\mathrm{TeV}\), \(k_{\rm BH}=0.7\)--\(1\), and
\(M_{\rm BH}=50~\mathrm{TeV}\), the temperature is approximately
\begin{equation}
 T_{\rm BH}\simeq1.4\text{--}2.0~\mathrm{TeV}.
 \label{eq:Tthreshold}
\end{equation}
All Standard-Model species are consequently kinematically accessible.
Nevertheless, the emission is not exactly democratic: the relative rates are
weighted by internal degrees of freedom and by spin-, frequency-, and
dimension-dependent greybody factors
\cite{Kanti:2004nr,Cardoso:2005vb,Ida:2002ez,Ida:2005ax,Kanti:2014dxa}.

The mean primary multiplicity can be estimated as
\begin{equation}
 \langle N\rangle
 \simeq
 \frac{M_{\rm BH}}{\chi T_{\rm BH}}
 =
 \frac{4\pi k_{\rm BH}}{3\chi}
 \left(\frac{M_{\rm BH}}{M_*}\right)^{4/3},
 \label{eq:multcoll}
\end{equation}
where \(\chi\simeq2.7\)--\(3.2\) represents the mean energy per emitted
quantum in units of the Hawking temperature.  At the benchmark threshold
\(M_{\rm BH}=50~\mathrm{TeV}\), Eq.~\eqref{eq:multcoll} gives
\begin{equation}
 \langle N\rangle\simeq8\text{--}13.
 \label{eq:Nthreshold}
\end{equation}
The larger estimate \(\langle N\rangle\sim20\)--\(35\), obtained at
\(M_{\rm BH}\sim100~\mathrm{TeV}\), describes the mass scaling but not a
realistically populated FCC-hh benchmark because that mass lies at the
hadronic endpoint.

The resulting inclusive topology consists of energetic jets, charged leptons,
photons, heavy Standard-Model particles, neutrinos, and possibly bulk
gravitational radiation.  Compared with ordinary QCD multijet production, the
most useful discriminants are expected to be the large scalar sum of
transverse momenta, moderately high object multiplicity, approximate event
isotropy, and the presence of energetic leptons or photons.  These statements
are qualitative until the Hawking decay is interfaced with parton showering
and a detector simulation.

\begin{figure}[t]
 \centering
 \includegraphics[width=\columnwidth]
 {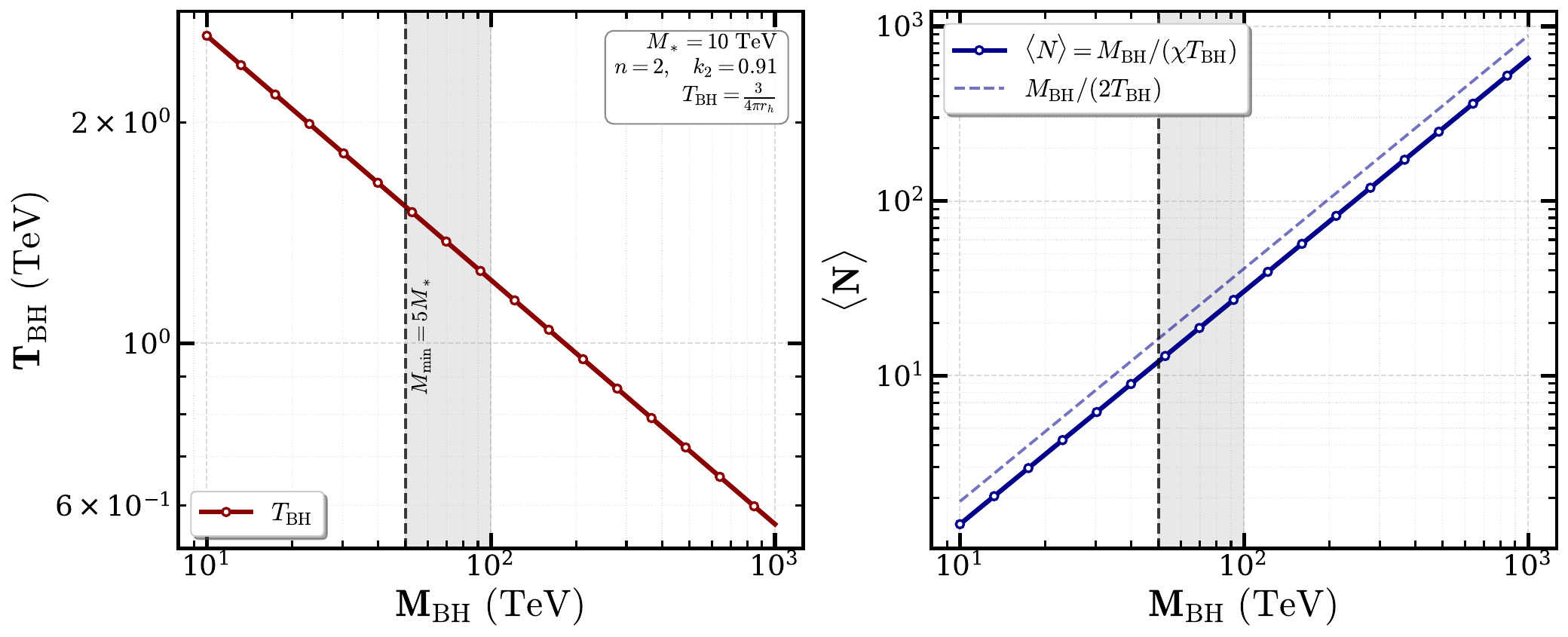}
 \caption{Hawking temperature and estimated primary multiplicity of a
 six-dimensional microscopic black hole.  The robust scalings are
 \(T_{\rm BH}\propto M_{\rm BH}^{-1/3}\) and
 \(\langle N\rangle\propto M_{\rm BH}^{4/3}\).  The normalization depends on
 \(k_{\rm BH}\), the mean emitted energy, and greybody factors.  Although the
 nominal kinematic range extends to \(M_{\rm BH}=100~\mathrm{TeV}\), the
 PDF-convoluted production rate is strongly concentrated toward the lower
 masses and vanishes at the endpoint.}
 \label{fig:T_mult}
\end{figure}

The complementary behavior of the temperature and multiplicity is displayed
in Fig.~\ref{fig:T_mult}.  The left panel decreases with mass according to
Eq.~\eqref{eq:Tcoll}: as the black-hole mass increases, its horizon grows and
its Hawking temperature falls.  The right panel rises because the larger mass
must then be distributed among quanta with a lower characteristic energy.
Combining these effects yields the stronger dependence
\(\langle N\rangle\propto M_{\rm BH}^{4/3}\).  Thus lower-mass black holes are
hotter and decay into fewer, harder primary quanta, whereas heavier black holes
are cooler and produce more highly populated final states.

In interpreting Fig.~\ref{fig:T_mult}, the region beginning at
\(M_{\rm min}=50~\mathrm{TeV}\) is the nominal
semiclassical domain for the benchmark adopted here.  The extension of the
curves to \(100~\mathrm{TeV}\) is useful for displaying the six-dimensional
scaling, but it should not be interpreted as a uniformly accessible FCC-hh
mass interval.  As demonstrated by Fig.~\ref{fig:BH_production_summary}, the
production probability decreases sharply across this range.  The experimentally
relevant events would therefore combine the relatively high temperatures and
moderate multiplicities found close to threshold.

\subsubsection{Bulk emission and missing transverse momentum}
\label{sec:MET}

The compactification scale is much larger than the microscopic horizon,
\(r_h\ll R\), while the Hawking temperature is much greater than the KK mass
spacing.  The black hole therefore resolves the higher-dimensional bulk, and
bulk graviton emission is not kinematically suppressed by the small KK mass
gap.  The power emitted into a species \(i\) can be expressed schematically as
\begin{equation}
 P_i
 =
 \sum_\ell\int_0^\infty\frac{d\omega}{2\pi}
 \frac{\omega\,\Gamma_{i\ell}(\omega)}
 {\exp(\omega/T_{\rm BH})\mp1},
 \label{eq:Pi}
\end{equation}
where \(\Gamma_{i\ell}\) denotes the appropriate greybody factor.  The
integrated invisible fraction is then
\begin{equation}
 f_{\rm inv}
 =
 \frac{P_{\rm bulk}+P_\nu}
 {P_{\rm SM}+P_{\rm bulk}},
 \qquad
 f_{\rm vis}
 =
 \frac{P_{\rm SM}-P_\nu}
 {P_{\rm SM}+P_{\rm bulk}},
 \label{eq:fvisfinv}
\end{equation}
where \(P_{\rm SM}\) includes all brane-localized Standard-Model species and
neutrino emission is included in the detector-level invisible contribution.

For a small number of extra dimensions, existing greybody studies generally
find that the large number of brane-localized Standard-Model degrees of freedom
keeps the visible channel dominant, although rotation can enhance bulk graviton
emission.  We may therefore use
\begin{equation}
 f_{\rm inv}=0.05\text{--}0.20
 \label{eq:finv_benchmark}
\end{equation}
as an illustrative nuisance-parameter range, rather than as a prediction of
the model.  A first-principles missing-transverse-momentum distribution would
require polarization- and spin-dependent greybody factors, the complete KK
spectrum, recoil and boost effects, parton showering, and detector acceptance.
\subsubsection{Illustrative polarization dependence of bulk emission}
\label{sec:scalar_tensor}
\begin{figure}[t]
 \centering
 \includegraphics[width=0.58\columnwidth]{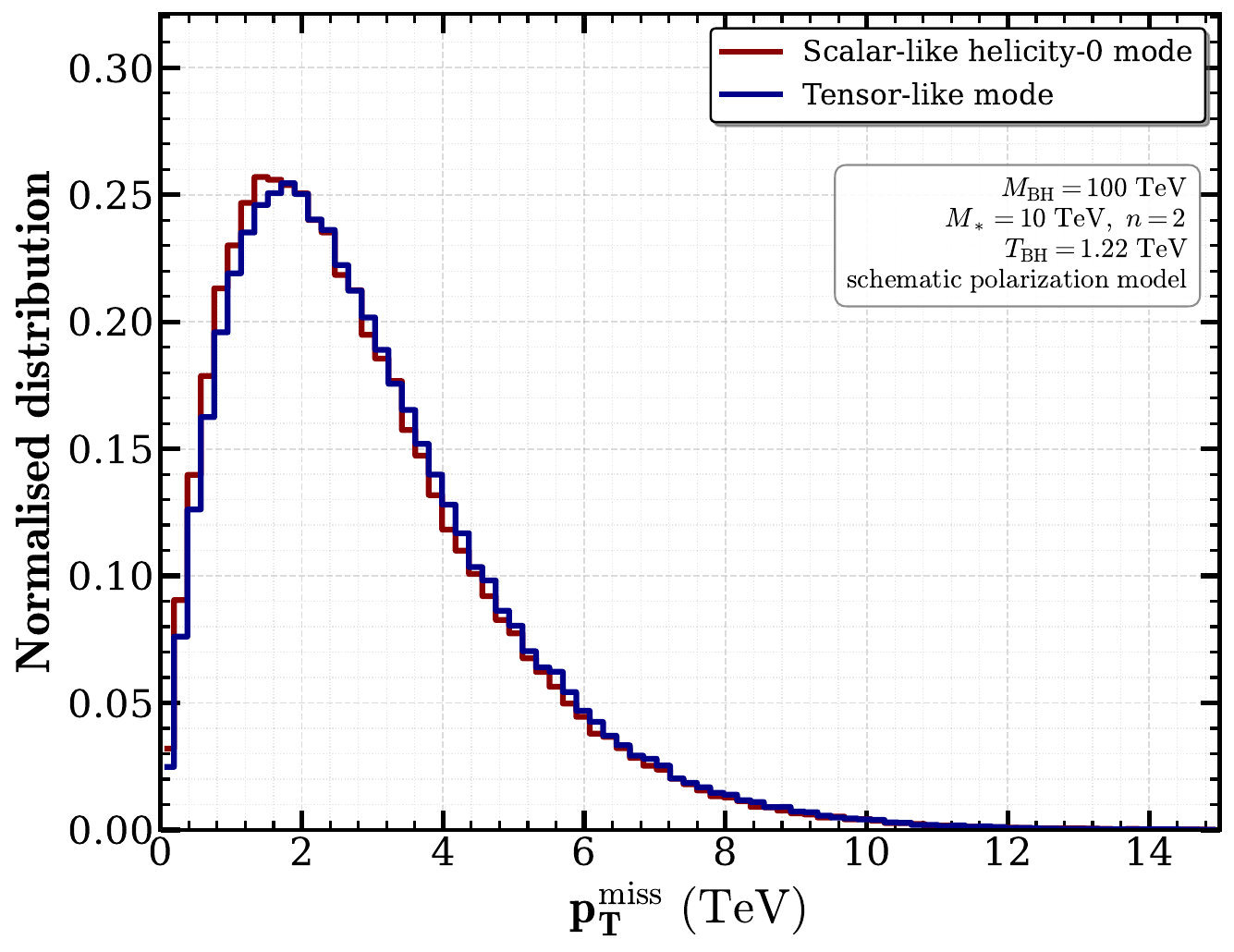}
 \caption{Illustrative missing-transverse-momentum templates for bulk
 graviton emission from a six-dimensional microscopic black hole.  The
 scalar-like and tensor-like curves follow from different angular-weighting
 hypotheses after a transverse boost of the decay system.  They are not
 first-principles KK-graviton predictions and do not include
 polarization-dependent greybody factors, the complete KK spectrum, parton
 showering, hadronization, or detector effects.}
 \label{fig:MET}
\end{figure}

From a four-dimensional viewpoint, a massive KK spin-2 excitation contains
helicity-\(\pm2\), helicity-\(\pm1\), and helicity-0 components.  Their relative
emission rates are determined by the corresponding couplings and
polarization-dependent greybody factors.  In principle, their different
angular distributions can modify the recoil of the visible Hawking system and
hence the missing-transverse-momentum spectrum.  This observation motivates
the comparison shown in Fig.~\ref{fig:MET}.

The two curves should not be interpreted as separate predictions for physical
KK eigenstates.  They are phenomenological templates obtained by imposing scalar-like and
tensor-like angular weights on the bulk emission and boosting the decay
system.  They illustrate the direction in which polarization-dependent
emission could affect the distribution.  A physical prediction would require
summing all KK polarizations with their correct couplings and greybody factors
and then including production kinematics, parton showering, and detector
response.  The figure therefore identifies a possible observable for a future
Monte Carlo study rather than a presently established discriminator.
Figure~\ref{fig:MET} should be read as a comparison of shapes rather than as a
rate calculation.  Any separation between the two curves originates from the
different assumed angular weights and from their mapping into transverse
momentum after the boost.  It demonstrates that bulk emission with the same
total invisible energy can lead to different recoil spectra if its angular or
polarization structure changes.  In an experimental analysis, however, this
effect would compete with neutrino missing momentum, fluctuations in the
visible Hawking decay, the distribution of black-hole boosts, and detector
resolution.  The physically meaningful question raised by the figure is
therefore whether a polarization-weighted signal template remains
distinguishable after all these contributions are included.

\subsubsection{Sensitivity to the number of extra dimensions}
\label{sec:n_determination}

For \(\mathfrak n\) extra dimensions, the temperature--mass relation has the
logarithmic slope
\begin{equation}
 \frac{d\log T_{\rm BH}}{d\log M_{\rm BH}}
 =
 -\frac{1}{\mathfrak n+1}.
 \label{eq:logT}
\end{equation}
A slope near \(-1/3\) would therefore be consistent with two extra dimensions.
In practice, the reconstructed mass is affected by invisible energy, while the
inferred temperature is distorted by greybody factors, black-hole rotation,
the finite multiplicity, and detector selection.  Equation~\eqref{eq:logT}
should consequently be viewed as a possible consistency test, rather than an
unambiguous determination of \(\mathfrak n\).  A quantitative discrimination
among different dimensionalities requires a likelihood analysis based on
simulated signal and background samples.

\begin{figure}[t]
 \centering
 \includegraphics[width=0.58\columnwidth]
 {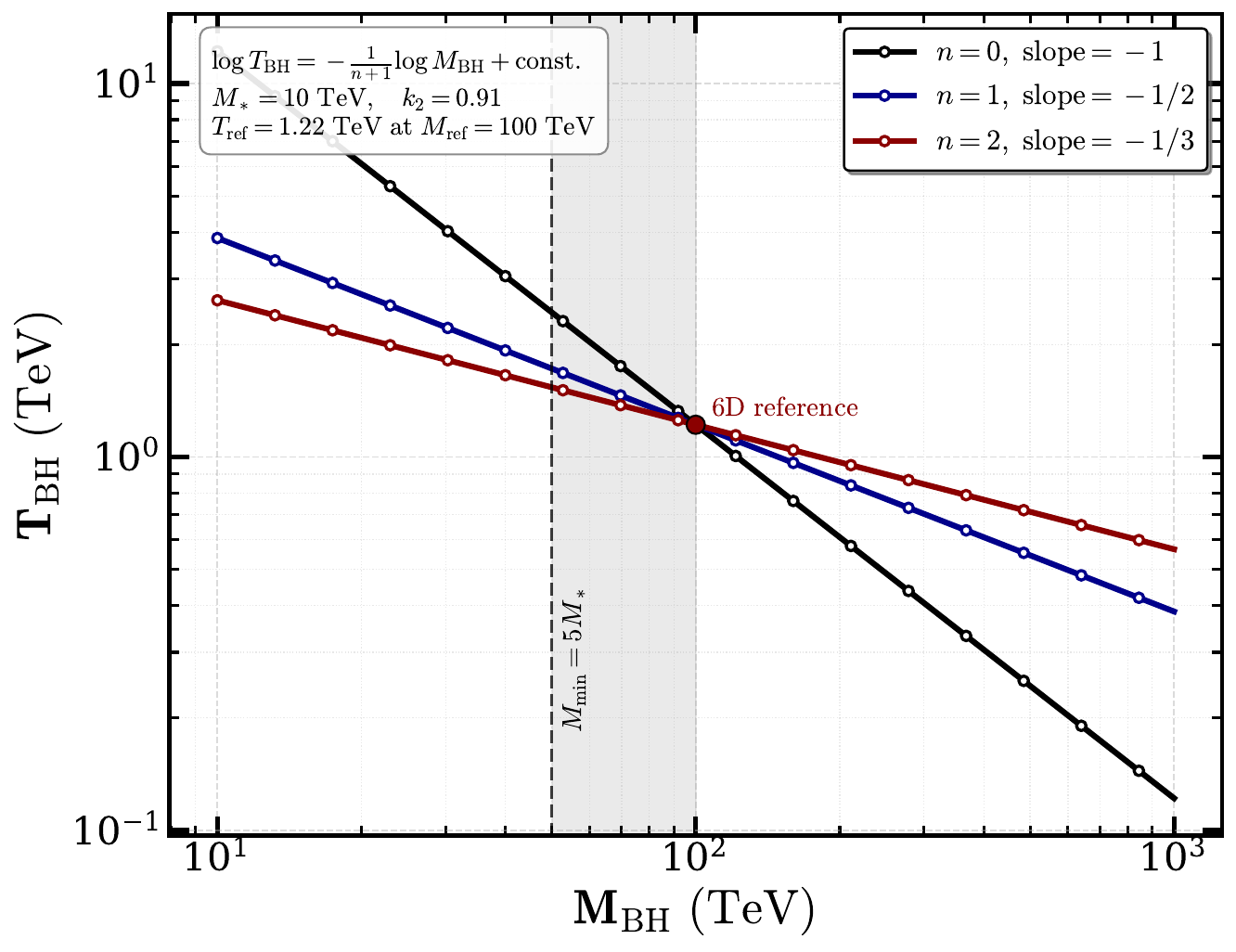}
 \caption{Theoretical logarithmic Hawking temperature--mass relation for
 different numbers of extra dimensions.  Its slope is
 \(d\log T_{\rm BH}/d\log M_{\rm BH}=-1/(\mathfrak n+1)\), giving \(-1/3\)
 for two extra dimensions.  The common normalization is chosen only to expose
 the slope dependence.  The curves do not include greybody factors,
 black-hole rotation, finite event statistics, missing-energy reconstruction,
 or detector uncertainties.}
 \label{fig:n_determination}
\end{figure}

Figure~\ref{fig:n_determination} displays this theoretical scaling for several
values of \(\mathfrak n\).  Normalizing the curves at one reference mass removes
the convention-dependent intercept and highlights the change in slope.  Their
visual separation does not represent an experimental confidence interval:
the attainable precision depends on the event yield, accessible mass lever
arm, missing-energy correction, and systematic uncertainties in the
reconstructed spectrum.

The ordering of the curves in Fig.~\ref{fig:n_determination} follows directly
from their slopes.  For \(\mathfrak n=0,1,2\), one obtains slopes
\(-1\), \(-1/2\), and \(-1/3\), respectively.  Increasing the number of extra
dimensions therefore makes the temperature decrease more slowly with mass.
This occurs because the higher-dimensional horizon itself grows more slowly,
\(r_h\propto M_{\rm BH}^{1/(\mathfrak n+1)}\).  The slope is more useful than
the absolute vertical normalization because the latter changes with the
definition of \(M_*\) and with rotational corrections.  Nevertheless, a
meaningful slope measurement requires events spanning a sufficiently broad
mass interval; a narrow sample concentrated near threshold would provide only
limited leverage for distinguishing the three idealized lines.

\subsection{Indirect probes of the fundamental gravity scale}
\label{sec:indirect}

Even if the semiclassical black-hole threshold is too high for an observable
production rate, real and virtual KK gravitons can probe the same fundamental
scale.  The monojet process
\begin{equation}
 pp\to j+G_{\rm KK}
\end{equation}
produces a hard jet recoiling against missing transverse momentum.  After
summing over the KK density of states, the inclusive real-emission rate for
two extra dimensions scales schematically as
\begin{equation}
 \sigma(pp\to j+G_{\rm KK})
 \propto
 \frac{1}{M_*^4},
 \label{eq:monojet_sigma}
\end{equation}
up to PDFs, phase-space cuts, the Planck-scale convention, and the treatment of
partonic events approaching the ultraviolet validity limit of the effective
theory \cite{Giudice:1998ck}.  Present monojet searches therefore impose
multi-TeV constraints on the fundamental scale, but the numerical bound is not
universal: it depends on the experimental analysis, convention for \(M_*\),
and EFT truncation prescription.  A dedicated FCC-hh projection must apply
realistic cuts and backgrounds rather than extrapolating Eq.~\eqref{eq:monojet_sigma}
alone.

\begin{figure}[t]
 \centering
 \includegraphics[width=0.58\columnwidth]{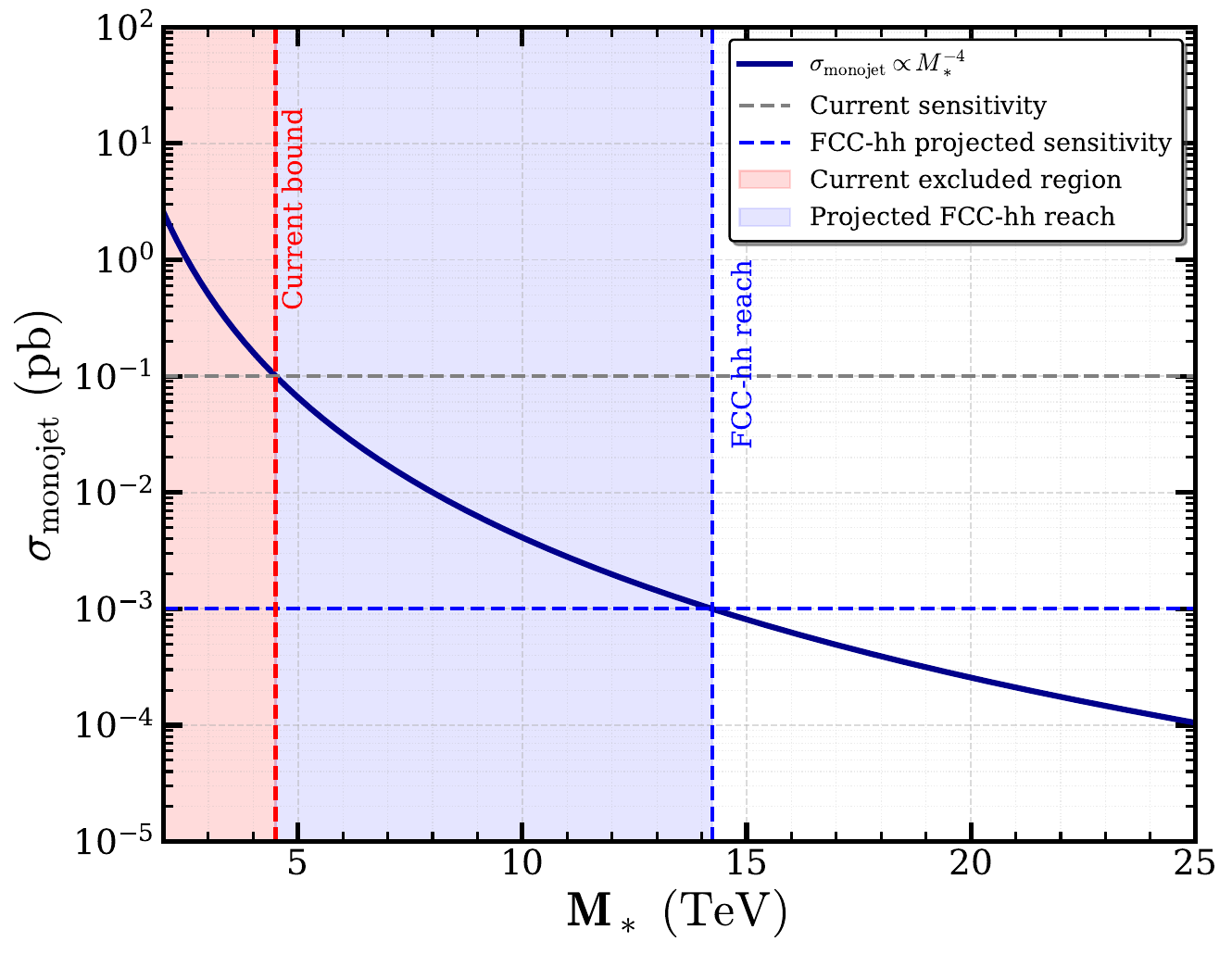}
 \caption{Schematic dependence of the monojet plus missing-momentum rate on
 the fundamental gravity scale in the two-extra-dimensions scenario.  The
 inclusive dimensional scaling is approximately
 \(\sigma_{\rm monojet}\propto M_*^{-4}\), while its normalization depends on
 PDFs, event selection, ultraviolet truncation, and the convention for
 \(M_*\).  The shaded regions illustrate present and prospective sensitivity
 domains; they are not confidence-level exclusions or a detector-level
 FCC-hh forecast.}
 \label{fig:monojet}
\end{figure}

The scaling is illustrated in Fig.~\ref{fig:monojet}.  Its shaded bands show
the qualitative separation between currently constrained scales and the
region to which a higher-energy collider could become sensitive.  Because no
experimental likelihood or detector-level recast is performed here, the band
edges should not be read as confidence-level limits.  Their purpose is to
display the rapid decrease of the rate with increasing \(M_*\) and the
potential gain from the larger FCC-hh parton luminosities.

The steep trend in Fig.~\ref{fig:monojet} is a direct consequence of the
dimension-eight gravitational interaction: at fixed collider energy and cuts,
raising \(M_*\) by one decade suppresses the schematic rate by four decades.
The region at smaller \(M_*\) is therefore the most readily tested, while the
high-scale region requires both increased collision energy and enhanced parton
luminosity.  The displayed FCC-hh band represents this qualitative gain in
reach.  Its location cannot be inferred from the \(M_*^{-4}\) scaling alone,
because the signal acceptance and the dominant \(Z(\nu\bar\nu)+j\),
\(W(\ell\nu)+j\), and instrumental backgrounds also change with the missing
momentum selection.  Accordingly, the figure supports the conclusion that an
FCC-hh can extend the sensitivity, but not a specific exclusion value without
a detector-level statistical analysis.

Virtual KK-graviton exchange can also modify high-mass dilepton and diphoton
production \cite{Giudice:1998ck,Han:1998sg,Hewett:1998sn}.  At energies below
the ultraviolet cutoff, its leading effect may be represented schematically by
a dimension-eight interaction,
\begin{equation}
 \mathcal L_{\rm eff}
 =
 \frac{C_{\rm KK}}{M_*^4}
 T^{\mu\nu}T_{\mu\nu},
 \label{eq:virtual_eff}
\end{equation}
where \(C_{\rm KK}\) depends on the KK summation and operator convention.  For
two extra dimensions, the virtual KK sum is logarithmically sensitive to the
ultraviolet cutoff.  Bounds from Drell--Yan or diphoton spectra must therefore
be quoted together with the cutoff prescription and cannot be interpreted as
strictly model-independent measurements of \(M_*\).

Direct black-hole searches, real graviton emission, and virtual exchange probe
different energy regimes and have different theoretical uncertainties.  A
combined analysis could test whether these channels admit a common value of
\(M_*\) and are compatible with \(\mathfrak n=2\), but it would not by itself
establish the ultraviolet completion of the model.

\subsection{Complementarity with primordial-black-hole and
gravitational-wave observations}
\label{sec:complementarity}

Collider and cosmological observations probe complementary aspects of the
same higher-dimensional framework.  For a compact two-dimensional space,
\begin{equation}
 M_{\rm Pl}^2=V_2M_*^4,
 \label{eq:Planckcollider}
\end{equation}
up to the convention used for the reduced Planck scale and the compactification
volume \(V_2\).  Collider processes are primarily sensitive to \(M_*\) and to
the effective dimensionality at short distances.  By contrast, PBH
evaporation and the associated scalar-induced gravitational-wave signal also
depend on the compactification scale, the PBH formation history, and the
assumed memory-burden dynamics.

An independent collider constraint on \(M_*\) would therefore fix one of the
microscopic inputs entering the six-dimensional PBH entropy, temperature, and
evaporation law.  Together with Eq.~\eqref{eq:Planckcollider}, it would also
constrain the compactification volume and hence the KK mass gap.  Conversely,
a gravitational-wave signal compatible with the PBH mass range discussed in
Sec.~\ref{sec:SIGW} would probe the early-Universe realization of the same
geometric framework.

The collider implications should thus be interpreted as consistency tests
rather than guaranteed discovery predictions.  Agreement among a collider
determination of \(M_*\), the compactification relation, and the PBH
gravitational-wave phenomenology would provide a nontrivial multimessenger test
of the two-dark-dimensions scenario.  Collider observations alone, however,
would neither demonstrate the memory-burden mechanism nor uniquely establish
the PBH origin of a stochastic gravitational-wave background.

\section{Conclusions}
\label{sec:conclusions}

In this work we explored the phenomenological consequences of the
two-dark-dimensions scenario, focusing on primordial black holes,
scalar-induced gravitational waves, and possible collider signatures of
low-scale higher-dimensional gravity.  The framework is characterized by
a fundamental gravity scale $M_* $ near the multi-TeV regime and a
compactification scale that fixes the KK mass gap.  Within this setup,
the quantum-gravitational memory burden can substantially extend the
lifetime of light primordial black holes.  For the benchmark case
$p=2$, this opens a broad PBH dark-matter window,
$10^{-3}\,{\rm g}\lesssim M_{\rm PBH}\lesssim10^{21}\,{\rm g}$, which
is qualitatively different from the standard four-dimensional Hawking
evaporation picture.

We showed that the scalar perturbations responsible for PBH formation
also source a stochastic gravitational-wave background.  A key point is
that the interpretation of this signal depends on the relation between
the PBH formation scale and the KK mass gap.  For the compactification
parameters used here, the boundary \(H_{\rm form}=m_{\rm KK}\)
corresponds to \(M_{\rm KK}\simeq8.5\times10^{23}\,{\rm g}\), or
equivalently, \(f_{\rm KK}^{\rm regime}\simeq1.2\times10^{-4}\,{\rm Hz}\).
Above this mass scale, the massive KK tower is kinematically
inaccessible, and the usual four-dimensional SIGW calculation is
self-consistent.  Below it, the KK tower can be excited and the
four-dimensional spectrum should be interpreted as the massless
zero-mode benchmark rather than the complete six-dimensional prediction.
The memory-burdened PBH dark-matter window lies entirely in this second
regime.

We therefore used the zero-mode SIGW spectrum as the robust observable
benchmark and introduced a phenomenological KK-tower extension only as
an illustrative model of possible six-dimensional effects.  A complete
calculation of the full six-dimensional tensor spectrum would require
the massive tensor Green functions, brane-to-bulk overlap coefficients,
the evolution of massive modes, and the detector response to massive
tensor polarizations.  The resulting gravitational-wave phenomenology
spans a wide frequency range: PBHs near the upper end of the
memory-burdened window can produce signals relevant for LISA, DECIGO,
and BBO, while lighter PBHs shift the peak toward high-frequency
gravitational-wave searches.  PTA-band signals correspond instead to
much larger PBH masses and belong to the regime where the
four-dimensional calculation is reliable.

We also discussed collider probes of the same higher-dimensional
framework.  If \(M_*\sim10~{\rm TeV}\), microscopic six-dimensional black
holes and KK graviton effects may be accessible at a future 100 TeV
proton--proton collider.  Using a consistent collider normalization for
the six-dimensional Schwarzschild radius with
\(k_2=(3/4)^{1/3}\simeq0.91\), we studied the geometric black-hole
production cross section, Hawking temperature, average multiplicity, and
schematic missing-energy signatures. The characteristic scalings
\(T_{\rm BH}\propto M_{\rm BH}^{-1/3}\) and
\(\langle N\rangle\propto M_{\rm BH}^{4/3}\) provide possible handles on
the number of extra dimensions.  We treated the collider estimates
conservatively, since precision predictions require modern PDFs,
greybody factors, parton showering, detector acceptance, and Standard
Model backgrounds.

The main message is that gravitational-wave and collider observables
probe complementary aspects of the two-dark-dimensions scenario.
Collider measurements would be sensitive to the short-distance gravity
scale \(M_*\) and the effective dimensionality of spacetime, while PBH
evaporation and scalar-induced gravitational waves probe the
early-Universe realization of the compactified theory.  A future
combination of collider evidence for low-scale higher-dimensional
gravity with a gravitational-wave signal compatible with the
memory-burdened PBH window would provide a strong consistency test of
the framework and would sharply constrain the connection between
TeV-scale gravity, primordial black holes, dark matter, and the
higher-dimensional structure of spacetime.

\appendix \label{ap1}
\section{Massive KK tensor modes and the full six-dimensional SIGW spectrum}
\label{app:6D_SIGW_derivation}

In this appendix, we derive the formal structure of the scalar-induced gravitational-wave spectrum in the six-dimensional two-dark-dimensions scenario. The purpose is to clarify how the usual four-dimensional result is recovered as the massless zero-mode contribution, and how the massive Kaluza--Klein tensor modes modify the full higher-dimensional prediction.

\subsection{KK decomposition of the tensor perturbation}

We consider a six-dimensional spacetime compactified on a square two-torus,
\begin{equation}
\mathcal M_4\times T^2 ,
\end{equation}
with compactification length
\begin{equation}
L\equiv 2\pi R .
\end{equation}
The tensor perturbation of the higher-dimensional metric can be decomposed into eigenmodes of the compact space. Schematically,
\begin{equation}
h_{ij}(x,y)
=
h_{ij}^{(0)}(x)
+
\sum_{\vec n\neq0}
h_{ij}^{(\vec n)}(x)\,Y_{\vec n}(y),
\label{eq:app_KK_decomposition}
\end{equation}
where \(x\) denotes the four-dimensional coordinates, \(y\) denotes the two compact coordinates, and \(Y_{\vec n}(y)\) are the internal wave functions on \(T^2\). For a square torus,
\begin{equation}
Y_{\vec n}(y)
\propto
\exp\left[
i\left(
\frac{n_1 y_1+n_2 y_2}{L}
\right)
\right],
\qquad
\vec n=(n_1,n_2).
\end{equation}
The corresponding KK masses are
\begin{equation}
m_{\vec n}
=
m_{\rm KK}\sqrt{n_1^2+n_2^2},
\qquad
m_{\rm KK}
=
\frac{1}{L}
=
\frac{1}{2\pi R}.
\label{eq:app_KK_masses}
\end{equation}
The \(\vec n=0\) mode is massless and corresponds to the ordinary four-dimensional graviton. The modes with \(\vec n\neq0\) behave as massive spin-2 tensor modes in the effective four-dimensional theory.

For each tensor polarization \(\lambda=+,\times\), the massless zero mode satisfies the usual induced tensor equation
\begin{equation}
h_{0}^{\lambda\prime\prime}
+
2\mathcal H h_{0}^{\lambda\prime}
+
k^2 h_{0}^{\lambda}
=
S_{0}^{\lambda}(k,\eta),
\label{eq:app_zero_mode_eq}
\end{equation}
where primes denote derivatives with respect to conformal time \(\eta\), and \(\mathcal H=a'/a\).

For a massive KK tensor mode, the effective four-dimensional equation contains an additional mass term:
\begin{equation}
h_{\vec n}^{\lambda\prime\prime}
+
2\mathcal H h_{\vec n}^{\lambda\prime}
+
\left(
k^2+a^2m_{\vec n}^2
\right)
h_{\vec n}^{\lambda}
=
S_{\vec n}^{\lambda}(k,\eta).
\label{eq:app_massive_mode_eq}
\end{equation}
The scalar-induced source \(S_{\vec n}^{\lambda}\) is obtained by projecting the second-order scalar source onto the \(\vec n\)-th KK tensor wave function. Schematically,
\begin{equation}
S_{\vec n}^{\lambda}(k,\eta)
=
\mathcal C_{\vec n}
\int
\frac{d^3p}{(2\pi)^3}
e_{ij}^{\lambda}(\mathbf k)\,
p^i p^j\,
\Phi(\mathbf p,\eta)
\Phi(\mathbf k-\mathbf p,\eta)
\,\mathcal T(p,|\mathbf k-\mathbf p|,\eta),
\label{eq:app_source_projected}
\end{equation}
where \(e_{ij}^{\lambda}\) is the polarization tensor, \(\Phi\) denotes the scalar perturbation, and \(\mathcal T\) is the scalar transfer-function combination appearing in the second-order source. The coefficient \(\mathcal C_{\vec n}\) encodes the overlap between the scalar source and the KK tensor wave function. If the scalar perturbations are localized on a brane, \(\mathcal C_{\vec n}\) depends on the brane thickness and on the brane-bulk coupling. Therefore it is model dependent.

The massive mode has the dispersion relation
\begin{equation}
\omega_{\vec n}^2(k,\eta)
=
k^2+a^2m_{\vec n}^2 .
\label{eq:app_dispersion}
\end{equation}
The solution of Eq.~(\ref{eq:app_massive_mode_eq}) can be written in terms of the massive Green function:
\begin{equation}
h_{\vec n}^{\lambda}(k,\eta)
=
\int^{\eta} d\tilde\eta\,
G_{\vec n}(k;\eta,\tilde\eta)\,
S_{\vec n}^{\lambda}(k,\tilde\eta).
\label{eq:app_green_solution}
\end{equation}
The Green function satisfies
\begin{equation}
G_{\vec n}^{\prime\prime}
+
2\mathcal H G_{\vec n}^{\prime}
+
\left(
k^2+a^2m_{\vec n}^2
\right)G_{\vec n}
=
\delta(\eta-\tilde\eta).
\label{eq:app_green_eq}
\end{equation}
For the zero mode, \(m_{\vec n}=0\), this reduces to the ordinary massless tensor Green function used in the standard four-dimensional scalar-induced gravitational-wave calculation. For \(\vec n\neq0\), the mass term changes the time kernel and therefore changes the induced tensor power spectrum.

We define the dimensionless tensor power spectrum for the \(\vec n\)-th mode by
\begin{equation}
\left\langle
h_{\vec n}^{\lambda}(\mathbf k,\eta)
h_{\vec n}^{\lambda'}(\mathbf k',\eta)
\right\rangle
=
\frac{2\pi^2}{k^3}
\delta_{\lambda\lambda'}
\delta^{(3)}(\mathbf k+\mathbf k')
\mathcal P_h^{(\vec n)}(k,\eta).
\label{eq:app_tensor_power_def}
\end{equation}
Using the Green-function solution, the induced tensor power spectrum has the schematic form
\begin{equation}
\mathcal P_h^{(\vec n)}(k,\eta)
=
4
\int_0^{\infty} dv
\int_{|1-v|}^{1+v}du\,
\mathcal K(u,v)\,
\mathcal C_{\vec n}^{\,2}\,
I_{\vec n}^{2}(u,v,k,\eta)\,
\mathcal P_{\mathcal R}(ku)\,
\mathcal P_{\mathcal R}(kv).
\label{eq:app_Ph_massive_formal}
\end{equation}
Here \(u\) and \(v\) are the standard dimensionless variables,
\begin{equation}
u=\frac{|\mathbf k-\mathbf p|}{k},
\qquad
v=\frac{p}{k},
\end{equation}
and
\begin{equation}
\mathcal K(u,v)
=
\left[
\frac{
4v^2-(1+v^2-u^2)^2
}
{4uv}
\right]^2
\label{eq:app_kernel}
\end{equation}
is the usual angular kernel from the transverse-traceless projection. The function \(I_{\vec n}(u,v,k,\eta)\) is the massive time integral. It is obtained by integrating the scalar source against the massive Green function \(G_{\vec n}\). Therefore, unlike the standard 4d kernel, \(I_{\vec n}\) depends on the KK mass through \(m_{\vec n}\).

For the zero mode,
\begin{equation}
m_{\vec n}=0,
\qquad
\mathcal C_{\vec n}=1,
\qquad
I_{\vec n}(u,v,k,\eta)\rightarrow I_0(u,v,k,\eta),
\end{equation}
and Eq.~(\ref{eq:app_Ph_massive_formal}) reduces to the standard four-dimensional scalar-induced tensor power spectrum.

\subsection{Energy density of a massive tensor mode}

The energy density carried by a tensor perturbation follows from the quadratic action. For a massive KK tensor mode, the kinetic, gradient, and mass terms give schematically
\begin{equation}
\rho_{\rm GW}^{(\vec n)}
\simeq
\frac{M_{\rm Pl}^2}{8a^2}
\sum_{\lambda}
\left[
\left|h_{\vec n}^{\lambda\prime}\right|^2
+
\left(
k^2+a^2m_{\vec n}^2
\right)
\left|h_{\vec n}^{\lambda}\right|^2
\right].
\label{eq:app_rho_massive}
\end{equation}
For an oscillating mode, one may average over several oscillations. The kinetic and potential pieces are then related by
\begin{equation}
\overline{
\left|h_{\vec n}^{\lambda\prime}\right|^2
}
\simeq
\left(
k^2+a^2m_{\vec n}^2
\right)
\overline{
\left|h_{\vec n}^{\lambda}\right|^2
}.
\label{eq:app_osc_avg}
\end{equation}
Using Eq.~(\ref{eq:app_tensor_power_def}), the energy density per logarithmic interval in \(k\) becomes proportional to
\begin{equation}
\frac{d\rho_{\rm GW}^{(\vec n)}}{d\ln k}
\propto
\frac{M_{\rm Pl}^2}{a^2}
\left(
k^2+a^2m_{\vec n}^2
\right)
\overline{\mathcal P_h^{(\vec n)}(k,\eta)}.
\label{eq:app_drho_massive}
\end{equation}
Dividing by the background density
\begin{equation}
\rho_{\rm tot}
=
3M_{\rm Pl}^2H^2,
\end{equation}
one obtains the fractional energy density
\begin{equation}
\Omega_{\rm GW}^{(\vec n)}(k,\eta)
\equiv
\frac{1}{\rho_{\rm tot}}
\frac{d\rho_{\rm GW}^{(\vec n)}}{d\ln k}.
\label{eq:app_Omega_def}
\end{equation}
With the conventional normalization used for tensor power spectra, this gives
\begin{equation}
\Omega_{\rm GW}^{(\vec n)}(k,\eta)
=
\frac{1}{24}
\left(
\frac{k}{aH}
\right)^2
\left(
1+\frac{a^2m_{\vec n}^2}{k^2}
\right)
\overline{\mathcal P_h^{(\vec n)}(k,\eta)}.
\label{eq:app_Omega_massive}
\end{equation}
The factor
\begin{equation}
1+\frac{a^2m_{\vec n}^2}{k^2}
\end{equation}
is a direct consequence of the massive dispersion relation. In the massless limit,
\begin{equation}
m_{\vec n}\rightarrow0,
\end{equation}
Eq.~(\ref{eq:app_Omega_massive}) reduces to
\begin{equation}
\Omega_{\rm GW}^{(0)}(k,\eta)
=
\frac{1}{24}
\left(
\frac{k}{aH}
\right)^2
\overline{\mathcal P_h^{(0)}(k,\eta)}.
\label{eq:app_Omega_zero}
\end{equation}

\subsection{Full six-dimensional spectrum}

The full six-dimensional gravitational-wave spectrum is obtained by summing over the massless zero mode and all massive KK modes:
\begin{equation}
\Omega_{\rm GW}^{6d}(k,\eta)
=
\Omega_{\rm GW}^{(0)}(k,\eta)
+
\sum_{\vec n\neq0}
\Omega_{\rm GW}^{(\vec n)}(k,\eta).
\label{eq:app_Omega_6D_sum}
\end{equation}
The zero-mode contribution is the standard four-dimensional scalar-induced gravitational-wave spectrum:
\begin{equation}
\Omega_{\rm GW,0}^{(0)}(k)
=
\Omega_{\rm GW,0}^{(4d)}(k).
\label{eq:app_zero_equals_4D}
\end{equation}
During radiation domination, the usual present-day expression is
\begin{align}
\Omega_{\rm GW,0}^{(4d)}(k)
&=
\frac{\Omega_{r,0}}{24}
\left(
\frac{g_{*,0}}{g_{*,c}}
\right)^{1/3}
\int_0^{\infty}dv
\int_{|1-v|}^{1+v}du\,
\mathcal K(u,v)\,
\overline{I_0^2(u,v)}
\mathcal P_{\mathcal R}(ku)
\mathcal P_{\mathcal R}(kv).
\label{eq:app_Omega_4D}
\end{align}
The massive-mode contribution is formally
\begin{align}
\Omega_{\rm GW,0}^{(\vec n)}(k)
&=
\frac{\Omega_{r,0}}{24}
\left(
\frac{g_{*,0}}{g_{*,c}}
\right)^{1/3}
\int_0^{\infty}dv
\int_{|1-v|}^{1+v}du\,
\mathcal K(u,v)\,
\mathcal C_{\vec n}^{\,2}\,
\overline{I_{\vec n}^2(u,v,k)}
\notag\\
&\qquad\qquad\times
\left(
1+\frac{a_c^2m_{\vec n}^2}{k^2}
\right)
\mathcal P_{\mathcal R}(ku)
\mathcal P_{\mathcal R}(kv),
\label{eq:app_Omega_massive_integral}
\end{align}
where \(a_c\) denotes the scale factor at the time when the source has saturated. The function \(I_{\vec n}\) is the massive Green-function kernel and differs from the standard massless kernel \(I_0\).

Therefore, the full present-day 6d result may be written formally as
\begin{equation}
\Omega_{\rm GW,0}^{6d}(k)
=
\Omega_{\rm GW,0}^{(4d)}(k)
+
\sum_{\vec n\neq0}
\Omega_{\rm GW,0}^{(\vec n)}(k).
\label{eq:app_full_6D_present}
\end{equation}

\begin{acknowledgments}
The authors thank Shabbar Raza for useful discussions.
\end{acknowledgments}

\bibliographystyle{unsrt}
\bibliography{main.bib}

@article{Vafa:2005ui,
  author        = {Vafa, Cumrun},
  title         = {The String Landscape and the Swampland},
  journal       = {},
  year          = {2005},
  eprint        = {hep-th/0509212},
  archivePrefix = {arXiv},
  primaryClass  = {hep-th},
  reportNumber  = {HUTP-05-A043},
  note          = {arXiv:hep-th/0509212}
}

@article{Ooguri:2006in,
    author = "Ooguri, Hirosi and Vafa, Cumrun",
    title = "{On the Geometry of the String Landscape and the Swampland}",
    eprint = "hep-th/0605264",
    archivePrefix = "arXiv",
    reportNumber = "CALT-68-2600, HUTP-06-A017",
    doi = "10.1016/j.nuclphysb.2006.10.033",
    journal = "Nucl. Phys. B",
    volume = "766",
    pages = "21--33",
    year = "2007"
}

@article{Palti:2019pca,
    author = "Palti, Eran",
    title = "{The Swampland: Introduction and Review}",
    eprint = "1903.06239",
    archivePrefix = "arXiv",
    primaryClass = "hep-th",
    reportNumber = "MPP-2019-53",
    doi = "10.1002/prop.201900037",
    journal = "Fortsch. Phys.",
    volume = "67",
    number = "6",
    pages = "1900037",
    year = "2019"
}

@article{vanBeest:2021lhn,
    author = "van Beest, Marieke and Calder{\'o}n-Infante, Jos{\'e} and Mirfendereski, Delaram and Valenzuela, Irene",
    title = "{Lectures on the Swampland Program in String Compactifications}",
    eprint = "2102.01111",
    archivePrefix = "arXiv",
    primaryClass = "hep-th",
    doi = "10.1016/j.physrep.2022.09.002",
    journal = "Phys. Rept.",
    volume = "989",
    pages = "1--50",
    year = "2022"
}

@article{Agmon:2022thq,
    author = "Agmon, Nathan Benjamin and Bedroya, Alek and Kang, Monica Jinwoo and Vafa, Cumrun",
    title = "{Lectures on the String Landscape and the Swampland}",
    journal = "{}",
    eprint = "2212.06187",
    archivePrefix = "arXiv",
    primaryClass = "hep-th",
    month = "12",
    year = "2022",
    note = "{arXiv:2212.06187 [hep-th]}"
}

@article{Myers:1986un,
    author = "Myers, Robert C. and Perry, M. J.",
    title = "{Black Holes in Higher Dimensional Space-Times}",
    reportNumber = "PRINT-86-0067 (PRINCETON)",
    doi = "10.1016/0003-4916(86)90186-7",
    journal = "Annals Phys.",
    volume = "172",
    pages = "304",
    year = "1986"
}

@article{Montero:2022prj,
    author = "Montero, Miguel and Vafa, Cumrun and Valenzuela, Irene",
    title = "{The dark dimension and the Swampland}",
    eprint = "2205.12293",
    archivePrefix = "arXiv",
    primaryClass = "hep-th",
    doi = "10.1007/JHEP02(2023)022",
    journal = "JHEP",
    volume = "02",
    pages = "022",
    year = "2023"
}

@article{Leontaris:2025piz,
    author = "Leontaris, George K. and Prampromis, George",
    title = "{5D rotating black holes as dark matter in dark dimension scenario: Hawking radiation versus the memory burden effect}",
    eprint = "2512.10381",
    archivePrefix = "arXiv",
    primaryClass = "hep-th",
    doi = "10.1088/1475-7516/2026/05/014",
    journal = "JCAP",
    volume = "05",
    pages = "014",
    year = "2026"
}

@article{Anchordoqui:2026hys,
    author = "Anchordoqui, Luis A. and Lust, Dieter",
    title = "{Breaking Free from the Swampland of Impossible Universes through the DESI Portal}",
    journal = "{}",
    eprint = "2605.10476",
    archivePrefix = "arXiv",
    primaryClass = "astro-ph.CO",
    reportNumber = "MPP-2026-84",
    month = "5",
    year = "2026",
    note = "{arXiv:2605.10476 [astro-ph.CO]}"
}

@article{Hawking:1975vcx,
    author = "Hawking, S. W.",
    editor = "Gibbons, G. W. and Hawking, S. W.",
    title = "{Particle Creation by Black Holes}",
    doi = "10.1007/BF02345020",
    journal = "Commun. Math. Phys.",
    volume = "43",
    pages = "199--220",
    year = "1975",
    note = "[Erratum: Commun.Math.Phys. 46, 206 (1976)]"
}

@article{Kanti:2004nr,
    author = "Kanti, Panagiota",
    title = "{Black holes in theories with large extra dimensions: A Review}",
    eprint = "hep-ph/0402168",
    archivePrefix = "arXiv",
    doi = "10.1142/S0217751X04018324",
    journal = "Int. J. Mod. Phys. A",
    volume = "19",
    pages = "4899--4951",
    year = "2004"
}

@article{Tangherlini:1963bw,
    author = "Tangherlini, F. R.",
    title = "{Schwarzschild field in n dimensions and the dimensionality of space problem}",
    doi = "10.1007/BF02784569",
    journal = "Nuovo Cim.",
    volume = "27",
    pages = "636--651",
    year = "1963"
}

@article{Carr:2020gox,
    author = "Carr, Bernard and Kohri, Kazunori and Sendouda, Yuuiti and Yokoyama, Jun'ichi",
    title = "{Constraints on primordial black holes}",
    eprint = "2002.12778",
    archivePrefix = "arXiv",
    primaryClass = "astro-ph.CO",
    reportNumber = "RESCEU-03/20; KEK-Cosmo-249; KEK-TH-2199; IPMU20-0024",
    doi = "10.1088/1361-6633/ac1e31",
    journal = "Rept. Prog. Phys.",
    volume = "84",
    number = "11",
    pages = "116902",
    year = "2021"
}

@article{Giudice:2001ce,
    author = "Giudice, Gian F. and Rattazzi, Riccardo and Wells, James D.",
    title = "{Transplanckian collisions at the LHC and beyond}",
    eprint = "hep-ph/0112161",
    archivePrefix = "arXiv",
    reportNumber = "CERN-TH-2001-306, DESY-01-206",
    doi = "10.1016/S0550-3213(02)00142-6",
    journal = "Nucl. Phys. B",
    volume = "630",
    pages = "293--325",
    year = "2002"
}

@article{Yoshino:2002tx,
    author = "Yoshino, Hirotaka and Nambu, Yasusada",
    title = "{Black hole formation in the grazing collision of high-energy particles}",
    eprint = "gr-qc/0209003",
    archivePrefix = "arXiv",
    reportNumber = "DPNU-02-26",
    doi = "10.1103/PhysRevD.67.024009",
    journal = "Phys. Rev. D",
    volume = "67",
    pages = "024009",
    year = "2003"
}

@article{Anchordoqui:2024akj,
    author = "Anchordoqui, Luis A. and Antoniadis, Ignatios and Lust, Dieter",
    title = "{Dark dimension, the swampland, and the dark matter fraction composed of primordial near-extremal black holes}",
    eprint = "2401.09087",
    archivePrefix = "arXiv",
    primaryClass = "hep-th",
    reportNumber = "LMU-ASC 01/24, MPP-2024-5",
    doi = "10.1103/PhysRevD.109.095008",
    journal = "Phys. Rev. D",
    volume = "109",
    number = "9",
    pages = "095008",
    year = "2024"
}

@article{Ida:2005ax,
    author = "Ida, Daisuke and Oda, Kin-ya and Park, Seong Chan",
    title = "{Rotating black holes at future colliders. II. Anisotropic scalar field emission}",
    eprint = "hep-th/0503052",
    archivePrefix = "arXiv",
    doi = "10.1103/PhysRevD.71.124039",
    journal = "Phys. Rev. D",
    volume = "71",
    pages = "124039",
    year = "2005"
}

@article{Bekenstein:1973ur,
    author = "Bekenstein, Jacob D.",
    title = "{Black holes and entropy}",
    doi = "10.1103/PhysRevD.7.2333",
    journal = "Phys. Rev. D",
    volume = "7",
    pages = "2333--2346",
    year = "1973"
}

@article{Dvali:2011aa,
    author = "Dvali, Gia and Gomez, Cesar",
    title = "{Black Hole's Quantum N-Portrait}",
    eprint = "1112.3359",
    archivePrefix = "arXiv",
    primaryClass = "hep-th",
    doi = "10.1002/prop.201300001",
    journal = "Fortsch. Phys.",
    volume = "61",
    pages = "742--767",
    year = "2013"
}

@article{Dvali:2020wft,
    author = "Dvali, Gia and Eisemann, Lukas and Michel, Marco and Zell, Sebastian",
    title = "{Black hole metamorphosis and stabilization by memory burden}",
    eprint = "2006.00011",
    archivePrefix = "arXiv",
    primaryClass = "hep-th",
    doi = "10.1103/PhysRevD.102.103523",
    journal = "Phys. Rev. D",
    volume = "102",
    number = "10",
    pages = "103523",
    year = "2020"
}

@article{Hawking:1971ei,
    author = "Hawking, Stephen",
    title = "{Gravitationally collapsed objects of very low mass}",
    doi = "10.1093/mnras/152.1.75",
    journal = "Mon. Not. Roy. Astron. Soc.",
    volume = "152",
    pages = "75",
    year = "1971"
}

@article{Carr:1974nx,
    author = "Carr, Bernard J. and Hawking, S. W.",
    title = "{Black holes in the early Universe}",
    doi = "10.1093/mnras/168.2.399",
    journal = "Mon. Not. Roy. Astron. Soc.",
    volume = "168",
    pages = "399--415",
    year = "1974"
}

@article{Klein:1926tv,
    author = "Klein, Oskar",
    editor = "Taylor, J. C.",
    title = "{Quantum Theory and Five-Dimensional Theory of Relativity. (In German and English)}",
    doi = "10.1007/BF01397481",
    journal = "Z. Phys.",
    volume = "37",
    pages = "895--906",
    year = "1926"
}

@article{Arkani-Hamed:1998jmv,
    author = "Arkani-Hamed, Nima and Dimopoulos, Savas and Dvali, G. R.",
    title = "{The Hierarchy problem and new dimensions at a millimeter}",
    eprint = "hep-ph/9803315",
    archivePrefix = "arXiv",
    reportNumber = "SLAC-PUB-7769, SU-ITP-98-13",
    doi = "10.1016/S0370-2693(98)00466-3",
    journal = "Phys. Lett. B",
    volume = "429",
    pages = "263--272",
    year = "1998"
}

@article{Antoniadis:1998ig,
    author = "Antoniadis, Ignatios and Arkani-Hamed, Nima and Dimopoulos, Savas and Dvali, G. R.",
    title = "{New dimensions at a millimeter to a Fermi and superstrings at a TeV}",
    eprint = "hep-ph/9804398",
    archivePrefix = "arXiv",
    reportNumber = "SLAC-PUB-7801, SU-ITP-98-28, CPTH-S608-0498, IC-98-39",
    doi = "10.1016/S0370-2693(98)00860-0",
    journal = "Phys. Lett. B",
    volume = "436",
    pages = "257--263",
    year = "1998"
}

@article{Kaluza:1921tu,
    author = "Kaluza, Th.",
    title = {{Zum Unit{\"a}tsproblem der Physik}},
    eprint = "1803.08616",
    archivePrefix = "arXiv",
    primaryClass = "physics.hist-ph",
    reportNumber = "HUPD-8401",
    doi = "10.1142/S0218271818700017",
    journal = "Sitzungsber. Preuss. Akad. Wiss. Berlin (Math. Phys. )",
    volume = "1921",
    pages = "966--972",
    year = "1921"
}

@article{Overduin:1997sri,
    author = "Overduin, J. M. and Wesson, P. S.",
    title = "{Kaluza-Klein gravity}",
    eprint = "gr-qc/9805018",
    archivePrefix = "arXiv",
    doi = "10.1016/S0370-1573(96)00046-4",
    journal = "Phys. Rept.",
    volume = "283",
    pages = "303--380",
    year = "1997"
}

@article{Ananda:2006af,
    author = "Ananda, Kishore N. and Clarkson, Chris and Wands, David",
    title = "{The Cosmological gravitational wave background from primordial density perturbations}",
    eprint = "gr-qc/0612013",
    archivePrefix = "arXiv",
    doi = "10.1103/PhysRevD.75.123518",
    journal = "Phys. Rev. D",
    volume = "75",
    pages = "123518",
    year = "2007"
}

@article{Baumann:2007zm,
    author = "Baumann, Daniel and Steinhardt, Paul J. and Takahashi, Keitaro and Ichiki, Kiyotomo",
    title = "{Gravitational Wave Spectrum Induced by Primordial Scalar Perturbations}",
    eprint = "hep-th/0703290",
    archivePrefix = "arXiv",
    doi = "10.1103/PhysRevD.76.084019",
    journal = "Phys. Rev. D",
    volume = "76",
    pages = "084019",
    year = "2007"
}

@article{Kohri:2024qpd,
    author = "Kohri, Kazunori and Terada, Takahiro and Yanagida, Tsutomu T.",
    title = "{Induced gravitational waves probing primordial black hole dark matter with the memory burden effect}",
    eprint = "2409.06365",
    archivePrefix = "arXiv",
    primaryClass = "astro-ph.CO",
    reportNumber = "KEK-TH-2654, KEK-Cosmo-0358",
    doi = "10.1103/PhysRevD.111.063543",
    journal = "Phys. Rev. D",
    volume = "111",
    number = "6",
    pages = "063543",
    year = "2025"
}

@book{Appelquist:1987nr,
    editor    = "Appelquist, T. and Chodos, A. and Freund, P. G. O.",
    title     = "{Modern Kaluza--Klein Theories}",
    publisher = "Addison-Wesley",
    address   = "Reading, MA",
    year      = "1987",
    isbn      = "9780201157677"
}

@article{Inomata:2018epa,
    author = "Inomata, Keisuke and Nakama, Tomohiro",
    title = "{Gravitational waves induced by scalar perturbations as probes of the small-scale primordial spectrum}",
    eprint = "1812.00674",
    archivePrefix = "arXiv",
    primaryClass = "astro-ph.CO",
    reportNumber = "IPMU 18-0200",
    doi = "10.1103/PhysRevD.99.043511",
    journal = "Phys. Rev. D",
    volume = "99",
    number = "4",
    pages = "043511",
    year = "2019"
}

@article{Dvali:2024hsb,
    author = "Dvali, Gia and Valbuena-Berm{\'u}dez, Juan Sebasti{\'a}n and Zantedeschi, Michael",
    title = "{Memory burden effect in black holes and solitons: Implications for PBH}",
    eprint = "2405.13117",
    archivePrefix = "arXiv",
    primaryClass = "hep-th",
    doi = "10.1103/PhysRevD.110.056029",
    journal = "Phys. Rev. D",
    volume = "110",
    number = "5",
    pages = "056029",
    year = "2024"
}

@article{NANOGrav:2023gor,
    author = "Agazie, Gabriella and others",
    collaboration = "NANOGrav",
    title = "{The NANOGrav 15 yr Data Set: Evidence for a Gravitational-wave Background}",
    eprint = "2306.16213",
    archivePrefix = "arXiv",
    primaryClass = "astro-ph.HE",
    doi = "10.3847/2041-8213/acdac6",
    journal = "Astrophys. J. Lett.",
    volume = "951",
    number = "1",
    pages = "L8",
    year = "2023"
}

@article{LISA:2017pwj,
    author = "Amaro-Seoane, Pau and others",
    collaboration = "LISA",
    title = "{Laser Interferometer Space Antenna}",
    journal = "{arXiv}",
    eprint = "1702.00786",
    archivePrefix = "arXiv",
    primaryClass = "astro-ph.IM",
    month = "2",
    year = "2017",
    note = "{arXiv:1702.00786 [astro-ph.IM]}"
}

@article{Seto:2001qf,
    author = "Seto, Naoki and Kawamura, Seiji and Nakamura, Takashi",
    title = "{Possibility of direct measurement of the acceleration of the universe using 0.1-Hz band laser interferometer gravitational wave antenna in space}",
    eprint = "astro-ph/0108011",
    archivePrefix = "arXiv",
    doi = "10.1103/PhysRevLett.87.221103",
    journal = "Phys. Rev. Lett.",
    volume = "87",
    pages = "221103",
    year = "2001"
}

@article{Anchordoqui:2025nmb,
    author = "Anchordoqui, Luis A. and Antoniadis, Ignatios and Lust, Dieter",
    title = "{Two Micron-Size Dark Dimensions}",
    eprint = "2501.11690",
    archivePrefix = "arXiv",
    primaryClass = "hep-th",
    reportNumber = "MPP-2025-5, LMU-ASC 02/25",
    doi = "10.1002/prop.70015",
    journal = "Fortsch. Phys.",
    volume = "73",
    number = "8",
    pages = "e70015",
    year = "2025"
}

@article{Leontaris:2026kvu,
    author = "Leontaris, George K. and Prampromis, George",
    title = "{Micron-sized Extra Dimensions and Primordial Black Holes: Charged, Rotating, and Memory Burdened}",
    journal = {},
    eprint = "2605.00252",
    archivePrefix = "arXiv",
    primaryClass = "hep-ph",
    month = "4",
    year = "2026",
    note = "{arXiv:2605.00252 [hep-ph]}"
}

@article{FCC:2025lpp,
    author = "Benedikt, M. and others",
    collaboration = "FCC",
    title = "{Future Circular Collider Feasibility Study Report: Volume 1, Physics, Experiments, Detectors}",
    eprint = "2505.00272",
    archivePrefix = "arXiv",
    primaryClass = "hep-ex",
    reportNumber = "CERN-FCC-PHYS-2025-0002",
    doi = "10.1140/epjc/s10052-025-15077-x",
    journal = "Eur. Phys. J. C",
    volume = "85",
    number = "12",
    pages = "1468",
    year = "2025"
}

@article{Ahmed:2026pjd,
    author = "Ahmed, Waqas and Leontaris, George K.",
    title = "{Secondary Gravitational Wave Signatures from 5D Rotating Primordial Black Holes in the Dark Dimension}",
    journal = {},
    eprint = "2605.12948",
    archivePrefix = "arXiv",
    primaryClass = "hep-ph",
    month = "5",
    year = "2026",
    note = "{arXiv:2605.12948 [hep-ph]}"
}

@article{King:2014nza,
    author = "King, Stephen F. and Merle, Alexander and Morisi, Stefano and Shimizu, Yusuke and Tanimoto, Morimitsu",
    title = "{Neutrino Mass and Mixing: from Theory to Experiment}",
    eprint = "1402.4271",
    archivePrefix = "arXiv",
    primaryClass = "hep-ph",
    doi = "10.1088/1367-2630/16/4/045018",
    journal = "New J. Phys.",
    volume = "16",
    pages = "045018",
    year = "2014"
}

@article{Giudice:1998ck,
    author = "Giudice, Gian F. and Rattazzi, Riccardo and Wells, James D.",
    title = "{Quantum gravity and extra dimensions at high-energy colliders}",
    eprint = "hep-ph/9811291",
    archivePrefix = "arXiv",
    reportNumber = "CERN-TH-98-354",
    doi = "10.1016/S0550-3213(99)00044-9",
    journal = "Nucl. Phys. B",
    volume = "544",
    pages = "3--38",
    year = "1999"
}

@article{Anchordoqui:2003ug,
    author = "Anchordoqui, Luis A. and Feng, Jonathan L. and Goldberg, Haim and Shapere, Alfred D.",
    title = "{Inelastic black hole production and large extra dimensions}",
    eprint = "hep-ph/0311365",
    archivePrefix = "arXiv",
    reportNumber = "NUB-3243-TH-03, UCI-TR-2003-30, UK-03-15",
    doi = "10.1016/j.physletb.2004.05.051",
    journal = "Phys. Lett. B",
    volume = "594",
    pages = "363--367",
    year = "2004"
}

@article{Montero:2019ekk,
    author = "Montero, Miguel and Van Riet, Thomas and Venken, Victoria",
    title = "{Festina Lente: EFT Constraints from Charged Black Hole Evaporation in de Sitter}",
    eprint = "1910.01648",
    archivePrefix = "arXiv",
    primaryClass = "hep-th",
    doi = "10.1007/JHEP01(2020)039",
    journal = "JHEP",
    volume = "01",
    pages = "039",
    year = "2020"
}

@article{Montero:2021otb,
    author = "Montero, Miguel and Vafa, Cumrun and Van Riet, Thomas and Venken, Victoria",
    title = "{The FL bound and its phenomenological implications}",
    eprint = "2106.07650",
    archivePrefix = "arXiv",
    primaryClass = "hep-th",
    reportNumber = "UUITP-26/2",
    doi = "10.1007/JHEP10(2021)009",
    journal = "JHEP",
    volume = "10",
    pages = "009",
    year = "2021"
}

@article{Nariai:1999iok,
    author = "Nariai, Hidekazu",
    title = "{On a New Cosmological Solution of Einstein's Field Equations of Gravitation}",
    doi = "10.1023/A:1026602724948",
    journal = "Gen. Rel. Grav.",
    volume = "31",
    number = "6",
    pages = "963--971",
    year = "1999"
}

@article{Bousso:1996au,
    author = "Bousso, Raphael and Hawking, Stephen W.",
    title = "{Pair creation of black holes during inflation}",
    eprint = "gr-qc/9606052",
    archivePrefix = "arXiv",
    reportNumber = "DAMTP-R-96-33",
    doi = "10.1103/PhysRevD.54.6312",
    journal = "Phys. Rev. D",
    volume = "54",
    pages = "6312--6322",
    year = "1996"
}

@article{Schwinger:1951nm,
    author = "Schwinger, Julian S.",
    editor = "Milton, K. A.",
    title = "{On gauge invariance and vacuum polarization}",
    doi = "10.1103/PhysRev.82.664",
    journal = "Phys. Rev.",
    volume = "82",
    pages = "664--679",
    year = "1951"
}

@article{Garriga:1994bm,
    author = "Garriga, J.",
    title = "{Pair production by an electric field in (1+1)-dimensional de Sitter space}",
    doi = "10.1103/PhysRevD.49.6343",
    journal = "Phys. Rev. D",
    volume = "49",
    pages = "6343--6346",
    year = "1994"
}

@article{Dimopoulos:2001hw,
    author = "Dimopoulos, Savas and Landsberg, Greg L.",
    title = "{Black holes at the LHC}",
    eprint = "hep-ph/0106295",
    archivePrefix = "arXiv",
    reportNumber = "SU-ITP-01-31",
    doi = "10.1103/PhysRevLett.87.161602",
    journal = "Phys. Rev. Lett.",
    volume = "87",
    pages = "161602",
    year = "2001"
}

@article{Giddings:2001bu,
    author = "Giddings, Steven B. and Thomas, Scott D.",
    title = "{High-energy colliders as black hole factories: The End of short distance physics}",
    eprint = "hep-ph/0106219",
    archivePrefix = "arXiv",
    reportNumber = "NSF-ITP-01-62, SU-ITP-01-30",
    doi = "10.1103/PhysRevD.65.056010",
    journal = "Phys. Rev. D",
    volume = "65",
    pages = "056010",
    year = "2002"
}

@article{Hewett:1998sn,
    author = "Hewett, JoAnne L.",
    title = "{Indirect collider signals for extra dimensions}",
    eprint = "hep-ph/9811356",
    archivePrefix = "arXiv",
    reportNumber = "SLAC-PUB-8001",
    doi = "10.1103/PhysRevLett.82.4765",
    journal = "Phys. Rev. Lett.",
    volume = "82",
    pages = "4765--4768",
    year = "1999"
}

@article{Arkani-Hamed:1998sfv,
    author = "Arkani-Hamed, Nima and Dimopoulos, Savas and Dvali, G. R.",
    title = "{Phenomenology, astrophysics and cosmology of theories with submillimeter dimensions and TeV scale quantum gravity}",
    eprint = "hep-ph/9807344",
    archivePrefix = "arXiv",
    reportNumber = "SLAC-PUB-7864, SU-ITP-98-142, IC-98-44",
    doi = "10.1103/PhysRevD.59.086004",
    journal = "Phys. Rev. D",
    volume = "59",
    pages = "086004",
    year = "1999"
}

@article{Domenech:2021ztg,
    author = "Dom{\`e}nech, Guillem",
    title = "{Scalar Induced Gravitational Waves Review}",
    eprint = "2109.01398",
    archivePrefix = "arXiv",
    primaryClass = "gr-qc",
    doi = "10.3390/universe7110398",
    journal = "Universe",
    volume = "7",
    number = "11",
    pages = "398",
    year = "2021"
}

@article{Han:1998sg,
    author = "Han, Tao and Lykken, Joseph D. and Zhang, Ren-Jie",
    title = "{On Kaluza-Klein states from large extra dimensions}",
    eprint = "hep-ph/9811350",
    archivePrefix = "arXiv",
    reportNumber = "MADPH-98-1092, FERMILAB-PUB-98-364",
    doi = "10.1103/PhysRevD.59.105006",
    journal = "Phys. Rev. D",
    volume = "59",
    pages = "105006",
    year = "1999"
}

@article{Cardoso:2005vb,
    author = "Cardoso, Vitor and Cavaglia, Marco and Gualtieri, Leonardo",
    title = "{Black Hole Particle Emission in Higher-Dimensional Spacetimes}",
    eprint = "hep-th/0512002",
    archivePrefix = "arXiv",
    doi = "10.1103/PhysRevLett.96.071301",
    journal = "Phys. Rev. Lett.",
    volume = "96",
    pages = "071301",
    year = "2006",
    note = "[Erratum: Phys.Rev.Lett. 96, 219902 (2006)]"
}

@article{Kanti:2014dxa,
    author = "Kanti, Panagiota and Pappas, Thomas and Pappas, Nikolaos",
    title = "{Greybody factors for scalar fields emitted by a higher-dimensional Schwarzschild{\textendash}de Sitter black hole}",
    eprint = "1409.8664",
    archivePrefix = "arXiv",
    primaryClass = "hep-th",
    doi = "10.1103/PhysRevD.90.124077",
    journal = "Phys. Rev. D",
    volume = "90",
    number = "12",
    pages = "124077",
    year = "2014"
}

@article{FCC:2018vvp,
    author = "Abada, A. and others",
    collaboration = "FCC",
    title = "{FCC-hh: The Hadron Collider}: {Future Circular Collider Conceptual Design Report Volume 3}",
    reportNumber = "CERN-ACC-2018-0058",
    doi = "10.1140/epjst/e2019-900087-0",
    journal = "Eur. Phys. J. ST",
    volume = "228",
    number = "4",
    pages = "755--1107",
    year = "2019"
}

@article{Benedikt:2022kan,
    author = "Benedikt, M. and others",
    title = "{Future Circular Hadron Collider FCC-hh: Overview and Status}",
    journal = {},
    eprint = "2203.07804",
    archivePrefix = "arXiv",
    primaryClass = "physics.acc-ph",
    reportNumber = "FERMILAB-CONF-22-182-AD",
    month = "3",
    year = "2022",
    note = "{arXiv:2203.07804 [physics.acc-ph]}"
}

@article{Tegmark:1996bz,
    author = "Tegmark, Max and Taylor, Andy and Heavens, Alan",
    title = "{Karhunen-Loeve eigenvalue problems in cosmology: How should we tackle large data sets?}",
    eprint = "astro-ph/9603021",
    archivePrefix = "arXiv",
    doi = "10.1086/303939",
    journal = "Astrophys. J.",
    volume = "480",
    pages = "22",
    year = "1997"
}

@article{Cutler:1994ys,
    author = "Cutler, Curt and Flanagan, Eanna E.",
    title = "{Gravitational waves from merging compact binaries: How accurately can one extract the binary's parameters from the inspiral wave form?}",
    eprint = "gr-qc/9402014",
    archivePrefix = "arXiv",
    reportNumber = "GRP-369",
    doi = "10.1103/PhysRevD.49.2658",
    journal = "Phys. Rev. D",
    volume = "49",
    pages = "2658--2697",
    year = "1994"
}

@article{Alexandre:2024nuo,
    author = "Alexandre, Ana and Dvali, Gia and Koutsangelas, Emmanouil",
    title = "{New mass window for primordial black holes as dark matter from the memory burden effect}",
    eprint = "2402.14069",
    archivePrefix = "arXiv",
    primaryClass = "hep-ph",
    doi = "10.1103/PhysRevD.110.036004",
    journal = "Phys. Rev. D",
    volume = "110",
    number = "3",
    pages = "036004",
    year = "2024"
}

@article{Saito:2008jc,
    author = "Saito, Ryo and Yokoyama, Jun'ichi",
    title = "{Gravitational wave background as a probe of the primordial black hole abundance}",
    eprint = "0812.4339",
    archivePrefix = "arXiv",
    primaryClass = "astro-ph",
    reportNumber = "RESCEU-63-08",
    doi = "10.1103/PhysRevLett.102.161101",
    journal = "Phys. Rev. Lett.",
    volume = "102",
    pages = "161101",
    year = "2009",
    note = "[Erratum: Phys.Rev.Lett. 107, 069901 (2011)]"
}

@article{Ando:2010zz,
    author = "Ando, Masaki and Ishidoshiro, Koji and Yamamoto, Kazuhiro and Yagi, Kent and Kokuyama, Wataru and Tsubono, Kimio and Takamori, Akiteru",
    title = "{Torsion-Bar Antenna for Low-Frequency Gravitational-Wave Observations}",
    doi = "10.1103/PhysRevLett.105.161101",
    journal = "Phys. Rev. Lett.",
    volume = "105",
    pages = "161101",
    year = "2010"
}

@article{Corbin:2005ny,
    author = "Corbin, Vincent and Cornish, Neil J.",
    title = "{Detecting the cosmic gravitational wave background with the big bang observer}",
    eprint = "gr-qc/0512039",
    archivePrefix = "arXiv",
    doi = "10.1088/0264-9381/23/7/014",
    journal = "Class. Quant. Grav.",
    volume = "23",
    pages = "2435--2446",
    year = "2006"
}

@article{Yoshino:2005hi,
    author = "Yoshino, Hirotaka and Rychkov, Vyacheslav S.",
    title = "{Improved analysis of black hole formation in high-energy particle collisions}",
    eprint = "hep-th/0503171",
    archivePrefix = "arXiv",
    reportNumber = "DPNU-05-01, ITFA-2005-09",
    doi = "10.1103/PhysRevD.71.104028",
    journal = "Phys. Rev. D",
    volume = "71",
    pages = "104028",
    year = "2005",
    note = "[Erratum: Phys.Rev.D 77, 089905 (2008)]"
}

@article{Emparan:2008eg,
    author = "Emparan, Roberto and Reall, Harvey S.",
    title = "{Black Holes in Higher Dimensions}",
    eprint = "0801.3471",
    archivePrefix = "arXiv",
    primaryClass = "hep-th",
    doi = "10.12942/lrr-2008-6",
    journal = "Living Rev. Rel.",
    volume = "11",
    pages = "6",
    year = "2008"
}

@article{Ida:2002ez,
    author = "Ida, Daisuke and Oda, Kin-ya and Park, Seong Chan",
    title = "{Rotating black holes at future colliders: Greybody factors for brane fields}",
    eprint = "hep-th/0212108",
    archivePrefix = "arXiv",
    reportNumber = "KEK-TH-828, KIAS-P02074, TUM-HEP-429-02, UT-ICEPP-02-07",
    doi = "10.1103/PhysRevD.67.064025",
    journal = "Phys. Rev. D",
    volume = "67",
    pages = "064025",
    year = "2003",
    note = "[Erratum: Phys.Rev.D 69, 049901 (2004)]"
}

@article{Casals:2006xp,
    author = "Casals, Marc and Dolan, S. R. and Kanti, P. and Winstanley, E.",
    title = "{Brane Decay of a (4+n)-Dimensional Rotating Black Hole. III. Spin-1/2 particles}",
    eprint = "hep-th/0608193",
    archivePrefix = "arXiv",
    doi = "10.1088/1126-6708/2007/03/019",
    journal = "JHEP",
    volume = "03",
    pages = "019",
    year = "2007"
}

@article{Maartens:2010ar,
    author = "Maartens, Roy and Koyama, Kazuya",
    title = "{Brane-World Gravity}",
    eprint = "1004.3962",
    archivePrefix = "arXiv",
    primaryClass = "hep-th",
    doi = "10.12942/lrr-2010-5",
    journal = "Living Rev. Rel.",
    volume = "13",
    pages = "5",
    year = "2010"
}

@article{Fierz:1939ix,
    author = "Fierz, M. and Pauli, W.",
    title = "{On relativistic wave equations for particles of arbitrary spin in an electromagnetic field}",
    doi = "10.1098/rspa.1939.0140",
    journal = "Proc. Roy. Soc. Lond. A",
    volume = "173",
    pages = "211--232",
    year = "1939"
}

@article{Hinterbichler:2011tt,
    author = "Hinterbichler, Kurt",
    title = "{Theoretical Aspects of Massive Gravity}",
    eprint = "1105.3735",
    archivePrefix = "arXiv",
    primaryClass = "hep-th",
    doi = "10.1103/RevModPhys.84.671",
    journal = "Rev. Mod. Phys.",
    volume = "84",
    pages = "671--710",
    year = "2012"
}

@article{deRham:2014zqa,
    author = "de Rham, Claudia",
    title = "{Massive Gravity}",
    eprint = "1401.4173",
    archivePrefix = "arXiv",
    primaryClass = "hep-th",
    doi = "10.12942/lrr-2014-7",
    journal = "Living Rev. Rel.",
    volume = "17",
    pages = "7",
    year = "2014"
}

@article{Kohri:2018awv,
    author = "Kohri, Kazunori and Terada, Takahiro",
    title = "{Semianalytic calculation of gravitational wave spectrum nonlinearly induced from primordial curvature perturbations}",
    eprint = "1804.08577",
    archivePrefix = "arXiv",
    primaryClass = "gr-qc",
    reportNumber = "KEK-TH-2046, KEK-COSMO-223",
    doi = "10.1103/PhysRevD.97.123532",
    journal = "Phys. Rev. D",
    volume = "97",
    number = "12",
    pages = "123532",
    year = "2018"
}

@article{Inomata:2019ivs,
    author = "Inomata, Keisuke and Kohri, Kazunori and Nakama, Tomohiro and Terada, Takahiro",
    title = "{Enhancement of Gravitational Waves Induced by Scalar Perturbations due to a Sudden Transition from an Early Matter Era to the Radiation Era}",
    eprint = "1904.12879",
    archivePrefix = "arXiv",
    primaryClass = "astro-ph.CO",
    reportNumber = "IPMU 19-0067, KEK-TH-2122, KEK-Cosmo-237",
    doi = "10.1103/PhysRevD.108.049901",
    journal = "Phys. Rev. D",
    volume = "100",
    pages = "043532",
    year = "2019",
    note = "[Erratum: Phys.Rev.D 108, 049901 (2023)]"
}

@article{Pi:2020otn,
    author = "Pi, Shi and Sasaki, Misao",
    title = "{Gravitational Waves Induced by Scalar Perturbations with a Lognormal Peak}",
    eprint = "2005.12306",
    archivePrefix = "arXiv",
    primaryClass = "gr-qc",
    reportNumber = "YITP-20-75, YITP-75, IPMU20-0054",
    doi = "10.1088/1475-7516/2020/09/037",
    journal = "JCAP",
    volume = "09",
    pages = "037",
    year = "2020"
}

@article{Nishizawa:2009bf,
    author = "Nishizawa, Atsushi and Taruya, Atsushi and Hayama, Kazuhiro and Kawamura, Seiji and Sakagami, Masa-aki",
    title = "{Probing non-tensorial polarizations of stochastic gravitational-wave backgrounds with ground-based laser interferometers}",
    eprint = "0903.0528",
    archivePrefix = "arXiv",
    primaryClass = "astro-ph.CO",
    doi = "10.1103/PhysRevD.79.082002",
    journal = "Phys. Rev. D",
    volume = "79",
    pages = "082002",
    year = "2009"
}

@article{EPTA:2023fyk,
    author = "Antoniadis, J. and others",
    collaboration = "EPTA, InPTA:",
    title = "{The second data release from the European Pulsar Timing Array - III. Search for gravitational wave signals}",
    eprint = "2306.16214",
    archivePrefix = "arXiv",
    primaryClass = "astro-ph.HE",
    doi = "10.1051/0004-6361/202346844",
    journal = "Astron. Astrophys.",
    volume = "678",
    pages = "A50",
    year = "2023"
}

@article{Robson:2018ifk,
    author = "Robson, Travis and Cornish, Neil J. and Liu, Chang",
    title = "{The construction and use of LISA sensitivity curves}",
    eprint = "1803.01944",
    archivePrefix = "arXiv",
    primaryClass = "astro-ph.HE",
    doi = "10.1088/1361-6382/ab1101",
    journal = "Class. Quant. Grav.",
    volume = "36",
    number = "10",
    pages = "105011",
    year = "2019"
}

@article{Yagi:2011wg,
    author = "Yagi, Kent and Seto, Naoki",
    title = "{Detector configuration of DECIGO/BBO and identification of cosmological neutron-star binaries}",
    eprint = "1101.3940",
    archivePrefix = "arXiv",
    primaryClass = "astro-ph.CO",
    doi = "10.1103/PhysRevD.83.044011",
    journal = "Phys. Rev. D",
    volume = "83",
    pages = "044011",
    year = "2011",
    note = "[Erratum: Phys.Rev.D 95, 109901 (2017)]"
}

@article{Thrane:2013oya,
    author = "Thrane, Eric and Romano, Joseph D.",
    title = "{Sensitivity curves for searches for gravitational-wave backgrounds}",
    eprint = "1310.5300",
    archivePrefix = "arXiv",
    primaryClass = "astro-ph.IM",
    doi = "10.1103/PhysRevD.88.124032",
    journal = "Phys. Rev. D",
    volume = "88",
    number = "12",
    pages = "124032",
    year = "2013"
}

@article{Gibbons:1977mu,
    author = "Gibbons, G. W. and Hawking, S. W.",
    title = "{Cosmological Event Horizons, Thermodynamics, and Particle Creation}",
    doi = "10.1103/PhysRevD.15.2738",
    journal = "Phys. Rev. D",
    volume = "15",
    pages = "2738--2751",
    year = "1977"
}


\end{document}